\documentclass[prl,aps,twocolumn,preprintnumbers,amsmath,amssymb,superscriptaddress,longbibliography]{revtex4-2}
\usepackage{bibunits}
\usepackage{ulem} 
\usepackage{amsmath,amssymb,bbm,mathrsfs,mathtools,multirow,diagbox,xcolor,%
graphicx,tabularx,comment,amsfonts,dsfont,lettrine,units,comment}
\usepackage{subfiles}
\usepackage{graphicx}
\usepackage{bookmark}
\usepackage{xcolor}
\usepackage{cases}
\usepackage{chngcntr}
\usepackage{nicefrac}
\usepackage{bm}
\usepackage{bbm}
\usepackage{cancel} 
\usepackage{orcidlink}
\usepackage{multirow}
\usepackage{orcidlink}

\defaultbibliographystyle{apsrev4-2}
\defaultbibliography{bib-file.bib}

\usepackage{enumerate}
\usepackage{wasysym}
\usepackage{stmaryrd}
\usepackage{microtype}

\usepackage{tikz}
\usepackage{tikzit}
\usetikzlibrary{backgrounds}

\tikzstyle{Green Node}=[fill={zx_green}, inner sep=0mm, minimum size=2mm, draw=black, shape=circle, tikzit fill=green, font={\footnotesize\boldmath}]
\tikzstyle{Red Node}=[fill={zx_red}, inner sep=0mm, minimum size=2mm, draw=black, shape=circle, tikzit fill=red, font={\footnotesize\boldmath}]
\tikzstyle{H}=[fill=yellow, draw=black, shape=rectangle, inner sep=0.6mm, minimum height=1.5mm, minimum width=1.5mm]
\tikzstyle{new style 0}=[fill={rgb,255: red,255; green,0; blue,4}, draw=black, shape=rectangle]
\tikzstyle{new style 1}=[fill=green, draw=black, shape=rectangle]
\tikzstyle{swirl}=[fill=white, draw=black, shape=circle]
\tikzstyle{new style 2}=[fill=black, draw=black, shape=circle]
\tikzstyle{gphase}=[rounded rectangle, rounded rectangle arc length=90, fill={zx_green}, inner sep=2pt]
\tikzstyle{rphase}=[rounded rectangle, rounded rectangle arc length=90, fill={zx_red}, inner sep=2pt]
\tikzstyle{box}=[shape=rectangle, text height=1.5ex, text depth=0.25ex, yshift=0.5mm, fill=white, draw=black, minimum height=5mm, yshift=-0.5mm, minimum width=5mm, font={\small}]
\tikzstyle{Z dot}=[Green Node, tikzit fill=green]
\tikzstyle{Z phase dot}=[minimum size=4mm, font={\footnotesize\boldmath}, shape=rectangle, rounded corners=1.5mm, inner sep=0.2mm, outer sep=-2mm, scale=0.8, tikzit shape=circle, draw=black, fill={zx_green}, tikzit draw=blue, tikzit fill=green]
\tikzstyle{X phase dot}=[Z phase dot, tikzit shape=circle, fill={zx_red}, font={\footnotesize\boldmath}, tikzit draw=blue, tikzit fill=red]
\tikzstyle{X dot}=[Red Node, tikzit fill=red]
\tikzstyle{hadamard}=[H, tikzit shape=rectangle, tikzit fill=yellow]
\tikzstyle{vertex}=[inner sep=0mm, minimum size=1mm, shape=circle, draw=black, fill=black]
\tikzstyle{vertex set}=[inner sep=0mm, minimum size=1mm, shape=circle, draw=black, fill=white, font={\footnotesize\boldmath}]

\tikzstyle{Edge}=[-]
\tikzstyle{new edge style 0}=[draw=black, {|->}]
\tikzstyle{new edge style 1}=[-, draw={rgb,255: red,191; green,0; blue,64}]
\tikzstyle{brace edge}=[-, tikzit draw=blue, decorate, decoration={brace,amplitude=1mm,raise=-1mm}]
\tikzstyle{hadamard edge}=[-, color=blue, dashed, dash pattern=on 3pt off 1.5pt, thick, tikzit draw=blue]
\tikzstyle{dashed gray}=[-, dashed, tikzit fill={rgb,255: red,191; green,191; blue,191}, fill={rgb,255: red,191; green,191; blue,191}, draw={rgb,255: red,128; green,128; blue,128}]

\input{ZX.tikzdefs}

\usepackage{hyperref}
\hypersetup{
    colorlinks=true,
    linkcolor=magenta,
    filecolor=magenta,      
    urlcolor=magenta,
    citecolor=magenta,
    pdftitle={xmastree},    
    pdfauthor={},     
    hypertexnames=false
}

\usepackage{amsthm}
\DeclareMathOperator{\Tr}{Tr}

\theoremstyle{definition}

\usepackage{amsmath}

\usepackage{amsthm}

\usepackage{comment}

\usepackage{bm}

\makeatletter
\renewcommand*\env@matrix[1][*\c@MaxMatrixCols c]{%
  \hskip -\arraycolsep
  \let\@ifnextchar\new@ifnextchar
  \array{#1}}
\makeatother

\begin{document}


\title{A graphical diagnostic of topological order using ZX calculus}

\author{Sergi Mas-Mendoza}
\affiliation{Univ. Grenoble Alpes, CNRS, Grenoble INP, Institut N\'eel, 38000 Grenoble, France}
\affiliation{Donostia International Physics Center (DIPC),
Paseo Manuel de Lardiz\'{a}bal 4, 20018, Donostia-San Sebasti\'{a}n, Spain}

\author{Richard D. P. East}%
\affiliation{Haiqu, Inc., 95 Third Street, San Francisco, CA 94103, USA%
}%

\author{Michele Filippone\orcidlink{0000-0003-1102-8335}}%
\affiliation{Univ. Grenoble Alpes, CEA, IRIG-MEM-L\_Sim, Grenoble, France
}%

\author{Adolfo G. Grushin\orcidlink{0000-0001-7678-7100}}
\affiliation{Univ. Grenoble Alpes, CNRS, Grenoble INP, Institut N\'eel, 38000 Grenoble, France}
\affiliation{Donostia International Physics Center (DIPC),
Paseo Manuel de Lardiz\'{a}bal 4, 20018, Donostia-San Sebasti\'{a}n, Spain}
\affiliation{IKERBASQUE, Basque Foundation for Science, Maria Diaz de Haro 3, 48013 Bilbao, Spain}%

\begin{abstract}
Establishing a universal diagnostic of topological order remains an open theoretical challenge.   
In particular, diagnosing long-range entanglement through the entropic area law suffers from spurious contributions, failing to unambiguously identify topological order.
Here we devise a protocol based on the ZX calculus, a graphical tensor network, to determine the topological order of a state circumventing entropy calculations.
The protocol takes as input real-space bipartitions of a state and returns a ZX  \textit{contour diagram}, $\mathcal{D}_{\partial A}$, displaying long-range graph connectivity only for long-range entangled states.
We validate the protocol by showing that the contour diagrams of the toric and color codes are equivalent except for the number of non-local nodes, which differentiates their topological order.
The number of these nodes is robust to the choice of the boundary and ground-state superposition, and they are absent for trivial  states, even those with spurious entropy contributions.
Our results single out ZX calculus as a tool to detect topological long-range entanglement by leveraging the advantages of diagrammatic reasoning against entropic diagnostics.
\end{abstract}

\maketitle


\textit{Introduction} -- 
Classifying quantum many-body phases of matter beyond symmetry-breaking
requires discerning different types of topological order~\cite{wen2004quantum}. 
Diagnosing topological order is often achieved by calculating the entanglement entropy between subsystems, $S$~\cite{hamma2005ground,LevinWen2006,KitaevPreskill2006,chen2010local}.
Topologically-ordered gapped systems have long-range entanglement. Thus, they satisfy the area law $S =\alpha L -\gamma$, with $L$  the length of the boundary between partitions, and $\alpha,\gamma > 0$  real constants.
Finding $\gamma\neq0$ points to
long-range entanglement because it is independent of $L$. 

Unfortunately, a non-zero $\gamma$
is not sufficient to declare that a phase is topologically ordered. 
Typically, $\gamma=\gamma_{\mathrm{top}}= \ln(D)$, a topological entanglement entropy set by the total quantum dimension $D$ of the anyons~\cite{hamma2005ground,LevinWen2006,KitaevPreskill2006}. 
This observation served to diagnose topological order, both numerically~\cite{Jiang2012,Cincio2013,Zaletel2013,Grushin2015,laflorencie2016quantum} and experimentally~\cite{Satzinger2021}.
However, a growing body of examples showcases that $\gamma$  suffers from spurious, non-topological contributions $\gamma_{\mathrm{spur}}$~\cite{Cano2015,Bravyi,Zou2016,Fliss2017,Santos2018,Williamson2019,Kato2020,Stephen2019}.
Applying finite depth unitary circuits can change $\gamma=\gamma_{\mathrm{top}}$ to $\gamma=\gamma_{\mathrm{top}}+\gamma_{\mathrm{spur}}$, and
even topologically trivial states can display $\gamma=\gamma_{\mathrm{spur}}\neq 0$~\cite{Cano2015,Bravyi,Zou2016,Fliss2017,Santos2018,Williamson2019,Kato2020,Stephen2019}. 
The spurious contribution $\gamma_{\mathrm{spur}}$ depends on the computation method used,
emerging typically, but not exclusively~\cite{Kato2020}, from sub-system symmetries.
A full understanding of when $\gamma_{\mathrm{spur}}\neq0$ is to our knowledge lacking,
even though it was shown that $\gamma\geq \gamma_{\mathrm{top}}$ for certain computation schemes~\cite{Kim2023,Levin2024}.
The  non-universality of $\gamma$ calls for
alternative and robust diagnostics of topological order.
\begin{figure*}
    \centering
    \hspace{-18cm}
    \includegraphics[width=\linewidth]{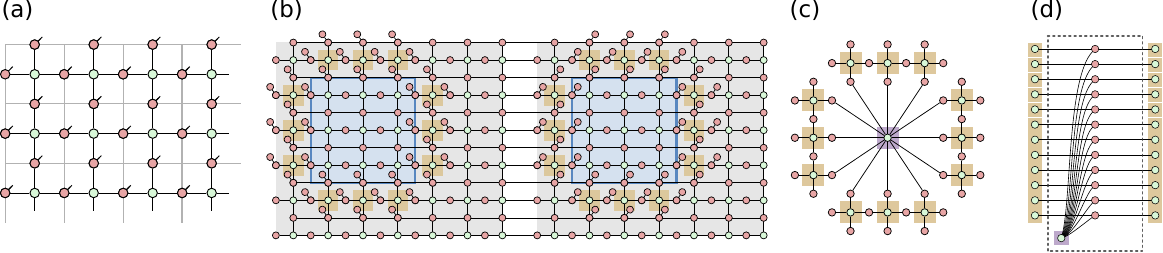}
\caption{
Protocol to find the contour diagram $\mathcal{D}_{\partial A}$ for the toric code on the square lattice.
(a) ZX diagram representing the toric-code ground state~\eqref{eq:groundstate00_toriccode_main}.
The qubits are marked by red spiders with outputs (\begin{tikzpicture}
	\begin{pgfonlayer}{nodelayer}
		\protect\node [style=X dot] (0) at (0, 0) {};
		\protect\node [style=none] (1) at (.3, .3) {};
	\end{pgfonlayer}
	\begin{pgfonlayer}{edgelayer}
		\protect\draw (0) to (1);
		\end{pgfonlayer}
\end{tikzpicture}) and sit on the edges of a square lattice shown in gray. The black lines denote ZX connections issued from the Pauli projectors~\eqref{eq:ZX_projetor}.
(b) ZX diagram representing $\mathrm{Tr}(\rho)$. It implements the trace of all qubits in (a) with their conjugate copies. 
Using the inverse fuse rule, we double (\textit{pin})  qubits with ZX connections to both regions $A$  and $B$, highlighted in blue and  gray respectively.
The orange squares highlight $\Sigma_{AB}=12$ green Z spiders  corresponding to Pauli projectors acting on the plaquettes connecting $A$ and $B$.
(c) Contour diagram $\mathcal D_{\partial A}$ obtained from simplifying the diagram (b), but without simplifying the pinned red X spiders at the boundary.
It features a single non-local spider ($N_{\mathrm{nl}}=1$)  connected to the  boundary plaquettes, whose number $\Sigma_{AB}$ scales  with the boundary length. 
(d) The contour diagram $\mathcal D_{\partial A}$ in panel (c) is simplified further to obtain a Pauli tree. It features an extension of the Pauli projector~\eqref{eq:ZX_projetor}, delimited by the  dotted box, acting on the $\Sigma_{AB}$ green Z spiders of each copy marked in orange in panel (b).
} 
    \label{fig:Fig2_TC}
\end{figure*}

In this work, we show the possibility to graphically identify topological long-range entanglement, bypassing the calculation of entanglement entropies $S$.
We propose a graphical diagnostic of topological order based on 
diagrammatic ZX calculus~\cite{coecke2008interacting,coecke2011interacting,CKbook,vandewetering2020zxcalculus}.
ZX diagrams are a type of tensor network that excels at visualizing entanglement.
By expressing quantum circuits as ZX diagrams, one can use graphical simplification rules, the ZX calculus, to find more compact,  equivalent circuits with fewer gates.
In these simplified diagrams only entangled qubits are graphically connected.
ZX-calculus reasoning has far-reaching applications 
in quantum information and circuit optimization~\cite{duncan2019graph,kissinger2020reducing,deBeaudrap2020Techniques,Cowtan2020phasegadget}, measurement-based quantum computation~\cite{duncan2010rewriting,kissinger2017MBQC,Backens2020extraction}  and error correction~\cite{horsman2017surgery,magicFactories,deBeaudrap2020Paulifusion,hanks2019effective,rodatz2024floquetifying,khesin2024equivalenceclassesquantumerrorcorrecting,bombin2023faulttolerant,khesin2023graphical,Bombin2024unifyingflavorsof}. It is used to study symmetry-protected topological phases \cite{east2022aklt}, loop quantum gravity \cite{east2021spin,east2022physique}, entanglement propagation~\cite{sommers2024zerotemperature} and deep-reinforcement learning~\cite{nagele2024optimizing}.

Here, we devise a ZX-based protocol that unambiguously  diagnoses topological long-range entanglement of  gapped states and is insensitive to spurious contributions.
Our protocol performs a graphical trace over \textit{all} degrees of freedom, delivering a \textit{number} whose ZX diagrammatic form, the \textit{contour diagram} $\mathcal{D}_{\partial A}$, informs  us about the entanglement between a chosen bipartition. 
We then conjecture that $\mathcal{D}_{\partial A}$ displays long-range graph connectivity if there is long-range entanglement.
We test this conjecture by computing $\mathcal{D}_{\partial A}$ for the toric code on  square  and hexagonal lattices~\cite{Kitaev2003} and for the color code~\cite{Bombin_colorcode,lacroix2024scalinglogiccolorcode}. 
In all these cases, $\mathcal{D}_{\partial A}$ displays a total number of non-local nodes that equals
$\gamma_\mathrm{top}$,
and can be cast into a \textit{Pauli-tree} form defined by the 
presence of Pauli projectors~\cite{KissingerWetering2024Book}.
Crucially, $\mathcal{D}_{\partial A}$ shows only local connectivity for trivial cluster states, even those with $\gamma_\mathrm{spur}>0$. 
We show that the contour diagram $\mathcal D_{\partial A}$ is robust to the choice of ground-state superposition, boundary deformation,  bipartition and to applying finite-depth unitaries.
Hence, our results put forward ZX diagrams as an advantageous tool to identify topological order.
%


\textit{Toric code: contour diagram $\mathcal{D}_{\partial A}$ and Pauli tree} -- 
To illustrate our method, we start by deriving the diagramatic representation of Kitaev's toric-code ground state on the square lattice~\cite{Kitaev2003}
\begin{equation}
    \label{eq:groundstate00_toriccode_main}
    \ket{\mathrm{GS}}_\mathrm{tc} = \frac{1}{\mathcal{N}}\prod_{\square} (I + P_{\square})\ket{0}^{\otimes N},
\end{equation}
where $N$ is the number of qubits, $\mathcal{N}$  the normalization constant and $P_{\square}{=\prod_{i \in \square} \sigma_{x}^{(i)}}$ acts  with four Pauli matrices $\sigma_{x}$ on each qubit $i$ at the edges of a plaquette $\square$, see Fig.~\ref{fig:Fig2_TC}a.
As the $P_{\square}$ commute with each other and leave invariant the state Eq.~\eqref{eq:groundstate00_toriccode_main}, they are stabilizer operators~\cite{Simon_TopologicalQuantum}.
Because the toric code is topologically ordered, the ground state is four-fold degenerate on a torus.
%

%
%

The building blocks of  ZX calculus are green Z and red X spiders defined, respectively, as
\begin{subequations} \label{eq:spiders}
\begin{equation}
\label{eq:greenspider}
\begin{tikzpicture}
	\begin{pgfonlayer}{nodelayer}
		\node [style=Z phase dot] (0) at (0, 0) {$\alpha$};
		\node [style=none] (1) at (1.25, 1) {};
		\node [style=none] (2) at (-1.25, 1) {};
		\node [style=none] (3) at (-1.25, -1) {};
		\node [style=none] (4) at (1.25, -1) {};
		\node [style=none] (5) at (1.25, 0.5) {};
		\node [style=none] (6) at (-1.25, 0.5) {};
		\node [style=none, rotate=90] (7) at (-1, -0.25) {...};
		\node [style=none, rotate=90] (8) at (1, -0.25) {...};
		\node [style=none] (18) at (-2, 1.25) {};
		\node [style=none] (19) at (-2, -1.25) {};
		\node [style=none] (20) at (-2.2, -1.25) {$_m$};
		\node [style=none] (21) at (2, 1.25) {};
		\node [style=none] (22) at (2, -1.25) {};
		\node [style=none] (23) at (2.2, -1.25) {$_n$};
	\end{pgfonlayer}
	\begin{pgfonlayer}{edgelayer}
		\draw [in=-141, out=0, looseness=0.75] (3.center) to (0);
		\draw [in=180, out=-39, looseness=0.75] (0) to (4.center);
		\draw [in=180, out=22, looseness=0.75] (0) to (5.center);
		\draw [in=180, out=39, looseness=0.75] (0) to (1.center);
		\draw [in=0, out=158, looseness=0.75] (0) to (6.center);
		\draw [in=141, out=0, looseness=0.75] (2.center) to (0);
		\draw [decorate, style=brace edge] (19.center) to (18.center);
		\draw [decorate, style=brace edge] (21.center) to (22.center);
	\end{pgfonlayer}
\end{tikzpicture}
= \ket{0}^{\otimes n}\bra{0}^{\otimes m} + e^{i\alpha}\ket{1}^{\otimes n}\bra{1}^{\otimes m},
\\
\end{equation}
\begin{equation}
\label{eq:Xspider}
\begin{tikzpicture}
	\begin{pgfonlayer}{nodelayer}
		\node [style=X phase dot] (0) at (0, 0) {$\alpha$};
		\node [style=none] (1) at (1.25, 1) {};
		\node [style=none] (2) at (-1.25, 1) {};
		\node [style=none] (3) at (-1.25, -1) {};
		\node [style=none] (4) at (1.25, -1) {};
		\node [style=none] (5) at (1.25, 0.5) {};
		\node [style=none] (6) at (-1.25, 0.5) {};
		\node [style=none, rotate=90] (7) at (-1, -0.25) {...};
		\node [style=none, rotate=90] (8) at (1, -0.25) {...};
		\node [style=none] (18) at (-2, 1.25) {};
		\node [style=none] (19) at (-2, -1.25) {};
		\node [style=none] (20) at (-2.2, -1.25) {$_m$};
		\node [style=none] (21) at (2, 1.25) {};
		\node [style=none] (22) at (2, -1.25) {};
		\node [style=none] (23) at (2.2, -1.25) {$_n$};
	\end{pgfonlayer}
	\begin{pgfonlayer}{edgelayer}
		\draw [in=-141, out=0, looseness=0.75] (3.center) to (0);
		\draw [in=180, out=-39, looseness=0.75] (0) to (4.center);
		\draw [in=180, out=22, looseness=0.75] (0) to (5.center);
		\draw [in=180, out=39, looseness=0.75] (0) to (1.center);
		\draw [in=0, out=158, looseness=0.75] (0) to (6.center);
		\draw [in=141, out=0, looseness=0.75] (2.center) to (0);
		\draw [decorate, style=brace edge] (19.center) to (18.center);
		\draw [decorate, style=brace edge] (21.center) to (22.center);
	\end{pgfonlayer}
\end{tikzpicture}
= \ket{+}^{\otimes n}\bra{+}^{\otimes m} + e^{i\alpha}\ket{-}^{\otimes n}\bra{-}^{\otimes m}
\end{equation}
\end{subequations}
The spiders define linear maps between $m$ input and $n$ output qubits, with $\ket{\pm} = \left(\ket{0} \pm \ket{1}\right)/\sqrt2$ and $\alpha \in [0,2\pi)$.
Spiders can be used to denote qubit states as
\begin{tikzpicture}
        \begin{pgfonlayer}{nodelayer}
            \node [style=Z phase dot] (0) at (-0.5, 0) {$\alpha$};
            \node [style=none] (1) at (0.5, 0) {};
        \end{pgfonlayer}
        \begin{pgfonlayer}{edgelayer}
            \draw (0) to (1.center);
        \end{pgfonlayer}
    \end{tikzpicture}~
$=\ket0+e^{i\alpha}\ket 1\,$ and
\begin{tikzpicture}
        \begin{pgfonlayer}{nodelayer}
            \node [style=X phase dot] (0) at (-0.5, 0) {$\alpha$};
            \node [style=none] (1) at (0.5, 0) {};
        \end{pgfonlayer}
        \begin{pgfonlayer}{edgelayer}
            \draw (0) to (1.center);
        \end{pgfonlayer}
\end{tikzpicture}
$~=\ket++e^{i\alpha}\ket -$. 
A spider without a phase assumes $\alpha=0$. Spiders with no legs encode multiplicative scalars. Unless otherwise noted, we will ignore such scalars, as they are fixed by normalization~\cite{SM,KissingerWetering2024Book}.
In the Supplemental Material (SM)~\cite{SM}, we offer a brief introduction to  ZX calculus, see also Refs.~\cite{vandewetering2020zxcalculus,KissingerWetering2024Book}.

To represent the state~\eqref{eq:groundstate00_toriccode_main} as a ZX diagram, we first superpose the identity with the plaquette operator $P_\square$ in order to build $(I + P_{\square})$. 
To do so, we represent $\sigma_{x}$ as 
\begin{tikzpicture}
	\begin{pgfonlayer}{nodelayer}
		\node [style=X phase dot] (0) at (0, 0) {$\pi$};
		\node [style=none] (1) at (-.7, 0) {};
		\node [style=none] (2) at (.7, 0) {};
	\end{pgfonlayer}
	\begin{pgfonlayer}{edgelayer}
		\draw (1.center) to (0);
		\draw (0) to (2.center);
	\end{pgfonlayer}
\end{tikzpicture}
and the identity as a spiderless wire 
\begin{tikzpicture}
	\begin{pgfonlayer}{nodelayer}
		\node [style=none] (0) at (0, 0) {};
		\node [style=none] (1) at (1, 0) {};
	\end{pgfonlayer}
	\begin{pgfonlayer}{edgelayer}
		\draw (0.center) to (1.center);
	\end{pgfonlayer}
\end{tikzpicture}
, which equals
\begin{tikzpicture}
	\begin{pgfonlayer}{nodelayer}
		\node [style=X  dot] (0) at (0, 0) {};
		\node [style=none] (1) at (-.7, 0) {};
		\node [style=none] (2) at (.7, 0) {};
	\end{pgfonlayer}
	\begin{pgfonlayer}{edgelayer}
		\draw (1.center) to (0);
		\draw (0) to (2.center);
	\end{pgfonlayer}
\end{tikzpicture}
$=$
\begin{tikzpicture}
	\begin{pgfonlayer}{nodelayer}
		\node [style=Z  dot] (0) at (0, 0) {};
		\node [style=none] (1) at (-.7, 0) {};
		\node [style=none] (2) at (.7, 0) {};
	\end{pgfonlayer}
	\begin{pgfonlayer}{edgelayer}
		\draw (1.center) to (0);
		\draw (0) to (2.center);
	\end{pgfonlayer}
\end{tikzpicture}
, see Eq.~\eqref{eq:spiders}.
As we illustrate in the SM~\cite{SM}, by applying (\textit i) the ZX equality
\begin{tikzpicture}
        \begin{pgfonlayer}{nodelayer}
            \node [style=X dot] (0) at (-0.5, 0) {};
            \node [style=none] (1) at (0.5, 0) {};
        \end{pgfonlayer}
        \begin{pgfonlayer}{edgelayer}
            \draw (0) to (1.center);
        \end{pgfonlayer}
    \end{tikzpicture}%
    \    \label{eq:ket-+} +
\begin{tikzpicture}
        \begin{pgfonlayer}{nodelayer}
            \node [style=X phase dot] (0) at (-0.5, 0) {$\pi$};
            \node [style=none] (1) at (0.5, 0) {};
        \end{pgfonlayer}
        \begin{pgfonlayer}{edgelayer}
            \draw (0) to (1.center);
        \end{pgfonlayer}
    \end{tikzpicture}%
    \     \label{eq:ket-+}
 $=$
\begin{tikzpicture}
        \begin{pgfonlayer}{nodelayer}
            \node [style=Z dot] (0) at (-0.5, 0) {};
            \node [style=none] (1) at (0.5, 0) {};
        \end{pgfonlayer}
        \begin{pgfonlayer}{edgelayer}
            \draw (0) to (1.center);
        \end{pgfonlayer}
    \end{tikzpicture}%
    \ $=  \ket{0} + \ket{1}$, (\textit{ii}) the spider fusion rule of phases, e.g.
\begin{tikzpicture}
	\begin{pgfonlayer}{nodelayer}
		\node [style=X phase dot] (0) at (1, 0) {$\alpha$};
		\node [style=X phase dot] (1) at (2, 0) {$\beta$};
		\node [style=none] (2) at (2.75, 0) {};
		\node [style=none] (3) at (0.5, 0.5) {};
		\node [style=none] (4) at (0.5, -0.5) {};
		\node [style=none] (5) at (0.5, 0.05) {\rotatebox[origin=c]{90}{\tiny \dots}};
	\end{pgfonlayer}
	\begin{pgfonlayer}{edgelayer}
		\draw (3.center) to (0);
		\draw (0) to (4.center);
		\draw (0) to (1);
		\draw (1) to (2.center);
	\end{pgfonlayer}
\end{tikzpicture}
=
\begin{tikzpicture}
	\begin{pgfonlayer}{nodelayer}
		\node [style=X phase dot] (0) at (1, 0) {$\alpha$+{$\beta$}};
		\node [style=none] (1) at (1.75, 0) {};
		\node [style=none] (2) at (0.5, 0.5) {};
		\node [style=none] (3) at (0.5, -0.5) {};
		\node [style=none] (4) at (0.3, 0) {\rotatebox[origin=c]{90}{\tiny \dots}};
	\end{pgfonlayer}
	\begin{pgfonlayer}{edgelayer}
		\draw (2.center) to (0);
		\draw (0) to (3.center);
		\draw (0) to (1.center);
	\end{pgfonlayer}
\end{tikzpicture}
, and (\textit{iii}) the copy rule
\begin{tikzpicture}
	\begin{pgfonlayer}{nodelayer}
		\node [style=Z dot] (0) at (1.25, 0) {};
		\node [style=none] (1) at (2, -0.5) {};
		\node [style=none] (2) at (0.5, -0.5) {};
		\node [style=X phase dot] (3) at (0.5, 0.25) {$a\pi$};
		\node [style=none] (4) at (1.3, -0.45) {\tiny \dots};
	\end{pgfonlayer}
	\begin{pgfonlayer}{edgelayer}
		\draw (3) to (0);
		\draw [bend right=15] (0) to (2.center);
		\draw [bend left=15] (0) to (1.center);
	\end{pgfonlayer}
\end{tikzpicture} =
\begin{tikzpicture}
	\begin{pgfonlayer}{nodelayer}
		\node [style=none] (3) at (0.5, -0.5) {};
		\node [style=none] (4) at (-0.5, -0.5) {};
		\node [style=X phase dot] (5) at (-0.5, 0.25) {$a\pi$};
		\node [style=none] (6) at (0.05, -0.4) {\tiny \dots};
		\node [style=X phase dot] (8) at (0.5, 0.25) {$a\pi$};
	\end{pgfonlayer}
	\begin{pgfonlayer}{edgelayer}
		\draw (5) to (4.center);
		\draw (8) to (3.center);
	\end{pgfonlayer}
\end{tikzpicture}
, we obtain the desired superposition of operators
\begin{equation}
I + P_{\square}\,=
\begin{tikzpicture}
	\begin{pgfonlayer}{nodelayer}
		\node [style=none] (5) at (1.25, 0.5) {};
		\node [style=none] (6) at (0.75, -0.5) {};
		\node [style=none] (7) at (0.25, 1.5) {};
		\node [style=none] (8) at (-0.25, 0.5) {};
		\node [style=none] (9) at (-0.75, 0.5) {};
		\node [style=none] (10) at (-1.25, -0.5) {};
		\node [style=none] (11) at (-0.25, -1.5) {};
		\node [style=none] (12) at (0.25, -0.5) {};
	\end{pgfonlayer}
	\begin{pgfonlayer}{edgelayer}
		\draw (11.center) to (12.center);
		\draw (6.center) to (5.center);
		\draw (8.center) to (7.center);
		\draw (10.center) to (9.center);
	\end{pgfonlayer}
\end{tikzpicture}
+
\begin{tikzpicture}
	\begin{pgfonlayer}{nodelayer}
		\node [style=X phase dot] (0) at (0, 1) {$\pi$};
		\node [style=X phase dot] (1) at (-1, 0) {$\pi$};
		\node [style=X phase dot] (2) at (0, -1) {$\pi$};
		\node [style=X phase dot] (3) at (1, 0) {$\pi$};
		\node [style=none] (5) at (1.25, 0.5) {};
		\node [style=none] (6) at (0.75, -0.5) {};
		\node [style=none] (7) at (0.25, 1.5) {};
		\node [style=none] (8) at (-0.25, 0.5) {};
		\node [style=none] (9) at (-0.75, 0.5) {};
		\node [style=none] (10) at (-1.25, -0.5) {};
		\node [style=none] (11) at (-0.25, -1.5) {};
		\node [style=none] (12) at (0.25, -0.5) {};
	\end{pgfonlayer}
	\begin{pgfonlayer}{edgelayer}
		\draw (8.center) to (0);
		\draw (0) to (7.center);
		\draw (1) to (10.center);
		\draw (1) to (9.center);
		\draw (11.center) to (2);
		\draw (2) to (12.center);
		\draw (3) to (5.center);
		\draw (3) to (6.center);
	\end{pgfonlayer}
\end{tikzpicture}
=
\begin{tikzpicture}
	\begin{pgfonlayer}{nodelayer}
		\node [style=Red Node] (0) at (0, 1) {};
		\node [style=Red Node] (1) at (-1, 0) {};
		\node [style=Red Node] (2) at (0, -1) {};
		\node [style=Red Node] (3) at (1, 0) {};
		\node [style=none] (5) at (1.25, 0.5) {};
		\node [style=none] (6) at (0.75, -0.5) {};
		\node [style=none] (7) at (0.25, 1.5) {};
		\node [style=none] (8) at (-0.25, 0.5) {};
		\node [style=none] (9) at (-0.75, 0.5) {};
		\node [style=none] (10) at (-1.25, -0.5) {};
		\node [style=none] (11) at (-0.25, -1.5) {};
		\node [style=none] (12) at (0.25, -0.5) {};
		\node [style=Green Node] (13) at (0, 0) {};
	\end{pgfonlayer}
	\begin{pgfonlayer}{edgelayer}
		\draw (8.center) to (0);
		\draw (0) to (7.center);
		\draw (1) to (10.center);
		\draw (1) to (9.center);
		\draw (11.center) to (2);
		\draw (2) to (12.center);
		\draw (3) to (5.center);
		\draw (3) to (6.center);
		\draw (1) to (13);
		\draw (13) to (0);
		\draw (13) to (3);
		\draw (13) to (2);
	\end{pgfonlayer}
\end{tikzpicture}\,.
\label{eq:ZX_projetor}
\end{equation}
Ignoring scalar prefactors, this operator is a Pauli projector~\cite{KissingerWetering2024Book}, a map  projecting  onto the +1 eigenspace of the plaquette operator  $P_{\square}$. 
Following Eq.~\eqref{eq:groundstate00_toriccode_main}, applying this operator to each plaquette of qubits initialized in $\ket{0}=$
\begin{tikzpicture}
        \begin{pgfonlayer}{nodelayer}
            \node [style=X dot] (0) at (-0.5, 0) {};
            \node [style=none] (1) at (0.5, 0) {};
        \end{pgfonlayer}
        \begin{pgfonlayer}{edgelayer}
            \draw (0) to (1.center);
        \end{pgfonlayer}
    \end{tikzpicture}
and using the fusion rule, we obtain the ZX representation of the toric-code ground state, see Fig.~\ref{fig:Fig2_TC}a~\cite{kissinger2022phasefreezxdiagramscss,khesin2024equivalenceclassesquantumerrorcorrecting}.

The standard diagnostic of the topological order of the state \eqref{eq:groundstate00_toriccode_main}, relies on the calculation of the subleading entanglement entropy $\gamma$~\cite{hamma2005ground,LevinWen2006,KitaevPreskill2006}.
It is extracted from the entanglement entropy $S$ using various prescriptions.
For instance, once chosen  a bipartition $(A,B)$, we consider
the $n$-th Renyi entropy $S_{n} = -\frac{1}{n-1}\log(\Tr(\rho_{A}^{n}))$ with $ n > 1$, where $\rho_{A} = \Tr_{B}(\rho)$ is the reduced density matrix of $\ket{\rm GS}_{\rm tc}$.
All the $n$-th Renyi entropies of the toric code, and the von Neumann entropy $S_{\mathrm{vN}}=\lim_{n\to 1} S_n$ satisfy the same area law $S=\Sigma_{AB}-\gamma$, with $\Sigma_{AB}$ the number of plaquette operators, or stabilizers, acting both on subsystems $A$ and $B$~\cite{Hamma,SM}.
Using finite size scaling with $\Sigma_{AB}$, one finds that for the toric code $\gamma=\gamma_\mathrm{top}=\ln(2)$~\cite{Hamma,SM}, which we write from now on as $\gamma_\mathrm{top}=1$ in base two.
Alternatively, a judicious combinations of entropies   extracts $\gamma$ directly, without requiring finite size scaling~\cite{LevinWen2006,KitaevPreskill2006}.

Because of  known counterexamples, where these methods of extracting $\gamma_\mathrm{top}$ fail~~\cite{Cano2015,Bravyi,Zou2016,Fliss2017,Santos2018,Williamson2019,Kato2020,Stephen2019,Kim2023,Levin2024},  
we devise a more direct strategy enabled by  diagrammatic ZX calculus: graphically isolating the long-range entanglement between bipartitions.
We start by representing the density matrix $\rho$ associated to the state~\eqref{eq:groundstate00_toriccode_main} as a ZX diagram.
This diagram is obtained using the ground-state diagram  in Fig.~\ref{fig:Fig2_TC}a and its conjugate-transpose.
Since all spiders have zero phases, conjugate-transposing $\ket{\mathrm{GS}}_{\rm tc}$ to obtain $\bra{\mathrm{GS}}_{\rm tc}$ reduces to interpreting input wires as output wires.
Connecting input and output wires between the two copies is equivalent to tracing over the corresponding qubits.
The diagrammatic representation of $\rho_A$ is obtained by using the ZX-calculus simplification rules to trace region $B$, see SM~\cite{SM}.

Instead of calculating $\rho_A$, consider tracing \textit{all} degrees of freedom (qubits), as if calculating $\mathrm{Tr}(\rho)=1$. 
The ZX diagram implementing this operation is a scalar (no inputs or outputs) shown in Fig.~\ref{fig:Fig2_TC}b. 
The advantage of ZX diagrams is that we can choose which parts of the diagram to simplify to gain graphical insight on the real-space entanglement structure. 
Specifically, we avoid simplifying those ZX connections linking the bipartitions in order to track their  entanglement.
Technically, we do so by doubling the spiders connected by ZX edges which link the qubits of region $A$ to qubits in region $B$. This is practically achieved relying on the inverse of the ZX fuse rule 
\begin{tikzpicture}
	\begin{pgfonlayer}{nodelayer}
		\node [style=X dot] (0) at (1, 0) {};
		\node [style=none] (1) at (0.5, 0.5) {};
		\node [style=none] (2) at (0.5, -0.5) {};
		\node [style=none] (3) at (0.5, 0.04) {\rotatebox[origin=c]{90}{\tiny \dots}};
	\end{pgfonlayer}
	\begin{pgfonlayer}{edgelayer}
		\draw (1.center) to (0);
		\draw (0) to (2.center);
	\end{pgfonlayer}
\end{tikzpicture} = 
\begin{tikzpicture}
	\begin{pgfonlayer}{nodelayer}
		\node [style=X dot] (0) at (1, 0) {};
		\node [style=X dot] (1) at (2, 0) {};
		\node [style=none] (2) at (0.5, 0.5) {};
		\node [style=none] (3) at (0.5, -0.5) {};
		\node [style=none] (4) at (0.5, 0.04) {\rotatebox[origin=c]{90}{\tiny \dots}};
	\end{pgfonlayer}
	\begin{pgfonlayer}{edgelayer}
		\draw (2.center) to (0);
		\draw (0) to (3.center);
		\draw (0) to (1);
	\end{pgfonlayer}
\end{tikzpicture}
, see Fig.~\ref{fig:Fig2_TC}b.
We refer to this operation as \textit{pinning the spider}, as we  leave the doubled spiders unsimplified.
Pinning is  a computational tool allowing us to  track  the entanglement across the boundary while using the existing python packages \textsc{PyZX}~\cite{kissinger2019Pyzx} and \textsc{ZX-Live}~\cite{Zxlive} for diagrammatic simplification.
The resulting diagram is a scalar that graphically encodes information about the entanglement between bipartitions and we call it the \textit{contour diagram} $\mathcal{D}_{\partial A}$.

We now simplify $\mathcal{D}_{\partial A}$ with the pinned spiders using the ZX-calculus rules.
The result is shown in Fig.~\ref{fig:Fig2_TC}c, for a detailed proof see SM~\cite{SM}.
In the case of the toric code, the pinning keeps the plaquette operators acting on both bipartitions unsimplified (orange squares in Fig.~\ref{fig:Fig2_TC}), and their number $\Sigma_{AB}$ grows with the length of the boundary~\cite{Hamma}.
Additionally, \textit{one} green Z spider connects to each boundary plaquette, independent of $\Sigma_{AB}$. 
In ZX calculus, we can reposition the spiders as long as we preserve connections, inputs and  outputs~\cite{KissingerWetering2024Book}.
We hence  place the Z spider at the center of the diagram to highlight its non-locality, as  it connects all boundary plaquettes, including those physically far away. 

The existence of such non-local spider is a signature of long-range topological order.
The total number of non-local green Z spiders connected to the boundary qubits ($N_{\rm nl}=1$ for the toric code) does not change with the length of the boundary~\cite{SM}, suggesting that it is an invariant distinguishing different  topological orders. 
%

To clarify the connection between contour diagrams and entanglement entropies, we pin the $\Sigma_{AB}$ green Z spiders at the center of the boundary plaquettes in Fig.~\ref{fig:Fig2_TC}b or c, indicated by orange squares. This operation corresponds to pinning the $\Sigma_{AB}$ stabilizers acting simultaneously on regions $A$ and $B$~\cite{Hamma}.
The resulting simplified ZX diagram~\cite{SM}, Fig.~\ref{fig:Fig2_TC}d, is a scalar that includes a generalization of the Pauli projector~\eqref{eq:ZX_projetor} with $\Sigma_{AB}$ branches, as indicated by the dotted box. 
We will refer to diagrams with the structure of Fig.~\ref{fig:Fig2_TC}d as \textit{Pauli trees}. 
In the SM~\cite{SM}, we show that the Pauli tree in Fig.~\ref{fig:Fig2_TC}d determines the entanglement entropies $S_n=\Sigma_{AB}-\gamma_{\mathrm{top}}$~\cite{Hamma}.
The height of the Pauli projector equals $\Sigma_{AB}$, while the number of non-local green Z spiders, $N_\mathrm{nl}$, equals $N_\mathrm{nl}=\gamma_\mathrm{top}=1$.
This calculation establishes a direct connection between the Pauli tree and the  entanglement entropies $S_n$.

In light of these results, we propose the following protocol to diagnose topological order. First, represent diagrammatically $\mathrm{Tr}(\rho)$ while pinning the spiders connected by ZX edges which link qubits in region A to qubits in region B, defining the contour diagram $\mathcal{D}_{\partial A}$, as in Fig. \ref{fig:Fig2_TC}b. Second,  simplify $\mathcal{D}_{\partial A}$ \textit{excluding the pinned spiders}. The resulting diagram visually reveals the long-range entanglement at the boundary through a number of non-local nodes $N_\mathrm{nl}$, as in Fig.~\ref{fig:Fig2_TC}c. 
Note that arriving to Fig.~\ref{fig:Fig2_TC}c does not require prior knowledge of the $\Sigma_{AB}$ independent stabilizers, and hence our protocol can be applied to any state written as a ZX diagram.
However, if known, this information can be used to pin the stabilizers leading, in our toric code example, to the Pauli tree Fig.~\ref{fig:Fig2_TC}d, and to explicitly connect it to entanglement entropies.

This protocol is advantageous because simplifying $\mathcal{D}_{\partial A}$ is as complex as tracing the full density matrix $\rho$, it bypasses entropy calculations, and does not require prior knowledge of the type of topological order. 
It suggests that it is not necessary to calculate topological entanglement entropies to extract $\gamma$ but only to simplify $\mathcal{D}_{\partial A}$ using the ZX calculus.
This is the main result of our work.

For $\mathcal{D}_{\partial A}$ to be a candidate universal diagnostic, it is important to show that (\textit{i}) the number of non-local green Z spiders, $N_\mathrm{nl}$, is robust to changes in boundary choices and ground state superpositions~\cite{Zou2016,Ashvin_MES}, (\textit{ii}) $N_\mathrm{nl}$  changes only for different classes of topological orders~\cite{Simon_TopologicalQuantum} (\textit{iii}) $N_\mathrm{nl}=0$ even when spurious contributions change the numerical value of $\gamma$~\cite{Zou2016,Fliss2017,Santos2018,Williamson2019,Kato2020,Stephen2019,Kim2023}. 
We show now that $\mathcal{D}_{\partial A}$ meets these requirements using several examples.

\begin{figure*}[t]
    \centering
    \hspace{-17.8cm}
    \includegraphics[width=\linewidth]{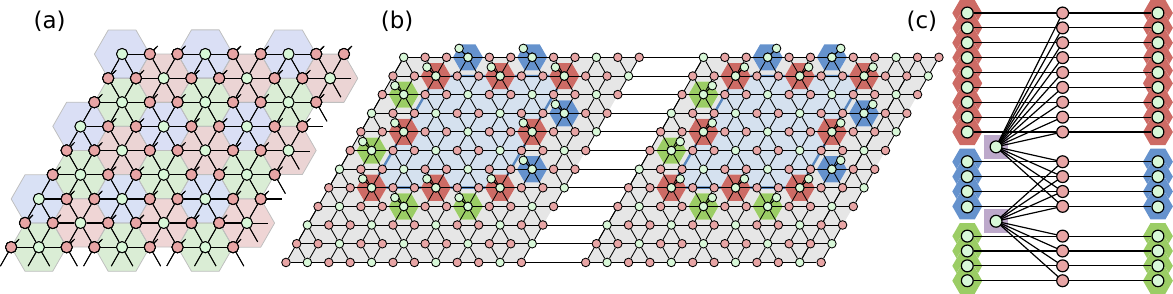}
    \caption{ Contour diagram $\mathcal{D}_{\rho_A}$ for the color code.
(a) ZX diagram representing a ground state of the color code. 
The color code is defined on an hexagonal lattice composed of  red, green and blue plaquettes. As in Fig.~\ref{fig:Fig2_TC}, the qubits are marked by red spiders with outputs (\begin{tikzpicture}
	\begin{pgfonlayer}{nodelayer}
		\protect\node [style=X dot] (0) at (0, 0) {};
		\protect\node [style=none] (1) at (.3, .3) {};
	\end{pgfonlayer}
	\begin{pgfonlayer}{edgelayer}
		\protect\draw (0) to (1);
		\end{pgfonlayer}
\end{tikzpicture}), but sit on the vertices of the hexagonal lattice. The black lines correspond to ZX connections issued from a generalization to six qubits of the Pauli projector~\eqref{eq:ZX_projetor}.
(b) ZX diagram representing $\mathrm{Tr}(\rho)$. The pinned spiders correspond to the $\Sigma_{AB}$ plaquette operators acting simultaneously on region $A$ and $B$, highlighted in blue and gray respectively.  This pinning simplifies directly to a Pauli-tree, shown in panel (c). The Pauli tree of the color code features $N_\mathrm{nl}=2$ non-local green spiders, in contrast to $N_\mathrm{nl}=1$ of the toric code (Fig.~\ref{fig:Fig2_TC}d).
}
    \label{fig:Fig4_TC}
\end{figure*}

%

\textit{Robustness of $\mathcal{D}_{\partial A}$ --}
To address point (\textit i), we consider the four degenerate ground states of the toric code, labelled as $\ket{\sigma\tau}_\mathrm{tc}$, where $\sigma,\,\tau=0,1$.
In this notation $\ket{\mathrm{GS}}_\mathrm{tc} \equiv \ket{00}_\mathrm{tc}$.
They can be obtained from one another by applying non-contractible strings of operators $\sigma_x$ winding around the torus, see Refs.~\cite{SM,Simon_TopologicalQuantum}.

Implementing the string operators in ZX weighted  by suitable  spiders, we construct arbitrary superpositions of toric-code ground states, $\ket\Psi= \sum_{\sigma,\tau}c_{\sigma\tau}\ket{\sigma\tau}_\mathrm{tc}$  with a single, compact diagram, see SM~\cite{SM}.
We  derive the contour diagram $\mathcal{D}_{\partial A}$ associated to this generic state following the  protocol described above. 
We obtain the same $\mathcal{D}_{\partial A}$ as shown in Fig.~\ref{fig:Fig2_TC}c and d, see SM~\cite{SM}.
Similarly, the non-local structure of $\mathcal{D}_{\partial A}$ persists irrespective of the contour defining the bipartition ($A,B$), see SM~\cite{SM}.

%
%

{\it Extension to the hexagonal toric and color codes} --
To address point (\textit{ii}) we show that the number of non-local spiders $N_\mathrm{nl}$  coincides with the intrinsic value of $\gamma_\mathrm{top}$ for two other exactly solvable models.

The first example is the hexagonal toric code, which is a generalization of the square toric code to an hexagonal lattice.
Its ground state can be written as Eq.~\eqref{eq:groundstate00_toriccode_main}, with the plaquette operators acting on the hexagons of the lattice.
Therefore, to represent the ground state as a ZX diagram, it suffices to generalize Eq.~\eqref{eq:ZX_projetor} to have six X spiders instead of four, all connected to a central Z spider to create the hexagonal lattice, see SM~\cite{SM}.
The ground state is in the same topological class as the square-lattice toric code, with $S=\Sigma_{AB}-1$, i.e. $\gamma=\gamma_{\mathrm{top}}=1$~\cite{Kitaev2003}.

By pinning the spiders associated to boundary qubits, as in Fig.~\ref{fig:Fig2_TC}b, and simplifying we obtain a $\mathcal D_{\partial A}$ that also features a single non-local green Z spider, as in Fig.~\ref{fig:Fig2_TC}c, connected to all the boundary (now hexagonal) plaquettes.
Pinning these $\Sigma_{AB}$ plaquettes we obtain \textit{the same} Pauli tree as in Fig.~\ref{fig:Fig2_TC}d, featuring one non-local green Z spider, $N_\mathrm{nl}=1$, and $\Sigma_{AB}$ branches.
These diagrams are thus robust to microscopic deformations of the lattice and hence identify the topological order of the state.

In contrast, the color code~\cite{Bombin_colorcode,lacroix2024scalinglogiccolorcode}, is a topologically ordered state with $S=\Sigma_{AB}-2$, i.e. $\gamma=\gamma_{\mathrm{top}}=2$~\cite{kargarian_tee_colorcode}.
It is defined by two plaquette operators $B^{x,z}_{\hexagon} = \prod_{i \in \hexagon} \sigma_{x,z}^{(i)}$ acting on three types of plaquettes, red, blue and green in Fig.~\ref{fig:Fig4_TC}a, see SM~\cite{SM}. Differently from the toric code, the qubits sit on the vertices of the plaquettes.
One of the 16 degenerate ground states of the color code is given by Eq.~\eqref{eq:groundstate00_toriccode_main}, with $P_\square$ replaced by $B^x_{\hexagon}$. 
Using similar reasoning as for the toric code, we find its ZX representation (see SM~\cite{SM}), shown in Fig.~\ref{fig:Fig4_TC}a. 
The corresponding, unsimplified, contour diagram $\mathcal{D}_{\partial A}$ is shown in Fig.~\ref{fig:Fig4_TC}b. 
To streamline our discussion, we discuss pinning the green Z spiders associated to the boundary plaquette operators $\Sigma_{AB}$, see Fig.~\ref{fig:Fig4_TC}b, instead of the boundary qubits~\footnote{Pinning the boundary qubits displays long-range entanglement, similar to Fig.~\ref{fig:Fig2_TC}b. However, pinning the plaquettes allows a more transparent comparison between models through crisp differences in their Pauli trees.}.
After simplification, we obtain the Pauli-tree diagram shown in Fig.~\ref{fig:Fig4_TC}c, see SM~\cite{SM}.
In the case of the color code, the Pauli tree features \textit{two} non-local Z spiders and $\Sigma_{AB}$ number of X spiders~\footnote{In the color code, $\Sigma_{AB}$ refers exclusively to the number of independent $B^x_{\hexagon}$ plaquette operators acting simultaneously on $A$ and $B$~\cite{kargarian_tee_colorcode}}.
We observe that $N_\mathrm{nl}=\gamma_{\rm top}=2$, confirming that the contour diagram  $\mathcal{D}_{\partial A}$ captures the topological difference with the toric code. 

{\it Absence of spurious spiders -- }
\begin{figure}
    \centering
    \hspace{-8.5cm}
    \includegraphics[width=\columnwidth]{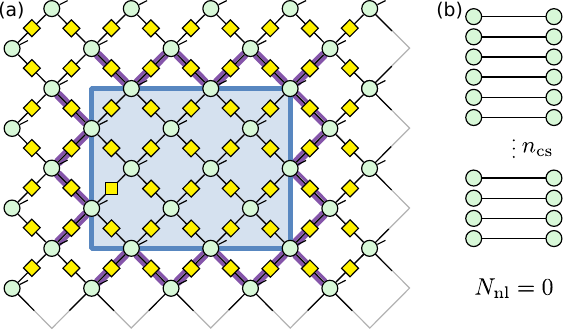}
    \caption{{Absence of non-local spiders in cluster states with spurious $\gamma$.}
        (a) ZX diagram of the cluster state considered in Ref.~\cite{Williamson2019}.        %
        All neighboring qubits on a square lattice in the \begin{tikzpicture}
	\begin{pgfonlayer}{nodelayer}
		\protect\node [style=Z dot] (0) at (0, 0) {};
		\protect\node [style=none] (1) at (.3, .3) {};
	\end{pgfonlayer}
	\begin{pgfonlayer}{edgelayer}
		\protect\draw (0) to (1);
		\end{pgfonlayer}
\end{tikzpicture}
$=\ket{+}$ state are connected by CZ =
\begin{tikzpicture}
	\begin{pgfonlayer}{nodelayer}
		\protect\node [style=Green Node] (0) at (0, 0) {};
		\protect\node [style=H] (1) at (.5, 0) {};
		\protect\node [style=Green Node] (2) at (1, 0) {};
		\protect\node [style=none] (3) at (0, 0.5) {};
		\protect\node [style=none] (4) at (0, -0.5) {};
		\protect\node [style=none] (5) at (1, 0.5) {};
		\protect\node [style=none] (6) at (1, -0.5) {};
	\end{pgfonlayer}
	\begin{pgfonlayer}{edgelayer}
		\protect\draw [style=Edge] (0) to (3.center);
		\protect\draw [style=Edge] (0) to (4.center);
		\protect\draw [style=Edge] (1) to (2);
		\protect\draw [style=Edge] (2) to (5.center);
		\protect\draw [style=Edge] (2) to (6.center);
		\protect\draw [style=Edge] (1) to (0);
	\end{pgfonlayer}
\end{tikzpicture} gates.
        Purple thicker lines mark the ZX links among the  $n_{\rm cs}=22$ qubits connecting regions $A$ (highlighted in blue) and $B$.
        (b) Contour diagram $\mathcal{D}_{\partial A}$ resulting from tracing the state in (a) while  pinning the $n_{\rm cs}$ qubits connecting regions $A$ and $B$. It features $n_{\rm cs}$ disconnected lines and its form is independent of the chosen bipartition. The absence of non-local spiders ($N_{\rm nl}=0$) indicates that the state has no long-range topological order.}
    \label{fig:Fig3_TC}
\end{figure}
Lastly we address point (\textit{iii}) by considering examples of topologically trivial  states   displaying  spurious topological entanglement entropy $\gamma\neq 0$~\cite{Cano2015,Bravyi,Zou2016,Fliss2017,Santos2018,Williamson2019,Kato2020,Stephen2019}.
A simple example is the two-dimensional cluster state~\cite{Williamson2019} represented in ZX in Fig.~\ref{fig:Fig3_TC}a. 
It is obtained by applying control-Z gates (in  ZX,  CZ =
\begin{tikzpicture}
	\begin{pgfonlayer}{nodelayer}
		\node [style=Green Node] (0) at (0, 0) {};
		\node [style=H] (1) at (.5, 0) {};
		\node [style=Green Node] (2) at (1, 0) {};
		\node [style=none] (3) at (0, 0.5) {};
		\node [style=none] (4) at (0, -0.5) {};
		\node [style=none] (5) at (1, 0.5) {};
		\node [style=none] (6) at (1, -0.5) {};
	\end{pgfonlayer}
	\begin{pgfonlayer}{edgelayer}
		\draw [style=Edge] (0) to (3.center);
		\draw [style=Edge] (0) to (4.center);
		\draw [style=Edge] (1) to (2);
		\draw [style=Edge] (2) to (5.center);
		\draw [style=Edge] (2) to (6.center);
		\draw [style=Edge] (1) to (0);
	\end{pgfonlayer}
\end{tikzpicture}
with 
\begin{tikzpicture}
	\begin{pgfonlayer}{nodelayer}
		\node [style=H] (1) at (.5, 0) {};
		\node [style=none] (3) at (0, 0) {};
		\node [style=none] (5) at (1, 0) {};
	\end{pgfonlayer}
	\begin{pgfonlayer}{edgelayer}
		\draw [style=Edge] (3.center) to (1);
		\draw [style=Edge] (1) to (5.center);
	\end{pgfonlayer}
\end{tikzpicture}
 the Hadamard matrix, see SM~\cite{SM}) to nearest neighbors of a square lattice with  qubits initialized in the state 
\begin{tikzpicture}
        \begin{pgfonlayer}{nodelayer}
            \node [style=Z dot] (0) at (-0.5, 0) {};
            \node [style=none] (1) at (0.2, 0) {};
        \end{pgfonlayer}
        \begin{pgfonlayer}{edgelayer}
            \draw (0) to (1.center);
        \end{pgfonlayer}
    \end{tikzpicture}%
    \ $= \ket{0} + \ket{1}=\sqrt2\ket+$.
Computing $S_n$ with ZX calculus, we recover the results of Ref.~\cite{Williamson2019}, namely a spurious subleading term $\gamma_\mathrm{spur}\neq 0$ depending on the chosen boundary, see SM~\cite{SM}.

In contrast, the simplified contour diagram  $\mathcal{D}_{\partial A}$, shown in Fig.~\ref{fig:Fig3_TC}b, does not feature non-local spiders ($N_{\mathrm{nl}}=0$), confirming the triviality of the state.
We also show that this is also the case when we apply a finite depth unitary circuit which further modifies $\gamma$, see SM~\cite{SM}. 

Using our protocol, we also  prove in the SM~\cite{SM} the triviality of Bravyi's example, defined as a 1D cluster state embedded in a 2D trivial state~\cite{Bravyi,Zou2016}. 
Specific entanglement cuts on such state  also lead to  $\gamma_\mathrm{spur}\neq 0$~\cite{Bravyi,Zou2016}, even when computed using the Kitaev-Preskill prescription~\cite{KitaevPreskill2006}. However,  we still find that $\mathcal{D}_{\partial A}$ does not feature non-local spiders, as expected for a trivial state.
This also suggest that the number of non-local spiders $N_{\rm nl}$ in the ZX contour diagram $\mathcal{D}_{\partial A}$ is a robust diagnostic of topological order, immune to spurious contributions.

\textit{Discussion --}
We have introduced a new protocol based on the ZX calculus to unambiguously diagnose long-range topological order. 
It is based on counting the number of non-local spiders $N_{\rm nl}$ in the simplified version of the contour diagram $\mathcal{D}_{\partial A}$, whose definition amounts to that of $\mathrm{Tr}(\rho)$. We have shown that the protocol correctly distinguishes among the topological orders of the toric code in the square and hexagonal lattices, and of the color code. 
This diagnostic, when applied to topologically trivial states with entanglement cuts generating spurious entanglement entropy, does not suffer from spurious non-local spiders and $N_{\rm nl}$ remains zero.

These results encourage the use of ZX diagrammatics as a diagnose of intrinsic long-range entanglement, beyond entropic methods~\cite{KitaevPreskill2006,LevinWen2006}. Our procedure bypasses the need to calculate the entanglement entropy, avoiding practical issues such as calculating logarithms of matrices, comparing with {fully mixed} reference states~\cite{Kato2020entropic} or performing finite size extrapolation~\cite{Cincio2013,Zaletel2013,Grushin2015}, all of which are numerically costly.
However, the contraction of a ZX diagram may result in a hard algorithmic problem for a generic state, which is the case for all tensor networks, including ZX diagrams. 
Finding efficient algorithms to count $N_{\mathrm{nl}}$ may be possible relying on graph-theory~\cite{khesin2023graphical,Khesin2025}.

Diagrammatic reasoning may improve the understanding of the physical properties of other solvable models, such as those with non-abelian or fracton excitations~\cite{He2018,Ma2018}.
ZX diagrams similar to the  Pauli trees discussed in our work appear in the ZX description of non-invertible symmetries~\cite{Gorantla2024}, a connection worth exploring further.
Lastly, ZX calculus is not limited by dimensionality or translational invariance, opening directions to diagnose topological order in higher dimensions, and non-crystalline lattices.

\textit{Acknowledgments -- }
We wish to acknowledge Niel De Beaudrap for discussions in the early stages of this work and J. H. Bardarson, B. A. Bernevig, J. N. F\"{u}chs, and N. Schuch for stimulating discussions. 
M.F. is grateful to ``Les Gustins'' summer school 2022 for providing the perfect occasion to lecture (and thus learn) about ZX calculus. 
A.G.G. and S. M. M. acknowledge financial support from the European Research Council (ERC) Consolidator grant under grant agreement No. 101042707 (TOPOMORPH). M.F. acknowledges support from  EPiQ
ANR-22-PETQ-0007 and from the HQI initiative (\hyperlink{https://www.hqi.fr/}{www.hqi.fr})  ANR-22-PNCQ-0002  financed by the French National Research
Agency and part of Plan France 2030.
All \textsc{ZXLive} proofs can be found in the Zenodo repository associated to this paper Ref.~\cite{Mas_mendoza_zenodo}.

\normalem
\bibliography{bib-file.bib}

\clearpage
\newpage
\onecolumngrid

\setcounter{secnumdepth}{5}
\renewcommand{\theparagraph}{\bf \thesubsubsection.\arabic{paragraph}}

\renewcommand{\thefigure}{S\arabic{figure}}
\setcounter{figure}{0} 


\begin{center}
{\bf \large Supplemental Material for ``A graphical diagnostic of topological order using ZX
calculus''}
\end{center}

\twocolumngrid

\section{Outline}
In this Supplemental Material, Section~\ref{sec:intro_ZX} provides a brief introduction to the basics of ZX calculus. In Section~\ref{sec:TEE}, we  discuss the spurious contributions affecting the diagnostics of long-range topological order based on entanglement. In Section~\ref{sec:toriccode}, we briefly review  Kitaev's toric code and its ground states. The ZX diagrams corresponding to these ground states are derived in Section~\ref{sec:ToricCode_ZX}. In Section~\ref{sec:results}, we show benchmarks of our ZX calculations for the toric code and provide details about the correspondence between contour diagrams, Pauli trees and topological entanglement. In Section~\ref{sec:hextoriccolorcode}, we shortly review the hexagonal toric and color codes and, in Section~\ref{sec:hexTC_CC_ZX}, we give details about the associated  ZX calculations. In Section~\ref{sec:universality_gamma}, we show that the ZX representation of the contour diagram  associated to some specific cluster states does not feature non-local spiders and that its structure is  robust to the application of local unitaries.
Along the text we will refer to \textsc{ZX-Live} proofs of the accompanying Zenodo repository, Ref.~\cite{Mas_mendoza_zenodo}.

\section{ZX calculus} \label{sec:intro_ZX}

ZX calculus is a diagrammatic language that  represents linear maps between two sets of qubits.
It was introduced as an efficient tool for quantum circuit optimization~\cite{coecke2008interacting,coecke2011interacting}.
By writing a quantum circuit as a ZX diagram, the ZX-calculus rules produce equivalent and simpler diagrams that perform the same operation with fewer gates.

The power of ZX calculus rests in its universality, soundness and completeness.
ZX diagrams are universal because they can represent any linear map $(\mathbb{C}^{2})^{\otimes n} \mapsto (\mathbb{C}^{2})^{\otimes m}$, from $n$ to $m$ qubits~\cite{KissingerWetering2024Book}. 
ZX calculus is complete because if two diagrams represent the same map, we can transform one diagram into another using ZX-calculus rules~\cite{KissingerWetering2024Book}.
It is also sound, because all equivalences given by ZX rules represent a correct equivalence between the operators they represent.

This Section introduces the basic concepts of ZX calculus, which are the spiders (Section~\ref{sec:spiders}) and their simplification rules (Section~\ref{sec:rules}). We illustrate how CNOT and CZ gates are represented in this language (Section~\ref{sec:ZXcontrolgates}) and how to perform operations such as transposition, conjugation and tracing (Section~\ref{sec:ZXtranspositionconjugationtracing}). This introduction equips the reader with the necessary ingredients to derive the results illustrated in the main text.

\subsection{Spiders \label{sec:spiders}}

Spiders are the building blocks of ZX diagrams, the basic linear operators.
There are two types of spiders, green (or Z) and red (or X), defined as~\cite{vandewetering2020zxcalculus}

\begin{subequations} \label{eq:spiders_def}
\begin{equation}
\label{eq:greenspider_def}
\begin{tikzpicture}
	\begin{pgfonlayer}{nodelayer}
		\node [style=Z phase dot] (0) at (0, 0) {$\alpha$};
		\node [style=none] (1) at (1.25, 1) {};
		\node [style=none] (2) at (-1.25, 1) {};
		\node [style=none] (3) at (-1.25, -1) {};
		\node [style=none] (4) at (1.25, -1) {};
		\node [style=none] (5) at (1.25, 0.5) {};
		\node [style=none] (6) at (-1.25, 0.5) {};
		\node [style=none, rotate=90] (7) at (-1, -0.25) {...};
		\node [style=none, rotate=90] (8) at (1, -0.25) {...};
		\node [style=none] (18) at (-2, 1.25) {};
		\node [style=none] (19) at (-2, -1.25) {};
		\node [style=none] (20) at (-2.2, -1.25) {$_m$};
		\node [style=none] (21) at (2, 1.25) {};
		\node [style=none] (22) at (2, -1.25) {};
		\node [style=none] (23) at (2.2, -1.25) {$_n$};
	\end{pgfonlayer}
	\begin{pgfonlayer}{edgelayer}
		\draw [in=-141, out=0, looseness=0.75] (3.center) to (0);
		\draw [in=180, out=-39, looseness=0.75] (0) to (4.center);
		\draw [in=180, out=22, looseness=0.75] (0) to (5.center);
		\draw [in=180, out=39, looseness=0.75] (0) to (1.center);
		\draw [in=0, out=158, looseness=0.75] (0) to (6.center);
		\draw [in=141, out=0, looseness=0.75] (2.center) to (0);
		\draw [decorate, style=brace edge] (19.center) to (18.center);
		\draw [decorate, style=brace edge] (21.center) to (22.center);
	\end{pgfonlayer}
\end{tikzpicture}
= \ket{0}^{\otimes n}\bra{0}^{\otimes m} + e^{i\alpha}\ket{1}^{\otimes n}\bra{1}^{\otimes m}
\end{equation}
\begin{equation}
\label{eq:redspider_def}
\begin{tikzpicture}
	\begin{pgfonlayer}{nodelayer}
		\node [style=X phase dot] (0) at (0, 0) {$\alpha$};
		\node [style=none] (1) at (1.25, 1) {};
		\node [style=none] (2) at (-1.25, 1) {};
		\node [style=none] (3) at (-1.25, -1) {};
		\node [style=none] (4) at (1.25, -1) {};
		\node [style=none] (5) at (1.25, 0.5) {};
		\node [style=none] (6) at (-1.25, 0.5) {};
		\node [style=none, rotate=90] (7) at (-1, -0.25) {...};
		\node [style=none, rotate=90] (8) at (1, -0.25) {...};
		\node [style=none] (18) at (-2, 1.25) {};
		\node [style=none] (19) at (-2, -1.25) {};
		\node [style=none] (20) at (-2.2, -1.25) {$_m$};
		\node [style=none] (21) at (2, 1.25) {};
		\node [style=none] (22) at (2, -1.25) {};
		\node [style=none] (23) at (2.2, -1.25) {$_n$};
	\end{pgfonlayer}
	\begin{pgfonlayer}{edgelayer}
		\draw [in=-141, out=0, looseness=0.75] (3.center) to (0);
		\draw [in=180, out=-39, looseness=0.75] (0) to (4.center);
		\draw [in=180, out=22, looseness=0.75] (0) to (5.center);
		\draw [in=180, out=39, looseness=0.75] (0) to (1.center);
		\draw [in=0, out=158, looseness=0.75] (0) to (6.center);
		\draw [in=141, out=0, looseness=0.75] (2.center) to (0);
		\draw [decorate, style=brace edge] (19.center) to (18.center);
		\draw [decorate, style=brace edge] (21.center) to (22.center);
	\end{pgfonlayer}
\end{tikzpicture}
= \ket{+}^{\otimes n}\bra{+}^{\otimes m} + e^{i\alpha}\ket{-}^{\otimes n}\bra{-}^{\otimes m}
\end{equation}
\end{subequations}
where $m$ and $n$ are the number of input and output qubits (two-level quantum degrees of freedom), $\alpha \in [0,2\pi)$ is a relative phase, and $\ket{\pm} =\left(\ket{0} \pm \ket{1}\right)/\sqrt2$.
The convention is to omit the phase $\alpha$ when it equals zero $(\begin{tikzpicture}
	\begin{pgfonlayer}{nodelayer}
		\node [style=X phase dot] (0) at (0, 0) {$0$};
	\end{pgfonlayer}
\end{tikzpicture}
= 
\begin{tikzpicture}
	\begin{pgfonlayer}{nodelayer}
		\node [style=X phase dot] (0) at (0, 0) {$~$};
	\end{pgfonlayer}
\end{tikzpicture}
= 
\begin{tikzpicture}
	\begin{pgfonlayer}{nodelayer}
		\node [style=X  dot] (0) at (0, 0) {};
	\end{pgfonlayer}
\end{tikzpicture}
)$. 

To illustrate better the definitions in Eqs.~\eqref{eq:spiders_def}, we provide some  simple examples. 
Quantum states, matrices, and scalars can be inferred directly from the definitions of the spiders:
\begin{subequations} \label{eq:zx_spider_examples}
\begin{align} 
\begin{tikzpicture}
	\begin{pgfonlayer}{nodelayer}
		\node [style=X phase dot] (0) at (0, 0) {$~$};
		\node [style=none] (1) at (1, 0) {};
	\end{pgfonlayer}
	\begin{pgfonlayer}{edgelayer}
		\draw (0) to (1.center);
	\end{pgfonlayer}
\end{tikzpicture}
&= \ket{+} + \ket{-} = \sqrt{2}\ket{0}\,,\label{eq:0_zx}\\
\begin{tikzpicture}
	\begin{pgfonlayer}{nodelayer}
		\node [style=X phase dot] (2) at (0, 0) {$\pi$};
		\node [style=none] (1) at (1, 0) {};
	\end{pgfonlayer}
	\begin{pgfonlayer}{edgelayer}
		\draw (2) to (1.center);
	\end{pgfonlayer}
\end{tikzpicture}
&= \ket{+} - \ket{-} = \sqrt{2}\ket{1}\,, \label{eq:1_zx}\\
\begin{tikzpicture}
	\begin{pgfonlayer}{nodelayer}
		\node [style=Z phase dot] (0) at (0, 0) {};
		\node [style=none] (1) at (1, 0) {};
	\end{pgfonlayer}
	\begin{pgfonlayer}{edgelayer}
		\draw (0) to (1.center);
	\end{pgfonlayer}
\end{tikzpicture}
&= \ket{0} + \ket{1} = \sqrt{2}\ket{+}\,,\label{eq:+_zx}\\
\begin{tikzpicture}
	\begin{pgfonlayer}{nodelayer}
		\node [style=none] (1) at (1, 0) {};
		\node [style=Z phase dot] (2) at (0, 0) {$\pi$};
	\end{pgfonlayer}
	\begin{pgfonlayer}{edgelayer}
		\draw (2) to (1.center);
	\end{pgfonlayer}
\end{tikzpicture}
&= \ket{0} - \ket{1} = \sqrt{2}\ket{-}\,,\label{eq:-_zx}\\
\begin{tikzpicture}
	\begin{pgfonlayer}{nodelayer}
		\node [style=none] (1) at (1, 0) {};
		\node [style=X phase dot] (2) at (0, 0) {$\pi$};
		\node [style=none] (3) at (-1, 0) {};
	\end{pgfonlayer}
	\begin{pgfonlayer}{edgelayer}
		\draw (2) to (1.center);
		\draw (2) to (3.center);
	\end{pgfonlayer}
\end{tikzpicture}
&= \ket{+}\bra{+} - \ket{-}\bra{-} =
\begin{pmatrix}
0 & 1 \\
1 & 0 
\end{pmatrix}
= \sigma_{x}\,,\label{eq:sigmax_zx}\\
\begin{tikzpicture}
	\begin{pgfonlayer}{nodelayer}
		\node [style=none] (1) at (1, 0) {};
		\node [style=Z phase dot] (2) at (0, 0) {$\pi$};
		\node [style=none] (3) at (-1, 0) {};
	\end{pgfonlayer}
	\begin{pgfonlayer}{edgelayer}
		\draw (2) to (1.center);
		\draw (2) to (3.center);
	\end{pgfonlayer}
\end{tikzpicture}
&= \ket{0}\bra{0} - \ket{1}\bra{1} =
\begin{pmatrix}
1 & 0 \\
0 & -1 
\end{pmatrix}
= \sigma_{z}\,,\label{eq:sigmaz_zx}\\
\begin{tikzpicture}
	\begin{pgfonlayer}{nodelayer}
		\node [style=Z phase dot] (0) at (0, 0) {};
		\node [style=none] (3) at (1, 0) {};
	\end{pgfonlayer}
\end{tikzpicture}
&= 1 + e^{i0} = 2\,,\\
\begin{tikzpicture}
	\begin{pgfonlayer}{nodelayer}
		\node [style=Z phase dot] (0) at (0, 0) {$\pi$};
		\node [style=none] (3) at (1, 0) {};	
	\end{pgfonlayer}
\end{tikzpicture}
&= 1 + e^{i\pi} = 0\,,
\end{align}
\end{subequations}
where we adopted the convention $(\dots)^{\otimes0} = 1$.

Unless otherwise stated, we will omit multiplicative scalar factors.
As our work concerns states and density matrices,  multiplicative scalars can be always fixed by making the appropriate normalization.
For example, we will simply write
\begin{tikzpicture}
	\begin{pgfonlayer}{nodelayer}
		\node [style=X phase dot] (0) at (0, 0) {};
		\node [style=none] (1) at (1, 0) {};
	\end{pgfonlayer}
	\begin{pgfonlayer}{edgelayer}
		\draw (0) to (1.center);
	\end{pgfonlayer}
\end{tikzpicture}
$= \ket{0}$, omitting the $\sqrt{2}$ factor in Eq.~\eqref{eq:0_zx}.

A ZX diagram is made  of compositions and tensor products of spiders~\cite{KissingerWetering2024Book}.
Suppose $D_{1},D_{2},...$ label different diagrams (e.g. spiders) describing a set of linear maps $M_{1},M_{2},...$, respectively.
We can tensor product two diagrams $D = D_{1} \otimes D_{2}$, to represent the diagram corresponding to $M_{1} \otimes M_{2}$.
We tensor product by stacking the two diagrams together, without any connection~\cite{vandewetering2020zxcalculus}.
For example, if $M_{1} = \sigma_{z}$ and $M_{2} = \sigma_{x}$, then,
\begin{equation}
\sigma_{z}\otimes\sigma_{x} =
\begin{tikzpicture}
	\begin{pgfonlayer}{nodelayer}
		\node [style=X phase dot] (12) at (0, -0.25) {$\pi$};
		\node [style=Z phase dot] (13) at (0, 0.5) {$\pi$};
		\node [style=none] (14) at (0.75, 0.5) {};
		\node [style=none] (15) at (-0.75, -0.25) {};
		\node [style=none] (16) at (-0.75, 0.5) {};
		\node [style=none] (17) at (0.75, -0.25) {};
	\end{pgfonlayer}
	\begin{pgfonlayer}{edgelayer}
		\draw (15.center) to (12);
		\draw (13) to (14.center);
		\draw (13) to (16.center);
		\draw (12) to (17.center);
	\end{pgfonlayer}
\end{tikzpicture}\,.
\end{equation}
Such procedure naturally extends to tensor products of states. Consider for instance $M_{1} = \ket{0}, M_{2} = \ket{1}, M_{3} = \ket{0}$, then,
\begin{equation}
\ket{010} = \ket{0}\otimes\ket{1}\otimes\ket{0} = 
\begin{tikzpicture}
	\begin{pgfonlayer}{nodelayer}
		\node [style=X phase dot] (0) at (0, 0.75) {$~$};
		\node [style=X phase dot] (1) at (0, -0.75) {$~$};
		\node [style=X phase dot] (2) at (0, 0) {$\pi$};
		\node [style=none] (3) at (1, 0.75) {};
		\node [style=none] (4) at (1, 0) {};
		\node [style=none] (5) at (1, -0.75) {};
	\end{pgfonlayer}
	\begin{pgfonlayer}{edgelayer}
		\draw (0) to (3.center);
		\draw (4.center) to (2);
		\draw (5.center) to (1);
	\end{pgfonlayer}
\end{tikzpicture}\,.
\end{equation}

We can also compose ZX diagrams, e.g. $D = D_{2} \circ D_{1}$, to represent the composition $M_{2} \circ M_{1}$.
We compose two diagrams by connecting the outputs of the first diagram $D_{1}$ with the inputs of the second diagram $D_{2}$~\cite{vandewetering2020zxcalculus}. 
For example, if $M_{1} = \sigma_{x}$ and $M_{2} = \sigma_{z}$, then,
\begin{equation} \label{eq:compositionzx_example}
\sigma_{z}\circ\sigma_{x} = 
\begin{tikzpicture}
	\begin{pgfonlayer}{nodelayer}
		\node [style=Z phase dot] (12) at (0.5, 0) {$\pi$};
		\node [style=X phase dot] (13) at (-0.5, 0) {$\pi$};
		\node [style=none] (16) at (-1.25, 0) {};
		\node [style=none] (17) at (1.25, 0) {};
	\end{pgfonlayer}
	\begin{pgfonlayer}{edgelayer}
		\draw (13) to (16.center);
		\draw (12) to (17.center);
		\draw (13) to (12);
	\end{pgfonlayer}
\end{tikzpicture}\,.
\end{equation}

The Hadamard gate, $H$, is a particularly useful operator defined as a composition of ZX spiders.
As a matrix, the Hadamard gate is defined as
\begin{equation}
    H = \frac{1}{\sqrt{2}}
\begin{pmatrix}
1 & 1 \\
1 & -1 
\end{pmatrix}.
\end{equation}
It changes between the Z $\{\ket{0},\ket{1}\}$ and the X $\{\ket{+},\ket{-}\}$ basis: $\ket{0}\leftrightarrow\ket{+},\ket{1}\leftrightarrow\ket{-}$.
Within the calculus, the Hadamard gate symbol is defined by the composition~\cite{vandewetering2020zxcalculus}
\begin{equation}
H =
\begin{tikzpicture}
	\begin{pgfonlayer}{nodelayer}
		\node [style=none] (1) at (0.75, 0) {};
		\node [style=none] (2) at (-0.75, 0) {};
		\node [style=none] (3) at (1.5, 0) {=};
		\node [style=none] (4) at (3.5, 0) {};
		\node [style=none] (5) at (7.5, 0) {};
		\node [style=Z phase dot] (6) at (4.5, 0) {$\frac{ \pi}{2}$};
		\node [style=Z phase dot] (7) at (6.5, 0) {$\frac{ \pi}{2}$};
		\node [style=X phase dot] (8) at (5.5, 0) {$\frac{ \pi}{2}$};
		\node [style=none] (9) at (2.75, 0.15) {$e^{-i\frac{\pi}{4}}$};
		\node [style=hadamard] (10) at (-3, 0) {};
		\node [style=none] (11) at (-2.25, 0) {};
		\node [style=none] (12) at (-3.75, 0) {};
		\node [style=none] (13) at (-1.5, 0) {=};
	\end{pgfonlayer}
	\begin{pgfonlayer}{edgelayer}
		\draw (4.center) to (6);
		\draw (6) to (8);
		\draw (8) to (7);
		\draw (7) to (5.center);
		\draw (12.center) to (10);
		\draw (10) to (11.center);
		\draw [style=hadamard edge] (1.center) to (2.center);
	\end{pgfonlayer}
\end{tikzpicture}\,,
\end{equation}
where we include the scalar factor $e^{-i\frac{\pi}{4}}$.
The dashed-blue line representation of a Hadamard gate in its definition is used in ZX-software packages PyZX~\cite{kissinger2019Pyzx} and ZX-live~\cite{Zxlive}.
In the main text and this supplemental we will represent it as a yellow box, as in the first equality.
Because the Hadamard implements a basis change, it amounts to interchange green to red colored spiders, and vice versa, see Fig.~\ref{fig:zx-rules}, rules $\HColorChange$ and $\HCopy$.
For this reason, it is convenient to include the Hadamard gate $H$ as a third element in the ZX calculus, see also Section~\ref{sec:universality_gamma}.

ZH calculus is an alternative language that uses a generalization of the Hadamard gate as a generator.
The ZH calculus is also universal, complete and sound~\cite{zh-calculus}.
The two generators of the ZH calculus are
    the Z spider as in Eq.~\eqref{eq:greenspider_def}, and 
    the H-box, defined as
\begin{equation}\label{eq:Hbox_def}
\begin{tikzpicture}
	\begin{pgfonlayer}{nodelayer}
		\node [style=hadamard] (0) at (0, 0) {$a$};
		\node [style=none] (1) at (1.25, 1) {};
		\node [style=none] (2) at (-1.25, 1) {};
		\node [style=none] (3) at (-1.25, -1) {};
		\node [style=none] (4) at (1.25, -1) {};
		\node [style=none] (5) at (1.25, 0.5) {};
		\node [style=none] (6) at (-1.25, 0.5) {};
		\node [style=none, rotate=90] (7) at (-1, -0.25) {...};
		\node [style=none, rotate=90] (8) at (1, -0.25) {...};
		\node [style=none] (18) at (-2, 1.25) {};
		\node [style=none] (19) at (-2, -1.25) {};
		\node [style=none] (20) at (-2.5, 0) {$m$};
		\node [style=none] (21) at (2, 1.25) {};
		\node [style=none] (22) at (2, -1.25) {};
		\node [style=none] (23) at (2.5, 0) {$n$};
	\end{pgfonlayer}
	\begin{pgfonlayer}{edgelayer}
		\draw [in=-141, out=0, looseness=0.75] (3.center) to (0);
		\draw [in=180, out=-39, looseness=0.75] (0) to (4.center);
		\draw [in=180, out=22, looseness=0.75] (0) to (5.center);
		\draw [in=180, out=39, looseness=0.75] (0) to (1.center);
		\draw [in=0, out=158, looseness=0.75] (0) to (6.center);
		\draw [in=141, out=0, looseness=0.75] (2.center) to (0);
		\draw [decorate, style=brace edge] (19.center) to (18.center);
		\draw [decorate, style=brace edge] (21.center) to (22.center);
	\end{pgfonlayer}
\end{tikzpicture}
= \sum_{I_{m},J_{n}} a^{I_{m}J_{n}} \ket{J_{n}}\bra{I_{m}},
\end{equation}
where $a \in \mathbb{C}$, and $I_{m} = i_1,\ldots, i_m$ and $J_{n} = j_1,\ldots, j_n$, $i_{k},j_{k}\in\{0,1\}$ are multi-indices.
The H-box generalizes the Hadamard gate to an arbitrary number of inputs and outputs.
It is a matrix where all entries are equal to 1, except for the bottom right element, which equals $a$.
In this work we will use the H-box to form diagrammatic superpositions, see Section ~\ref{sec:genericgroundstateZX}.

\subsection{Simplification Rules \label{sec:rules}}
ZX calculus is equipped with a set of graphical rewrite rules.
By using these rules, a given diagram  can be simplified without referring to the underlying tensors: the diagram is the calculation~\cite{KissingerWetering2024Book,east2022aklt}. 

\begin{figure*}
    \centering
\begin{tikzpicture}
	\begin{pgfonlayer}{nodelayer}
		\node [style=X phase dot] (0) at (-4.625, -6) {$\alpha$};
		\node [style=none] (1) at (-3.375, -5) {};
		\node [style=none] (2) at (-5.875, -5) {};
		\node [style=none] (3) at (-5.875, -7) {};
		\node [style=none] (4) at (-3.375, -7) {};
		\node [style=none] (5) at (-3.375, -5.625) {};
		\node [style=none] (6) at (-5.875, -5.625) {};
		\node [style=none, rotate=90] (7) at (-5.625, -6.25) {...};
		\node [style=none, rotate=90] (8) at (-3.625, -6.25) {...};
		\node [style=H] (9) at (-6.375, -5) {};
		\node [style=H] (10) at (-6.375, -5.625) {};
		\node [style=H] (11) at (-6.375, -7) {};
		\node [style=H] (12) at (-2.875, -5) {};
		\node [style=H] (13) at (-2.875, -5.625) {};
		\node [style=H] (14) at (-2.875, -7) {};
		\node [style=none] (15) at (-2.375, -5) {};
		\node [style=none] (16) at (-2.375, -5.625) {};
		\node [style=none] (17) at (-2.375, -7) {};
		\node [style=none] (18) at (-6.875, -5) {};
		\node [style=none] (19) at (-6.875, -5.625) {};
		\node [style=none] (20) at (-6.875, -7) {};
		\node [style=Z phase dot] (21) at (6.375, -6) {$\alpha$};
		\node [style=none] (22) at (7.625, -5) {};
		\node [style=none] (23) at (5.125, -5) {};
		\node [style=none] (24) at (5.125, -7) {};
		\node [style=none] (25) at (7.625, -7) {};
		\node [style=none] (26) at (7.625, -5.625) {};
		\node [style=none] (27) at (5.125, -5.625) {};
		\node [style=none, rotate=90] (28) at (5.375, -6.25) {...};
		\node [style=none, rotate=90] (29) at (7.375, -6.25) {...};
		\node [style=none] (30) at (0, -6) {=};
		\node [style=none] (31) at (1.875, -6) {$\sqrt{2}^{n+m}$};
		\node [style=none] (32) at (-1.75, -4.75) {};
		\node [style=none] (33) at (-1.75, -7.25) {};
		\node [style=none] (34) at (-7.5, -4.75) {};
		\node [style=none] (35) at (-7.5, -7.25) {};
		\node [style=none] (36) at (-1.25, -6) {$n$};
		\node [style=none] (37) at (-8, -6) {$m$};
		\node [style=none] (38) at (8.25, -4.75) {};
		\node [style=none] (39) at (8.25, -7.25) {};
		\node [style=none] (40) at (8.75, -6) {$n$};
		\node [style=none] (41) at (4.5, -4.75) {};
		\node [style=none] (42) at (4.5, -7.25) {};
		\node [style=none] (43) at (4, -6) {$m$};
		\node [style=Z phase dot] (44) at (-2.25, 5.125) {$\beta$};
		\node [style=none] (45) at (-1.25, 5.375) {};
		\node [style=none] (46) at (-4, 5.375) {};
		\node [style=none] (47) at (-4, 4.125) {};
		\node [style=none] (48) at (-1.25, 4.125) {};
		\node [style=none, rotate=90] (49) at (-4.25, 4.75) {...};
		\node [style=none, rotate=90] (50) at (-1.25, 4.75) {...};
		\node [style=Z phase dot] (51) at (-3.75, 6.625) {$\alpha$};
		\node [style=none] (52) at (-1.75, 7.375) {};
		\node [style=none] (53) at (-4.75, 7.375) {};
		\node [style=none] (54) at (-1.75, 6.125) {};
		\node [style=none, rotate=90] (55) at (-1.75, 6.75) {...};
		\node [style=none] (56) at (-4.75, 6.125) {};
		\node [style=none, rotate=90] (57) at (-4.5, 6.75) {...};
		\node [style=none] (58) at (-1.25, 6.125) {};
		\node [style=none] (59) at (-1.25, 7.375) {};
		\node [style=none] (60) at (-4.75, 4.125) {};
		\node [style=none] (61) at (-4.75, 5.375) {};
		\node [style=none] (62) at (0, 6) {$=$};
		\node [style=none, rotate=90] (63) at (-3.1, 5.8) {...};
		\node [style=none] (64) at (1.25, 7) {};
		\node [style=none, rotate=90] (65) at (1.75, 5.75) {...};
		\node [style=none] (66) at (4.75, 4.5) {};
		\node [style=none, rotate=90] (67) at (4.5, 5.75) {...};
		\node [style=Z phase dot] (68) at (3, 5.75) {$\ \alpha\!+\!\beta\ $};
		\node [style=none] (69) at (4.75, 7) {};
		\node [style=none] (70) at (1.25, 4.5) {};
		\node [style=none] (71) at (0, 6.75) {\spiderrule};
		\node [style=Z phase dot] (72) at (19.5, 5.75) {$\ -\alpha\ $};
		\node [style=none] (73) at (21.75, 6.875) {};
		\node [style=none] (74) at (17, 5.75) {$=$};
		\node [style=none] (75) at (18.25, 5.75) {};
		\node [style=none] (76) at (21.75, 4.625) {};
		\node [style=X phase dot] (77) at (21, 4.625) {$\pi$};
		\node [style=X phase dot] (78) at (12.75, 5.75) {$\pi$};
		\node [style=none] (79) at (15.25, 6.875) {};
		\node [style=Z phase dot] (80) at (13.75, 5.75) {$\alpha$};
		\node [style=none, rotate=90] (81) at (15, 5.75) {...};
		\node [style=none, rotate=90] (82) at (21, 5.75) {...};
		\node [style=none] (83) at (12, 5.75) {};
		\node [style=none] (84) at (15.25, 4.625) {};
		\node [style=X phase dot] (85) at (21, 6.875) {$\pi$};
		\node [style=none] (86) at (17, 6.5) {\pirule};
		\node [style=X phase dot] (87) at (20.25, 2.625) {$a\pi$};
		\node [style=none, rotate=90] (88) at (21, 1.75) {...};
		\node [style=Z phase dot] (89) at (13.5, 1.75) {$\alpha$};
		\node [style=none] (90) at (14.75, 0.875) {};
		\node [style=none] (91) at (17, 1.75) {$=$};
		\node [style=none] (92) at (21.25, 0.875) {};
		\node [style=none] (93) at (21.25, 2.625) {};
		\node [style=none, rotate=90] (94) at (14.5, 1.75) {...};
		\node [style=X phase dot] (95) at (20.25, 0.875) {$a\pi$};
		\node [style=none] (96) at (17, 2.5) {\copyrule};
		\node [style=none] (97) at (14.75, 2.625) {};
		\node [style=X phase dot] (98) at (12.5, 1.75) {$a\pi$};
		\node [style=Z dot] (99) at (15.25, -1.75) {};
		\node [style=none] (100) at (14.5, -1.75) {};
		\node [style=none] (101) at (16, -1.75) {};
		\node [style=none] (102) at (18, -1.75) {};
		\node [style=none] (103) at (19, -1.75) {};
		\node [style=none] (104) at (17, -1) {\identityrule};
		\node [style=none] (105) at (17, -1.75) {$=$};
		\node [style=none] (106) at (0, 2.5) {\bialgrule};
		\node [style=none] (107) at (19.25, 4.5) {$e^{i\alpha}$};
		\node [style=none] (108) at (19, 1.75) {$\frac{e^{ia\alpha}}{\sqrt{2}^{n-1}}$};
		\node [style=X dot] (109) at (-4.75, 1.75) {};
		\node [style=Z dot] (110) at (-3.75, 1.75) {};
		\node [style=none] (111) at (-2.25, 3) {};
		\node [style=none] (112) at (-2.25, 0.5) {};
		\node [style=none] (113) at (-3, 1.75) {\ldots};
		\node [style=none] (114) at (-1.25, 1.75) {$n$};
		\node [style=none] (115) at (0, 1.5) {=};
		\node [style=none] (116) at (4.75, -0.5) {$\left(\sqrt{2}\right)^{(n-1)(m-1)}$};
		\node [style=none] (117) at (-6, 3) {};
		\node [style=none] (118) at (-6, 0.5) {};
		\node [style=none] (119) at (-7.25, 1.75) {$m$};
		\node [style=none] (120) at (-5.5, 1.75) {\ldots};
		\node [style=X dot] (121) at (6.25, 3) {};
		\node [style=X dot] (122) at (6.25, 0.5) {};
		\node [style=Z dot] (123) at (3.5, 3) {};
		\node [style=Z dot] (124) at (3.5, 0.5) {};
		\node [style=none] (125) at (7.25, 3) {};
		\node [style=none] (126) at (7.25, 0.5) {};
		\node [style=none] (127) at (2.5, 3) {};
		\node [style=none] (128) at (2.5, 0.5) {};
		\node [style=none] (129) at (6.25, 1.75) {\ldots};
		\node [style=none] (130) at (3.5, 1.75) {\ldots};
		\node [style=none] (131) at (-6.5, 0.25) {};
		\node [style=none] (132) at (-6.5, 3.25) {};
		\node [style=none] (133) at (-1.75, 0.25) {};
		\node [style=none] (134) at (-1.75, 3.25) {};
		\node [style=none] (135) at (1.25, 1.75) {$m$};
		\node [style=none] (136) at (2, 0.25) {};
		\node [style=none] (137) at (2, 3.25) {};
		\node [style=none] (138) at (8, 1.75) {$n$};
		\node [style=none] (139) at (7.5, 0.25) {};
		\node [style=none] (140) at (7.5, 3.25) {};
		\node [style=none] (141) at (15.75, 1.75) {$n$};
		\node [style=none] (142) at (15.25, 0.5) {};
		\node [style=none] (143) at (15.25, 3) {};
		\node [style=none] (144) at (-4.5, -2.75) {};
		\node [style=none] (145) at (-1.25, -2.75) {};
		\node [style=Z dot] (146) at (-3.5, -2.75) {};
		\node [style=X dot] (147) at (-2, -2.75) {};
		\node [style=none] (148) at (2.25, -2.75) {};
		\node [style=none] (149) at (4.75, -2.75) {};
		\node [style=Z dot] (150) at (3, -2.75) {};
		\node [style=X dot] (151) at (4, -2.75) {};
		\node [style=none] (152) at (0, -2.75) {=};
		\node [style=none] (153) at (1.5, -2.75) {1/2};
		\node [style=none] (154) at (0, -1.75) {\hopfrule};
		\node [style=H] (155) at (14, -4.25) {};
		\node [style=none] (156) at (13.25, -4.25) {};
		\node [style=none] (157) at (15.75, -4.25) {};
		\node [style=none] (158) at (19.25, -4.25) {};
		\node [style=none] (159) at (20.75, -4.25) {};
		\node [style=none] (160) at (17, -3.5) {\hhrule};
		\node [style=none] (161) at (17, -4.25) {$=$};
		\node [style=H] (162) at (15, -4.25) {};
		\node [style=none] (163) at (18.25, -4.25) {$2$};
		\node [style=hadamard] (164) at (15, -6.75) {};
		\node [style=X phase dot] (165) at (14, -6.75) {$a\pi$};
		\node [style=none] (166) at (16, -6.75) {};
		\node [style=none] (167) at (17, -6.75) {=};
		\node [style=Z phase dot] (168) at (18.25, -6.75) {$a\pi$};
		\node [style=none] (169) at (19.25, -6.75) {};
		\node [style=none] (170) at (17, -6) {\hcopy};
		\node [style=none] (171) at (0, -5) {\hcolorchange};
	\end{pgfonlayer}
	\begin{pgfonlayer}{edgelayer}
		\draw [in=-141, out=0, looseness=0.75] (3.center) to (0);
		\draw [in=180, out=-39, looseness=0.75] (0) to (4.center);
		\draw [in=180, out=22, looseness=0.75] (0) to (5.center);
		\draw [in=180, out=39, looseness=0.75] (0) to (1.center);
		\draw [in=0, out=158, looseness=0.75] (0) to (6.center);
		\draw [in=141, out=0, looseness=0.75] (2.center) to (0);
		\draw (20.center) to (3.center);
		\draw (19.center) to (6.center);
		\draw (18.center) to (2.center);
		\draw (1.center) to (15.center);
		\draw (5.center) to (16.center);
		\draw (4.center) to (17.center);
		\draw [in=-141, out=0, looseness=0.75] (24.center) to (21);
		\draw [in=180, out=-39, looseness=0.75] (21) to (25.center);
		\draw [in=180, out=22, looseness=0.75] (21) to (26.center);
		\draw [in=180, out=39, looseness=0.75] (21) to (22.center);
		\draw [in=0, out=158, looseness=0.75] (21) to (27.center);
		\draw [in=141, out=0, looseness=0.75] (23.center) to (21);
		\draw [decorate, style=brace edge] (32.center) to (33.center);
		\draw [decorate, style=brace edge] (35.center) to (34.center);
		\draw [decorate, style=brace edge] (38.center) to (39.center);
		\draw [decorate, style=brace edge] (42.center) to (41.center);
		\draw [in=-141, out=0, looseness=0.75] (47.center) to (44);
		\draw [in=180, out=-39, looseness=0.75] (44) to (48.center);
		\draw [in=180, out=39, looseness=0.75] (44) to (45.center);
		\draw [in=180, out=0] (46.center) to (44);
		\draw [in=-141, out=0, looseness=0.75] (56.center) to (51);
		\draw [in=180, out=0] (51) to (54.center);
		\draw [in=180, out=39, looseness=0.75] (51) to (52.center);
		\draw [in=141, out=0, looseness=0.75] (53.center) to (51);
		\draw [bend left=45] (51) to (44);
		\draw (52.center) to (59.center);
		\draw (54.center) to (58.center);
		\draw (61.center) to (46.center);
		\draw (60.center) to (47.center);
		\draw [bend left=45] (44) to (51);
		\draw [in=-120, out=0] (70.center) to (68);
		\draw [in=180, out=-60] (68) to (66.center);
		\draw [in=180, out=60] (68) to (69.center);
		\draw [in=120, out=0] (64.center) to (68);
		\draw [in=180, out=-75, looseness=0.75] (80) to (84.center);
		\draw [in=180, out=75, looseness=0.75] (80) to (79.center);
		\draw [in=180, out=0, looseness=0.50] (78) to (80);
		\draw (83.center) to (78);
		\draw [in=-60, out=180, looseness=0.75] (77) to (72);
		\draw [in=0, out=180, looseness=0.75] (72) to (75.center);
		\draw [in=60, out=180, looseness=0.75] (85) to (72);
		\draw (73.center) to (85);
		\draw (76.center) to (77);
		\draw [in=180, out=-75, looseness=0.75] (89) to (90.center);
		\draw [in=180, out=75, looseness=0.75] (89) to (97.center);
		\draw [in=180, out=0, looseness=0.50] (98) to (89);
		\draw (93.center) to (87);
		\draw (92.center) to (95);
		\draw (100.center) to (101.center);
		\draw (102.center) to (103.center);
		\draw [in=105, out=-15, looseness=0.75] (51) to (44);
		\draw [in=60, out=-180] (111.center) to (110);
		\draw [in=180, out=-60] (110) to (112.center);
		\draw (109) to (110);
		\draw [in=360, out=90, looseness=0.75] (109) to (117.center);
		\draw [in=0, out=-90, looseness=0.75] (109) to (118.center);
		\draw (125.center) to (121);
		\draw (122) to (126.center);
		\draw (123) to (127.center);
		\draw (124) to (128.center);
		\draw (123) to (122);
		\draw (121) to (123);
		\draw (124) to (121);
		\draw (122) to (124);
		\draw [style=brace edge] (131.center) to (132.center);
		\draw [style=brace edge] (134.center) to (133.center);
		\draw [style=brace edge] (136.center) to (137.center);
		\draw [style=brace edge] (140.center) to (139.center);
		\draw [style=brace edge] (143.center) to (142.center);
		\draw (144.center) to (146);
		\draw (147) to (145.center);
		\draw [bend left] (146) to (147);
		\draw [bend right, looseness=1.25] (146) to (147);
		\draw (148.center) to (150);
		\draw (151) to (149.center);
		\draw (156.center) to (157.center);
		\draw (158.center) to (159.center);
		\draw (164) to (165);
		\draw (164) to (166.center);
		\draw (168) to (169.center);
	\end{pgfonlayer}
\end{tikzpicture}
\caption{{ The rules of  ZX calculus.} The dots (``\dots'') refer to the extension of the rule to diagrams with 0 or more lines. The rule names stand, respectively, for $\FusionRule$usion, $\PiRule$-commutation, $\BialgRule$ialgebra, $\CopyRule$-copy, $\HopfRule$pf, $\IdRule$entity, $\HHRule$ Hadamard identity, $\HColorChange$ Hadamard color change and $\HCopy$ Hadamard copy. The addition between the phases $\alpha$ and $\beta$ in \FusionRule is to be understood modulo $2\pi$. The right-hand side of \BialgRule is a fully connected bipartite graph. }   \label{fig:zx-rules}
\end{figure*}

Figure~\ref{fig:zx-rules} shows the rewrite rules of ZX calculus, where the name of each rule is  indicated in parenthesis.
In this work, we will refer to the name of the rule  as we use them in each step of the diagrammatic simplifications.
For more details on the derivation of these rules, we refer the reader to  Refs.~\cite{vandewetering2020zxcalculus, KissingerWetering2024Book}.

In addition to the rewrite rules in Fig.~\ref{fig:zx-rules}, ZX diagrams are equivalent independently of the orientation of the wires, as long as the order of the inputs and of the outputs is preserved~\cite{vandewetering2020zxcalculus}.
This property, known as \textit{only connectivity matters} \cite{KissingerWetering2024Book}, is illustrated in the following example
\begin{equation}
\begin{tikzpicture}
	\begin{pgfonlayer}{nodelayer}
		\node [style=none] (0) at (-2, 2) {};
		\node [style=none] (1) at (-2, 0) {};
		\node [style=none] (2) at (-2, -2) {};
		\node [style=none] (3) at (2, 2) {};
		\node [style=none] (4) at (2, 0) {};
		\node [style=none] (5) at (2, -2) {};
		\node [style=Z phase dot] (6) at (0.75, -2) {$\frac{\pi}{2}$};
		\node [style=Z phase dot] (7) at (0.75, 0) {$\beta$};
		\node [style=Z phase dot] (8) at (0, 1) {$\pi$};
		\node [style=X phase dot] (9) at (-0.75, 2) {$\alpha$};
		\node [style=X phase dot] (10) at (-0.75, 0) {};
		\node [style=hadamard] (11) at (-0.75, -2) {};
		\node [style=none] (12) at (4, 2) {};
		\node [style=none] (13) at (4, 0) {};
		\node [style=none] (14) at (4, -2) {};
		\node [style=none] (15) at (8, 2) {};
		\node [style=none] (16) at (8, 0) {};
		\node [style=none] (17) at (8, -2) {};
		\node [style=Z phase dot] (18) at (6.75, 0) {$\frac{\pi}{2}$};
		\node [style=Z phase dot] (19) at (5.5, -0.75) {$\beta$};
		\node [style=Z phase dot] (20) at (5.25, 0.25) {$\pi$};
		\node [style=X phase dot] (21) at (4.75, 1.25) {$\alpha$};
		\node [style=X phase dot] (22) at (6.5, 1.25) {};
		\node [style=hadamard] (23) at (5.25, -1.75) {};
		\node [style=none] (24) at (3, 0) {=};
        \node [style=none] (25) at (3, 0.55) {};
	\end{pgfonlayer}
	\begin{pgfonlayer}{edgelayer}
		\draw [style=none] (2.center) to (11);
		\draw [style=none] (11) to (6);
		\draw [style=none] (6) to (5.center);
		\draw [style=none] (6) to (10);
		\draw [style=none] (10) to (7);
		\draw [style=none] (7) to (4.center);
		\draw [style=none] (10) to (1.center);
		\draw [style=none] (10) to (8);
		\draw [style=none] (8) to (9);
		\draw [style=none] (9) to (3.center);
		\draw [style=none] (9) to (0.center);
		\draw [style=none] (14.center) to (23);
		\draw [style=none, bend right=15] (23) to (18);
		\draw [style=none, bend right=15, looseness=0.50] (18) to (17.center);
		\draw [style=none] (18) to (22);
		\draw [style=none] (22) to (19);
		\draw [style=none, bend right=15] (19) to (16.center);
		\draw [style=none, bend right=15] (22) to (13.center);
		\draw [style=none] (22) to (20);
		\draw [style=none] (20) to (21);
		\draw [style=none, bend left=15] (21) to (15.center);
		\draw [style=none, bend right, looseness=0.75] (21) to (12.center);
	\end{pgfonlayer}
\end{tikzpicture}\,.
\end{equation}
In our work, we will often rearrange diagrams in this way to simplify their physical interpretation and we will not indicate explicitly  the use of this rule.
Lastly, we note that the rules shown in Fig.~\ref{fig:zx-rules} hold also when green and red spiders are swapped, when inputs and outputs are swapped, or if all the phases are conjugated.

We conclude by mentioning two additional  relations, which are useful to define superpositions of diagrams.  Up to scalar factors, these superposition $\SupRule$ rules read 
\begin{align} \label{eq:ZXsuperposition}
\begin{tikzpicture}
    \begin{pgfonlayer}{nodelayer}
        \node [style=Z phase dot] (0) at (-0.5, 0) {$~$};
        \node [style=none] (1) at (0.5, 0) {};
    \end{pgfonlayer}
    \begin{pgfonlayer}{edgelayer}
        \draw (0) to (1.center);
    \end{pgfonlayer}
\end{tikzpicture} &\stackrel{\SupRule}{=}
\begin{tikzpicture}
    \begin{pgfonlayer}{nodelayer}
        \node [style=X phase dot] (0) at (-0.5, 0) {$~$};
        \node [style=none] (1) at (0.5, 0) {};
    \end{pgfonlayer}
    \begin{pgfonlayer}{edgelayer}
        \draw (0) to (1.center);
    \end{pgfonlayer}
\end{tikzpicture}%
+
\begin{tikzpicture}
    \begin{pgfonlayer}{nodelayer}
        \node [style=X phase dot] (0) at (-0.5, 0) {$\pi$};
        \node [style=none] (1) at (0.5, 0) {};
    \end{pgfonlayer}
    \begin{pgfonlayer}{edgelayer}
        \draw (0) to (1.center);
    \end{pgfonlayer}
\end{tikzpicture}\,,&
\begin{tikzpicture}
    \begin{pgfonlayer}{nodelayer}
        \node [style=Z phase dot] (0) at (-0.5, 0) {$\pi$};
        \node [style=none] (1) at (0.5, 0) {};
    \end{pgfonlayer}
    \begin{pgfonlayer}{edgelayer}
        \draw (0) to (1.center);
    \end{pgfonlayer}
\end{tikzpicture} &\stackrel{\SupRule}{=}
\begin{tikzpicture}
    \begin{pgfonlayer}{nodelayer}
        \node [style=X phase dot] (0) at (-0.5, 0) {$~$};
        \node [style=none] (1) at (0.5, 0) {};
    \end{pgfonlayer}
    \begin{pgfonlayer}{edgelayer}
        \draw (0) to (1.center);
    \end{pgfonlayer}
\end{tikzpicture}%
-
\begin{tikzpicture}
    \begin{pgfonlayer}{nodelayer}
        \node [style=X phase dot] (0) at (-0.5, 0) {$\pi$};
        \node [style=none] (1) at (0.5, 0) {};
    \end{pgfonlayer}
    \begin{pgfonlayer}{edgelayer}
        \draw (0) to (1.center);
    \end{pgfonlayer}
\end{tikzpicture}\,.%
\end{align}
They directly follow from the diagrammatic representation of $\ket{0}$, $\ket{1}$, $\ket{+}$ and $\ket{-}$ in Eq.~\eqref{eq:zx_spider_examples} and also hold if we interchange the two colors.

Since the rules~\eqref{eq:ZXsuperposition} are not rigorously part of the ZX-calculus rules (there are no formal `sums' of diagrams), they are not included in Fig.~\ref{fig:zx-rules}. However, we will rely heavily on them, see for instance Eq.~(3) in the main text, where they are used to construct the ZX representation of the ground states of the  toric and color codes.


\subsection{CNOT and CZ gates \label{sec:ZXcontrolgates}}

To illustrate the ZX-calculus rules, we discuss the examples of the control NOT (CNOT) and control Z  (CZ) gates~\cite{KissingerWetering2024Book}, also used in this work.

The CNOT operation  flips the state of a target qubit $q_t$ conditional to the state of a control qubit $q_c$. Namely, $\sigma_x$ is applied on $q_t$ only if $\ket {q_c}=\ket 1$. In the computational basis $\{\ket{q_cq_t}\}=\{\ket{00},\ket{01},\ket{10},\ket{11}\}$, the CNOT operation is  represented by the matrix
\begin{equation}
\mathrm{CNOT} = 
\begin{pmatrix}
1 & 0 & 0 & 0\\
0 & 1 & 0 & 0\\
0 & 0 & 0 & 1\\
0 & 0 & 1 & 0
\end{pmatrix}.
\end{equation}
As a ZX diagram, the CNOT is represented  as~\cite{KissingerWetering2024Book}
\begin{equation}
\mathrm{CNOT} =
\begin{tikzpicture}
	\begin{pgfonlayer}{nodelayer}
		\node [style=Z dot] (0) at (0, 0.625) {};
		\node [style=X dot] (1) at (0, -0.625) {};
		\node [style=none] (3) at (1.25, 0.625) {};
		\node [style=none] (4) at (1.25, -0.625) {};
		\node [style=none] (5) at (-1.25, -0.625) {};
		\node [style=none] (6) at (-1.25, 0.625) {};
	\end{pgfonlayer}
	\begin{pgfonlayer}{edgelayer}
		\draw [style=none] (0) to (1);
		\draw [style=none] (0) to (3.center);
		\draw [style=none] (1) to (4.center);
		\draw [style=none] (1) to (5.center);
		\draw [style=none] (6.center) to (0);
	\end{pgfonlayer}
\end{tikzpicture}\,.
\end{equation}
We can prove that this diagrammatic representation is correct   by using the  $\FusionRule, \CopyRule, \IdentityRule$ rules in Fig.~\ref{fig:zx-rules}.
We first apply  the upper  external leg to the state $\ket{0}$:
\begin{subequations}
    \begin{equation}\label{eq:proofCNOT1}
        \begin{tikzpicture}
	\begin{pgfonlayer}{nodelayer}
		\node [style=Z phase dot] (0) at (0, 0.5) {};
		\node [style=X phase dot] (1) at (0, -0.75) {};
		\node [style=none] (3) at (1.25, 0.5) {};
		\node [style=none] (4) at (1.25, -0.75) {};
		\node [style=none] (5) at (-1.25, -0.75) {};
		\node [style=X phase dot] (6) at (-1.5, 0.5) {};
		\node [style=X phase dot] (8) at (4, -0.75) {};
		\node [style=none] (9) at (5.25, 0.5) {};
		\node [style=none] (10) at (5.25, -0.75) {};
		\node [style=none] (11) at (2.75, -0.75) {};
		\node [style=X phase dot] (12) at (4.5, 0.5) {};
		\node [style=X phase dot] (13) at (4, 0) {};
		\node [style=X phase dot] (14) at (8, -0.75) {};
		\node [style=none] (15) at (9.25, 0.5) {};
		\node [style=none] (16) at (9.25, -0.75) {};
		\node [style=none] (17) at (6.75, -0.75) {};
		\node [style=X phase dot] (18) at (8.5, 0.5) {};
		\node [style=none] (20) at (12.75, 0.5) {};
		\node [style=none] (21) at (12.75, -0.75) {};
		\node [style=none] (22) at (10.75, -0.75) {};
		\node [style=X phase dot] (23) at (12, 0.5) {};
		\node [style=none] (24) at (2, -0.25) {=};
		\node [style=none] (25) at (6, -0.25) {=};
		\node [style=none] (26) at (10, -0.25) {=};
		\node [style=none] (27) at (2, 0.25) {\copyrule};
		\node [style=none] (28) at (6, 0.25) {\fusionrule};
		\node [style=none] (29) at (10, 0.25) {\idrule};
	\end{pgfonlayer}
	\begin{pgfonlayer}{edgelayer}
		\draw (0) to (1);
		\draw (0) to (3.center);
		\draw (1) to (4.center);
		\draw (1) to (5.center);
		\draw (6) to (0);
		\draw (8) to (10.center);
		\draw (8) to (11.center);
		\draw (12) to (9.center);
		\draw (13) to (8);
		\draw (14) to (16.center);
		\draw (14) to (17.center);
		\draw (18) to (15.center);
		\draw (23) to (20.center);
		\draw (22.center) to (21.center);
	\end{pgfonlayer}
\end{tikzpicture}\,.
    \end{equation}
The last diagram shows that the top qubit remains unchanged in the state $\ket{0}$, while a naked wire acts upon the bottom qubit, which is by definition the identity matrix.
Similarly, we apply  the top external leg  to the state $\ket{1}$:
    \begin{equation}
        \begin{tikzpicture}
	\begin{pgfonlayer}{nodelayer}
		\node [style=Z phase dot] (30) at (0, 0.5) {};
		\node [style=X phase dot] (31) at (0, -0.75) {};
		\node [style=none] (32) at (1.25, 0.5) {};
		\node [style=none] (33) at (1.25, -0.75) {};
		\node [style=none] (34) at (-1.25, -0.75) {};
		\node [style=X phase dot] (36) at (4, -0.75) {};
		\node [style=none] (37) at (5.25, 0.5) {};
		\node [style=none] (38) at (5.25, -0.75) {};
		\node [style=none] (39) at (2.75, -0.75) {};
		\node [style=none] (43) at (9.25, 0.5) {};
		\node [style=none] (44) at (9.25, -0.75) {};
		\node [style=none] (45) at (6.75, -0.75) {};
		\node [style=none] (51) at (2, -0.25) {=};
		\node [style=none] (52) at (6, -0.25) {=};
		\node [style=none] (54) at (2, 0.25) {\copyrule};
		\node [style=none] (55) at (6, 0.25) {\fusionrule};
		\node [style=X phase dot] (57) at (-1.5, 0.5) {$\pi$};
		\node [style=X phase dot] (58) at (4.5, 0.5) {$\pi$};
		\node [style=X phase dot] (59) at (4, 0) {$\pi$};
		\node [style=X phase dot] (60) at (8, -0.75) {$\pi$};
		\node [style=X phase dot] (61) at (8.5, 0.5) {$\pi$};
	\end{pgfonlayer}
	\begin{pgfonlayer}{edgelayer}
		\draw (30) to (31);
		\draw (30) to (32.center);
		\draw (31) to (33.center);
		\draw (31) to (34.center);
		\draw (36) to (38.center);
		\draw (36) to (39.center);
		\draw (57) to (30);
		\draw (59) to (36);
		\draw (58) to (37.center);
		\draw (45.center) to (60);
		\draw (60) to (44.center);
		\draw (61) to (43.center);
	\end{pgfonlayer}
\end{tikzpicture}\,.
    \end{equation}
\end{subequations}
This time, the top qubit remains in the state $\ket{1}$, while the bottom qubit is flipped by the action of $\sigma_x$, see Eq.~\eqref{eq:sigmax_zx}.

We can reason similarly to represent the CZ gate as a ZX diagram. The CZ gate works as a CNOT, but acts conditionally on the target qubit with $\sigma_z$ instead of $\sigma_x$. In matrix form, the CZ gate reads
\begin{equation}\label{eq:CZ}
\mathrm{CZ} = 
\begin{pmatrix}
1 & 0 & 0 & 0\\
0 & 1 & 0 & 0\\
0 & 0 & 1 & 0\\
0 & 0 & 0 & -1
\end{pmatrix}\,.
\end{equation}
As a ZX diagram, the CZ gate is represented by~\cite{KissingerWetering2024Book}
\begin{equation}\label{eq:CZZX}
\mathrm{CZ} =
\begin{tikzpicture}
	\begin{pgfonlayer}{nodelayer}
		\node [style=Z dot] (0) at (0, 0.625) {};
		\node [style=Z dot] (1) at (0, -0.625) {};
		\node [style=none] (3) at (1.25, 0.625) {};
		\node [style=none] (4) at (1.25, -0.625) {};
		\node [style=none] (5) at (-1.25, -0.625) {};
		\node [style=none] (6) at (-1.25, 0.625) {};
		\node [style=Z dot] (8) at (3.5, 0.625) {};
		\node [style=Z dot] (9) at (3.5, -0.625) {};
		\node [style=none] (10) at (4.75, 0.625) {};
		\node [style=none] (11) at (4.75, -0.625) {};
		\node [style=none] (12) at (2.25, -0.625) {};
		\node [style=none] (13) at (2.25, 0.625) {};
		\node [style=hadamard] (14) at (3.5, 0) {};
		\node [style=none] (15) at (1.75, 0) {=};
	\end{pgfonlayer}
	\begin{pgfonlayer}{edgelayer}
		\draw [style=none] (0) to (3.center);
		\draw [style=none] (1) to (4.center);
		\draw [style=none] (1) to (5.center);
		\draw [style=none] (6.center) to (0);
		\draw [style=none] (8) to (10.center);
		\draw [style=none] (9) to (11.center);
		\draw [style=none] (9) to (12.center);
		\draw [style=none] (13.center) to (8);
		\draw (8) to (14);
		\draw (14) to (9);
		\draw [style=hadamard edge] (0) to (1);
	\end{pgfonlayer}
\end{tikzpicture}\,.
\end{equation}
The equivalence between Eqs.~\eqref{eq:CZ} and~\eqref{eq:CZZX} is shown by  applying the $\FusionRule, \CopyRule, \HCopy, \IdentityRule$ rules in Fig.~\ref{fig:zx-rules}. Proceeding analogously as for the CNOT, we  have:
\begin{subequations}
    \begin{equation}
    \label{eq:CZa}
\begin{tikzpicture}
	\begin{pgfonlayer}{nodelayer}
		\node [style=Z phase dot] (0) at (2, 0.75) {};
		\node [style=none] (1) at (3.25, 0.75) {};
		\node [style=none] (2) at (3.25, -1.25) {};
		\node [style=none] (3) at (0.75, -1.25) {};
		\node [style=X phase dot] (4) at (0.5, 0.75) {};
		\node [style=none] (5) at (6.75, 0.75) {};
		\node [style=none] (6) at (6.75, -1.25) {};
		\node [style=none] (7) at (4.25, -1.25) {};
		\node [style=X phase dot] (8) at (6, 0.75) {};
		\node [style=X phase dot] (9) at (5.5, 0.25) {};
		\node [style=none] (14) at (14.5, 0.75) {};
		\node [style=none] (15) at (14.5, -1.25) {};
		\node [style=none] (16) at (12, -1.25) {};
		\node [style=X phase dot] (17) at (13.75, 0.75) {};
		\node [style=none] (18) at (3.75, -0.25) {=};
		\node [style=none] (19) at (11.25, -0.25) {=};
		\node [style=none] (21) at (3.75, 0.25) {\copyrule};
		\node [style=none] (22) at (11.25, 0.25) {\idrule \fusionrule};
		\node [style=hadamard] (24) at (2, -0.25) {};
		\node [style=hadamard] (25) at (5.5, -0.5) {};
		\node [style=Z phase dot] (26) at (2, -1.25) {};
		\node [style=Z phase dot] (27) at (5.5, -1.25) {};
		\node [style=none] (29) at (10.25, 0.75) {};
		\node [style=none] (30) at (10.25, -1.25) {};
		\node [style=none] (31) at (7.75, -1.25) {};
		\node [style=X phase dot] (32) at (9.5, 0.75) {};
		\node [style=Z phase dot] (33) at (9, -1.25) {};
		\node [style=Z phase dot] (34) at (9, -0.25) {};
		\node [style=none] (35) at (7.25, -0.25) {=};
		\node [style=none] (36) at (7.25, 0.25) {\hcopy};
	\end{pgfonlayer}
	\begin{pgfonlayer}{edgelayer}
		\draw (0) to (1.center);
		\draw (4) to (0);
		\draw (8) to (5.center);
		\draw (17) to (14.center);
		\draw (16.center) to (15.center);
		\draw (0) to (24);
		\draw (24) to (26);
		\draw (26) to (3.center);
		\draw (26) to (2.center);
		\draw (7.center) to (27);
		\draw (27) to (6.center);
		\draw (25) to (27);
		\draw (9) to (25);
		\draw (32) to (29.center);
		\draw (31.center) to (33);
		\draw (33) to (30.center);
		\draw (34) to (33);
	\end{pgfonlayer}
\end{tikzpicture}\,,
    \end{equation}
    \begin{equation}
     \label{eq:CZb}
\begin{tikzpicture}
	\begin{pgfonlayer}{nodelayer}
		\node [style=Z phase dot] (0) at (0, 0.75) {};
		\node [style=none] (3) at (1.25, 0.75) {};
		\node [style=none] (4) at (1.25, -1.25) {};
		\node [style=none] (5) at (-1.25, -1.25) {};
		\node [style=none] (9) at (4.75, 0.75) {};
		\node [style=none] (10) at (4.75, -1.25) {};
		\node [style=none] (11) at (2.25, -1.25) {};
		\node [style=none] (15) at (11.75, 0.75) {};
		\node [style=none] (16) at (11.75, -1.25) {};
		\node [style=none] (17) at (9.25, -1.25) {};
		\node [style=none] (24) at (1.75, -0.25) {=};
		\node [style=none] (25) at (8.75, -0.25) {=};
		\node [style=none] (27) at (1.75, 0.25) {\copyrule};
		\node [style=none] (28) at (8.75, 0.25) {\fusionrule};
		\node [style=hadamard] (30) at (0, -0.25) {};
		\node [style=hadamard] (31) at (3.5, -0.5) {};
		\node [style=Z phase dot] (32) at (0, -1.25) {};
		\node [style=Z phase dot] (33) at (3.5, -1.25) {};
		\node [style=none] (35) at (8.25, 0.75) {};
		\node [style=none] (36) at (8.25, -1.25) {};
		\node [style=none] (37) at (5.75, -1.25) {};
		\node [style=Z phase dot] (41) at (7, -1.25) {};
		\node [style=none] (43) at (5.25, -0.25) {=};
		\node [style=none] (44) at (5.25, 0.25) {\hcopy};
		\node [style=X phase dot] (45) at (-1.5, 0.75) {$\pi$};
		\node [style=X phase dot] (46) at (4, 0.75) {$\pi$};
		\node [style=X phase dot] (47) at (3.5, 0.25) {$\pi$};
		\node [style=X phase dot] (48) at (7.5, 0.75) {$\pi$};
		\node [style=X phase dot] (49) at (11, 0.75) {$\pi$};
		\node [style=Z phase dot] (51) at (7, -0.25) {$\pi$};
		\node [style=Z phase dot] (52) at (10.5, -1.25) {$\pi$};
	\end{pgfonlayer}
	\begin{pgfonlayer}{edgelayer}
		\draw (0) to (3.center);
		\draw (0) to (30);
		\draw (30) to (32);
		\draw (32) to (5.center);
		\draw (32) to (4.center);
		\draw (11.center) to (33);
		\draw (33) to (10.center);
		\draw (31) to (33);
		\draw (37.center) to (41);
		\draw (41) to (36.center);
		\draw (45) to (0);
		\draw (47) to (31);
		\draw (46) to (9.center);
		\draw (48) to (35.center);
		\draw (49) to (15.center);
		\draw (17.center) to (52);
		\draw (52) to (16.center);
		\draw (51) to (41);
	\end{pgfonlayer}
\end{tikzpicture}\,.
    \end{equation}
\end{subequations}
Equation~\eqref{eq:CZa} shows that, when the first qubit is in the $\ket{0}$ state, no qubit changes, as for the CNOT, Eq.~\eqref{eq:proofCNOT1}. 
Equation~\eqref{eq:CZb} preserves the top qubit  unchanged in state $\ket{1}$, while $\sigma_z$ acts on the bottom qubit, see Eq.~\eqref{eq:sigmaz_zx}.


\subsection{Transposition, Conjugation and Tracing}\label{sec:ZXtranspositionconjugationtracing}
For the purpose of this paper, we illustrate the transposition, conjugation and tracing of ZX diagrams~\cite{KissingerWetering2024Book}.

{\it Transposition -- }To transpose a diagram, 
we convert its inputs as outputs and vice versa.
For example,
\begin{equation}
\begin{tikzpicture}
	\begin{pgfonlayer}{nodelayer}
		\node [style=none] (24) at (2, 0) {$\rightarrow$};
		\node [style=none] (25) at (2, 0.5) {$^{t}$};
		\node [style=X phase dot] (26) at (0, 0.75) {$~$};
		\node [style=X phase dot] (27) at (0, -0.75) {$~$};
		\node [style=X phase dot] (28) at (0, 0) {$\pi$};
		\node [style=none] (29) at (1, 0.75) {};
		\node [style=none] (30) at (1, 0) {};
		\node [style=none] (31) at (1, -0.75) {};
		\node [style=X phase dot] (32) at (4, 0.75) {$~$};
		\node [style=X phase dot] (33) at (4, -0.75) {$~$};
		\node [style=X phase dot] (34) at (4, 0) {$\pi$};
		\node [style=none] (35) at (3, 0.75) {};
		\node [style=none] (36) at (3, 0) {};
		\node [style=none] (37) at (3, -0.75) {};
	\end{pgfonlayer}
	\begin{pgfonlayer}{edgelayer}
		\draw (26) to (29.center);
		\draw (30.center) to (28);
		\draw (27) to (31.center);
		\draw (32) to (35.center);
		\draw (36.center) to (34);
		\draw (33) to (37.center);
	\end{pgfonlayer}
\end{tikzpicture}\,,
\end{equation}
or
\begin{equation}
\begin{tikzpicture}
	\begin{pgfonlayer}{nodelayer}
		\node [style=X phase dot] (12) at (0.75, 0) {$\pi$};
		\node [style=Z phase dot] (13) at (0, 0) {$~$};
		\node [style=none] (16) at (-0.75, 0.5) {};
		\node [style=none] (17) at (1.5, 0) {};
		\node [style=none] (18) at (-0.75, -0.5) {};
		\node [style=X phase dot] (19) at (4.25, 0) {$\pi$};
		\node [style=Z phase dot] (20) at (5, 0) {$~$};
		\node [style=none] (21) at (5.75, 0.5) {};
		\node [style=none] (22) at (3.5, 0) {};
		\node [style=none] (23) at (5.75, -0.5) {};
		\node [style=none] (24) at (2.5, 0) {$\rightarrow$};
		\node [style=none] (25) at (2.5, 0.5) {$^{t}$};
	\end{pgfonlayer}
	\begin{pgfonlayer}{edgelayer}
		\draw [bend right] (13) to (16.center);
		\draw (12) to (17.center);
		\draw (13) to (12);
		\draw [bend left] (13) to (18.center);
		\draw [bend left] (20) to (21.center);
		\draw (19) to (22.center);
		\draw (20) to (19);
		\draw [bend right] (20) to (23.center);
	\end{pgfonlayer}
\end{tikzpicture}\,.
\end{equation}

{\it Conjugation -- }The complex conjugation of diagrams is achieved by conjugating the phase of each spider,  e.g.
\begin{equation}
\begin{tikzpicture}
	\begin{pgfonlayer}{nodelayer}
		\node [style=X phase dot] (12) at (1, 0) {$\frac{\pi}{2}$};
		\node [style=Z phase dot] (13) at (0, 0) {$\alpha$};
		\node [style=none] (16) at (-0.75, 0.5) {};
		\node [style=none] (17) at (1.75, 0) {};
		\node [style=none] (18) at (-0.75, -0.5) {};
		\node [style=none] (24) at (2.75, 0) {$\rightarrow$};
		\node [style=none] (25) at (2.75, 0.5) {$\overline{*}$};
		\node [style=X phase dot] (26) at (5.5, 0) {$-\frac{\pi}{2}$};
		\node [style=Z phase dot] (27) at (4.5, 0) {$-\alpha$};
		\node [style=none] (28) at (3.75, 0.5) {};
		\node [style=none] (29) at (6.25, 0) {};
		\node [style=none] (30) at (3.75, -0.5) {};
	\end{pgfonlayer}
	\begin{pgfonlayer}{edgelayer}
		\draw [bend right] (13) to (16.center);
		\draw (12) to (17.center);
		\draw (13) to (12);
		\draw [bend left] (13) to (18.center);
		\draw [bend right] (27) to (28.center);
		\draw (26) to (29.center);
		\draw (27) to (26);
		\draw [bend left] (27) to (30.center);
	\end{pgfonlayer}
\end{tikzpicture}\,,
\end{equation}
where $\overline{*}$ denotes the complex conjugation of the left diagram.
By  transposing and conjugating, we turn bras into kets, and obtain the Hermitian conjugate of a diagram.

{\it Tracing -- }Consider a many-qubit state $\ket{A,B}$, where the qubits are partitioned in two sets $A$ and $B$, and $\rho=\ket{A,B}\bra{A,B}$ is the corresponding density matrix (the extension to mixed states is straightforward). The reduced density matrix $\rho_A=\mbox{Tr}_B(\rho)$ is obtained by tracing over the qubits in the set $B$.  In diagrammatic language, this trace is performed by connecting the input wires corresponding to the $B$ qubits of $\bra{A,B}$ to the output wires corresponding to the $B$ qubits in $\ket{A,B}$.  
For example, consider the representation of the $N$-qubit GHZ state $\ket{\mbox{GHZ}}=(\ket{0\ldots0}+\ket{1\ldots1})/\sqrt2$ in ZX
\begin{equation}
\ket{\mbox{GHZ}}=
\begin{tikzpicture}
	\begin{pgfonlayer}{nodelayer}
		\node [style=Z dot] (0) at (0.5, 0) {};
		\node [style=none] (1) at (1.5, 2) {};
		\node [style=none] (2) at (1.5, 1) {};
		\node [style=none] (3) at (1.5, 0) {};
		\node [style=none] (4) at (1.5, -0.75) {$\vdots$};
		\node [style=none] (5) at (1.5, -2) {};
		\node [style=none] (6) at (3, 0) {A};
		\node [style=none] (7) at (2.5, -2) {};
		\node [style=none] (8) at (2.5, -0.75) {};
		\node [style=none] (9) at (2.5, 0.75) {};
		\node [style=none] (10) at (2.5, 2) {};
		\node [style=none] (11) at (3, 1.5) {B};
		\node [style=none] (12) at (3, -1.5) {B};
		\node [style=none] (13) at (2.5, 0.25) {};
		\node [style=none] (14) at (2.5, -0.25) {};
	\end{pgfonlayer}
	\begin{pgfonlayer}{edgelayer}
		\draw [bend left=15, looseness=0.50] (0) to (1.center);
		\draw [bend left=15, looseness=0.50] (0) to (2.center);
		\draw (0) to (3.center);
		\draw [bend right=15, looseness=0.50] (0) to (5.center);
		\draw [style=gray] (10.center) to (9.center);
		\draw [style=gray] (8.center) to (7.center);
		\draw [style=gray] (13.center) to (14.center);
	\end{pgfonlayer}
\end{tikzpicture}\,,
\end{equation}
where $A$ and $B$ contain one and $N-1$ qubits respectively.  Using  ZX calculus, one can readily  show that the  reduced density matrix $\rho_A$ equals the identity matrix
\begin{equation}
\begin{tikzpicture}
	\begin{pgfonlayer}{nodelayer}
		\node [style=Z dot] (15) at (0, 0) {};
		\node [style=none] (16) at (-1, 2) {};
		\node [style=none] (17) at (-1, 1) {};
		\node [style=none] (18) at (-1, 0) {};
		\node [style=none] (19) at (-1, -0.75) {$\vdots$};
		\node [style=none] (20) at (-1, -2) {};
		\node [style=Z dot] (21) at (1, 0) {};
		\node [style=none] (22) at (2, 2) {};
		\node [style=none] (23) at (2, 1) {};
		\node [style=none] (24) at (2, 0) {};
		\node [style=none] (25) at (2, -0.75) {$\vdots$};
		\node [style=none] (26) at (2, -2) {};
		\node [style=none] (27) at (3.25, 0) {$\rightarrow$};
		\node [style=Z dot] (28) at (5.5, 0) {};
		\node [style=none] (29) at (4.5, 0) {};
		\node [style=Z dot] (30) at (6.5, 0) {};
		\node [style=none] (31) at (7.5, 0) {};
		\node [style=none] (32) at (6, -1.25) {$\vdots$};
		\node [style=none] (33) at (8.5, 0) {=};
		\node [style=none] (34) at (9.5, 0) {};
		\node [style=none] (35) at (11, 0) {};
		\node [style=Z dot] (36) at (10.25, 0) {};
		\node [style=none] (37) at (12, 0) {=};
		\node [style=none] (38) at (13, 0) {};
		\node [style=none] (39) at (14.5, 0) {};
		\node [style=none] (40) at (3.25, 0.75) {$\Tr_{B}(\rho)$};
		\node [style=none] (41) at (8.5, 0.75) {$\FusionRule$};
		\node [style=none] (42) at (12, 0.75) {$\IdRule$};
		\node [style=none] (43) at (6.75, 1) {};
		\node [style=none] (44) at (7, 0.5) {};
		\node [style=none] (45) at (5.25, 1) {};
		\node [style=none] (46) at (5, 0.5) {};
		\node [style=none] (47) at (5, -0.5) {};
		\node [style=none] (48) at (5.25, -1) {};
		\node [style=none] (49) at (7, -0.5) {};
		\node [style=none] (50) at (6.75, -1) {};
		\node [style=none] (51) at (7.25, 1.5) {};
		\node [style=none] (52) at (7, 2) {};
		\node [style=none] (53) at (4.75, 1.5) {};
		\node [style=none] (54) at (5, 2) {};
		\node [style=none] (55) at (4.75, -1.5) {};
		\node [style=none] (56) at (5, -2) {};
		\node [style=none] (57) at (7, -2) {};
		\node [style=none] (58) at (7.25, -1.5) {};
	\end{pgfonlayer}
	\begin{pgfonlayer}{edgelayer}
		\draw [bend right=15, looseness=0.50] (15) to (16.center);
		\draw [bend right=15, looseness=0.50] (15) to (17.center);
		\draw (15) to (18.center);
		\draw [bend left=15, looseness=0.50] (15) to (20.center);
		\draw [bend left=15, looseness=0.50] (21) to (22.center);
		\draw [bend left=15, looseness=0.50] (21) to (23.center);
		\draw (21) to (24.center);
		\draw [bend right=15, looseness=0.50] (21) to (26.center);
		\draw (28) to (29.center);
		\draw (30) to (31.center);
		\draw (34.center) to (36);
		\draw (38.center) to (39.center);
		\draw (36) to (35.center);
		\draw (28) to (46.center);
		\draw [in=180, out=135, looseness=2.00] (46.center) to (45.center);
		\draw (45.center) to (43.center);
		\draw [in=45, out=0, looseness=2.00] (43.center) to (44.center);
		\draw (44.center) to (30);
		\draw (28) to (47.center);
		\draw [in=180, out=-135, looseness=2.00] (47.center) to (48.center);
		\draw (48.center) to (50.center);
		\draw [in=-45, out=0, looseness=2.00] (50.center) to (49.center);
		\draw (49.center) to (30);
		\draw (30) to (58.center);
		\draw [in=0, out=-60, looseness=1.50] (58.center) to (57.center);
		\draw (57.center) to (56.center);
		\draw [in=-120, out=180, looseness=1.50] (56.center) to (55.center);
		\draw (55.center) to (28);
		\draw (28) to (53.center);
		\draw [in=-180, out=120, looseness=1.50] (53.center) to (54.center);
		\draw (54.center) to (52.center);
		\draw [in=60, out=0, looseness=1.50] (52.center) to (51.center);
		\draw (51.center) to (30);
	\end{pgfonlayer}
\end{tikzpicture}\,.
\end{equation}


\section{Topological Entanglement Entropy} \label{sec:TEE}

\subsection{Entanglement entropy of gapped ground states}

In this Section, we  review for completeness the definition of the topological entanglement entropy~\cite{KitaevPreskill2006, LevinWen2006} 
and how it is used to diagnose topological order.
Topologically ordered systems are many body systems characterized by long-range entanglement.
They feature a ground state degeneracy that grows exponentially with the genus of the manifold on which they are defined, and anyonic excitations with statistics different from those of bosons and fermions~\cite{wen2004quantum}.
The long-range entanglement characteristic of a gapped, topologically ordered, ground state results in a contribution to the entanglement entropy between bipartitions of a system known as {\it topological entanglement entropy}~\cite{Wen_QuantumInfo}.

\begin{figure}[t]
    \centering
    \hspace{-8.7cm}
    \includegraphics[width=\linewidth]{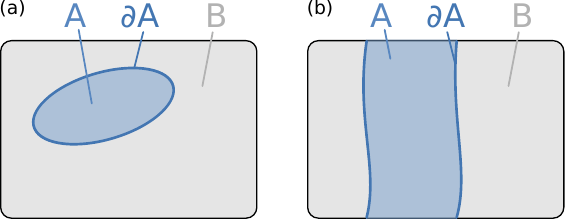}
    \caption{
    Schematic of a bipartition of a system with periodic boundary conditions (torus) into two subsystems: $A$, colored in blue, and $B$, colored in gray.  
    The boundary between $A$ and $B$, $\partial A$, is of length $L = |\partial A|$ and highlighted in dark blue.
    The boundary $\partial A$ is considered part of $A$. 
    The panels (a) and (b) show examples of a contractible and a non-contractible bipartition, respectively. 
    The difference between (a) and (b) is that in (b) the regions $A$ and $B$  both wind around one of the non-contractible cycles of the torus, while in (a) it is just $B$.
    }
    \label{fig:bipartitions}
\end{figure}

To define the topological entanglement entropy, consider
a bipartition of a system into two subsystems, $A$ and $B$, 
schematically represented in Fig.~\ref{fig:bipartitions}.
We will consider first bipartitions where $A$ does not wind around any of the cycles of the torus, as in Fig.~\ref{fig:bipartitions}a.
The von Neumann entanglement entropy, or simply entanglement entropy, between $A$ and $B$ is defined as~\cite{Simon_TopologicalQuantum}
\begin{equation}
    \label{eq:entanglement_entropy_app}
    S_{\mathrm{vN}} = -\Tr(\rho_{A}\log\rho_{A})\,,
\end{equation}
where $\rho_{A} = \Tr_{B}(\rho)$ is the partial density matrix obtained from the partial trace over region $B$ of the total density matrix $\rho$ $=\ket{\psi}\bra{\psi}$ of a quantum state $\ket{\psi}$. 

Equation~\eqref{eq:entanglement_entropy_app} requires computing the logarithm of the operator $\rho_A$, which is usually computationally impractical.
For this reason, it is often  convenient to work with  Renyi entropies~\cite{HammaWen_RenyiEntropy,Halasz2012}.
The $n$-th Renyi entanglement entropy is defined as~\cite{HammaWen_RenyiEntropy}
\begin{equation}
    \label{eq:renyi_entropy_app}
    S_{n} = -\frac{1}{n-1}\log(\Tr(\rho_{A}^{n}))\,, \qquad n \in \mathbb{N}_{>1}\,,
\end{equation}
where $\mathbb{N}_{>1}$ is the set of natural numbers larger than  one.
In Eq.~\eqref{eq:renyi_entropy_app}, the logarithm acts on the trace of the $n$-th power of $\rho_{A}$, which is a number.
Since $S_{2}$ is the Renyi entropy with the smallest power of $\rho_A$, it is practical and common to focus on $S_2$, as we do in this work.

The von Neumann entanglement entropy is recovered from the Renyi entropies taking the limit~\cite{HammaWen_RenyiEntropy}
\begin{equation}
    S_{\mathrm{vN}} = S_{n \rightarrow 1} = \lim_{n \rightarrow 1} S_{n}\,.
\end{equation}

In this work, we are interested in two-dimensional gapped ground-states.
For these, both the von Neumann and Renyi entropies follow an area law~\cite{Simon_TopologicalQuantum}
\begin{equation}
    S_n = \alpha_{n}L\,,
\end{equation}
where $\alpha_n$ is a positive constant that generally depends on $n$~\cite{HammaWen_RenyiEntropy,Halasz2012}, and $L = |\partial A|$ is the size of the boundary between the two subsystems, see Fig.~\ref{fig:bipartitions}.
The area law represents the fact that, in the absence of topological order, the gapped system is short-range entangled.
Intuitively, a topologically trivial subsystem $A$ is entangled to the rest of the system, $B$, only by the degrees of freedom close to the boundary~\cite{Wen_QuantumInfo}.
How close is determined by the correlation length, which is inversely proportional to the many-body gap of the system.
The area law is modified if the ground state of a gapped system is topologically ordered.
In this case the entanglement of the ground state follows the area law with a sub-leading constant term $\gamma$~\cite{LevinWen2006,KitaevPreskill2006}
\begin{equation}
    \label{eq:topologicalarealaw}
    S_{n} = \alpha_{n} L - \gamma\,, \qquad \gamma > 0\,.
\end{equation}
Because $\gamma$ is independent of $L$ by definition, a $\gamma\neq 0$ suggests that the ground state is long-range entangled.

Levin and Wen~\cite{LevinWen2006}, and Kitaev and Preskill~\cite{KitaevPreskill2006} proposed judicious combinations of system tripartitions and their entanglement entropies to isolate $\gamma$ in Eq.~\eqref{eq:topologicalarealaw}.
The sub-leading term was called topological entanglement entropy~\cite{KitaevPreskill2006}. It was proposed to be a universal quantity determined by the total quantum dimension $D$ of the  anyons of the system~\cite{LevinWen2006,KitaevPreskill2006,Simon_TopologicalQuantum}
\begin{equation} \label{eq:gamma_and_D}
    \gamma = \gamma_{\mathrm{top}} \equiv \log D\,,
\end{equation}
where $D = \sqrt{\sum_{i}d_{i}^{2}}$ and $d_{i}$ is the quantum dimension of an anyon $i$.
The anyon dimension $d_{i}$ is a measure of how the Hilbert space of the anyon $i$ grows when fused with itself.
A single boson has $d_{i}=1$ and hence $ \gamma_{\mathrm{top}}=0$.
A Laughlin state at filling fraction $\nu=1/m$ has $D^2=m$ and hence $\gamma_{\mathrm{top}}= \frac{1}{2}\log(m)$.
Because of the relation~\eqref{eq:gamma_and_D}, the topological entanglement entropy has been used extensively to characterize topological order, see, for example, Refs.~\cite{laflorencie2016quantum,Isakov2011,Jiang2012,Cincio2015,Grushin2015}. 
%


\subsection{Spurious contributions to $\gamma$ \label{sec:spurious}}
References~\cite{Cano2015,Bravyi,Zou2016,Fliss2017,Santos2018,Williamson2019,Kato2020,Stephen2019} established that $\gamma$, as defined in Eq.~\eqref{eq:topologicalarealaw}, is not always determined  by the anyon quantum dimensions through Eq.~\eqref{eq:gamma_and_D}.
In general, $\gamma \neq \gamma_{\mathrm{top}}$ and it can even be non-zero for topologically trivial ground states, with short range entanglement and no anyons~\cite{Cano2015,Bravyi,Zou2016,Fliss2017,Santos2018,Williamson2019,Kato2020,Stephen2019}.
The deviations that render Eq.~\eqref{eq:gamma_and_D} not valid are called spurious contributions or $\gamma_{\mathrm{spur}}$, and we define them as
$\gamma = \gamma_{\mathrm{top}}+\gamma_{\mathrm{spur}}$.
Thus, we say that  $\gamma$ is no longer universal, in the sense that Eq.~\eqref{eq:gamma_and_D} is not always valid.

The necessary conditions causing  $\gamma \neq \gamma_{\mathrm{top}}$ are not fully understood, but there are several known instances where $\gamma$ deviates from $\gamma_\mathrm{top}$ in Eq.~\eqref{eq:gamma_and_D}.
A relatively well understood spurious contribution to $\gamma$ is that arising from choosing a non-contractible bipartition, shown schematically in Fig.~\ref{fig:bipartitions}b.
For non-contractible bipartitions, $\gamma$ depends on the specific ground state, the type of non-contractible bipartition, and the Renyi entropy considered~\cite{Ashvin_MES}.

In hindsight, we can expect that $\gamma$ deviates from $\gamma_{\mathrm{top}}$ when considering non-contractible bipartitions because these bipartitions are defined by paths that wind around the full system.
These paths access non-local information, important to determine the topological class of a system.

Topologically ordered ground states are degenerate on the torus, and hence their superpositions are also ground states.
By studying the dependence on the ground state superposition, Ref.~\cite{Ashvin_MES} 
introduced the concept of minimally entangled states.
Minimally entangled states are defined as those ground states whose spurious contributions to $\gamma$ are minimal.
These states are a basis to obtain entries of the modular $S$ matrix, that describes certain modular transformations on the degenerate ground states of a topologically ordered system.

Another source of spurious contributions to $\gamma$ are subsystem symmetries.
If the boundary $\partial A$ or the region $A$ realizes a symmetry protected phase, the subleading term $\gamma$ may differ from $\gamma_\mathrm{top}$~\cite{Cano2015,Bravyi,Zou2016,Fliss2017,Santos2018,Williamson2019,Kato2020,Stephen2019}.
The existence of a subsystem symmetry protected phase can be proven to be a sufficient condition for a non-zero spurious contribution to exist~\cite{Zou2016}.
However, it does not seem to be a necessary condition, as Ref.~\cite{Kato2020} provided a numerical example of a spurious contribution to $\gamma$ unrelated to a subsystem symmetry protected phase.
The authors of Ref.~\cite{Stephen2019} suggested that spurious contributions can be used to diagnose subsystem symmetry-protected phases.

A more concerning fact is that even the Levin-Wen and Kitaev-Preskill prescriptions proposed to uniquely isolate the universal $\gamma_{\mathrm{top}}$~\cite{LevinWen2006,KitaevPreskill2006}, based on contractible multipartite bipartitions and their relative entanglement entropies, do not avoid spurious contributions.
References~\cite{Bravyi,Zou2016,Williamson2019} showed examples where these prescriptions suffer from spurious contributions associated to subsystem symmetries.
These references provide examples of cluster states with $\gamma > 0$. However, they are  topologically trivial ($\gamma_\mathrm{top} = 0$), as they are connected to product states by a finite depth unitary circuit.
We discuss some of these cluster states in the main text and in Section~\ref{sec:universality_gamma}.

More recently, Ref.~\cite{Kim2023} proved that, for the tripartition considered in Ref.~\cite{KitaevPreskill2006}, the sub-leading correction to the area law $\gamma$ satisfies
\begin{equation}
    \label{eq:gammabound}
    \gamma \geq \gamma_\mathrm{top}\,,
\end{equation}
suggesting that the anyon quantum dimension establishes a lower bound to the topological entanglement entropy.

Lastly, it is worth noting that a possible solution to isolate $\gamma_\mathrm{top}$ was put forward in Ref.~\cite{Kato2020entropic}. 
It relies on calculating a relative entropy between the state in question and a reference fully mixed state.  
The aim of our work is to provide a diagnostic that does not require comparison with another state. 
In the main text, we focus on  bipartitions of the type shown in Fig.~\ref{fig:bipartitions}a, which are topologically equivalent to a disk.
These bipartitions circumvent the problems related to the non-contractible bipartitions, which are relatively well understood.
Nonetheless, we have benchmarked our methods on non-contractible bipartitions of the type shown in Fig.~\ref{fig:bipartitions}b, and recover known results for $\gamma$~\cite{Hamma}. 
Hence, for the reminder of this Supplemental Material, we will mainly focus on contractible bipartitions. 

\section{Toric Code } \label{sec:toriccode}

In this Section, we introduce basic concepts about the toric code~\cite{Simon_TopologicalQuantum,Kitaev2003,herringer2020toric}, that we use in the main text. We also discuss results concerning entanglement entropies of the toric code's ground states, mostly derived in Ref.~\cite{Hamma}. These results  are useful to benchmark our ZX calculations and to understand the connection between  topological entanglement entropies and  the Pauli trees discussed in the main text. 

\subsection{Model definition and ground state}\label{sec:toriccode_def}
\begin{figure}
    \centering
    \hspace{-8.7cm}
    \includegraphics[width=\linewidth]{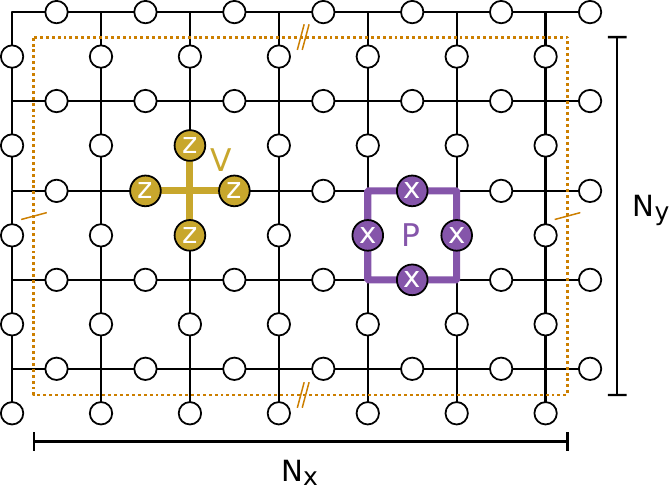}
    \caption{
    Schematic representation of the toric code model on the square lattice, with periodic boundary conditions.    %
    The qubits (white circles) sit on each edge of the lattice. 
    The qubits are acted upon by the vertex (yellow) and plaquette (purple) operators, which are products of $\sigma_z$  and $\sigma_x$ Pauli operators respectively, see Eqs.~\eqref{eq:vertexop_def_toriccode} and~\eqref{eq:plaquetteop_def_toriccode}. 
    Identifying the dotted orange lines of opposite edges implements periodic boundary conditions.
   }
    \label{fig:toriccode_def}
\end{figure}

The toric code is an exactly solvable model with long-range topological order~\cite{Kitaev2003}.
It is defined on a $N_{x}\times N_{y}$ square lattice with periodic boundary conditions, which has thus the topology of a torus, with $N = 2 N_{x}N_{y}$ qubits sitting at each edge of the lattice, see Fig.~\ref{fig:toriccode_def}. 

\begin{figure*}[t]
    \centering
    \hspace{-17cm}
    \includegraphics[width=.95\linewidth]{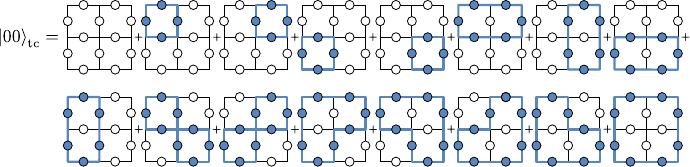}
    \caption{    
    Representation of the toric code ground state~\eqref{eq:groundstate00_toriccode} as an equal superposition of closed strings on a $2\times2$ lattice. To define a string, we assign a color to each edge (thick blue line) hosting a qubit in the $\ket 1$ state (blue-filled circles). To ease the representation, we assume open boundary conditions. }\label{fig:toriccode_groundstate_example}
\end{figure*}

The Hamiltonian of the toric code reads~\cite{Simon_TopologicalQuantum}
\begin{equation} \label{eq:hamiltonian_toriccode}
  H_\mathrm{TC} = -\sum_{+}V_{+} -\sum_{\square}P_{\square}\,,
\end{equation}  
where the sums run over all the vertices ``+'' and square plaquettes ``$\square$'' of the lattice. The vertex, or star, operators $V_{+}$ are products of four Pauli $\sigma_{z}$ matrices acting on each of the adjacent qubits of a vertex
\begin{equation} \label{eq:vertexop_def_toriccode}
    V_{+} = \prod_{i \in +} \sigma_{z}^{(i)}\,.
\end{equation}
Analogously, the plaquette operators $P_{\square}$ are products of four Pauli $\sigma_{x}$ matrices
acting on each of the qubits of a plaquette $\square$
\begin{equation} \label{eq:plaquetteop_def_toriccode}
    P_{\square} = \prod_{i \in \square} \sigma_{x}^{(i)}\,.
\end{equation}
The toric code thus features $N_x\cdot N_y$ vertex and plaquette operators. They commute with each other and square to the identity operator $I$. So do their products 
\begin{align}
\label{eq:constraints_operators_toriccode_P}
\prod_+ V_{+} &= I\,, &        \prod_\square P_{\square} &= I\,.
\end{align}
As a consequence, the number of independent conserved quantities is $2N_x N_y - 2$, leaving two degrees of freedom and thus each eigenstate is four-fold degenerate~\cite{Simon_TopologicalQuantum}.  We will focus here on the ground state
\begin{equation}
\label{eq:groundstate00_toriccode}
    {\ket{\rm GS}_{\rm tc}}\equiv\ket{00}_{\rm tc} = \frac{1}{{2^{(n_p+1)/2}}}\prod_{\square} (I + P_{\square})\ket{0}^{\otimes N}\, ,
\end{equation}
where ${n_p=}N/2$ indicates the total number of plaquettes $\square$. We introduce here the notation $\ket{00}_{\rm tc}$  for the ground state, to distinguish among the four degenerate ground states, including $\ket{01}_{\rm tc}$, $\ket{10}_{\rm tc}$ and $\ket{11}_{\rm tc}$, see discussion below.

If we assign a color to  lattice edges whose qubits are in the $\ket 1$ state and call such colored edges ``strings'', the state~\eqref{eq:groundstate00_toriccode} describes an equal superposition of all the states featuring closed contractible strings (loops) on the lattice, see Fig.~\ref{fig:toriccode_groundstate_example} for an example. To verify that the state~\eqref{eq:groundstate00_toriccode} is the ground state of Eq.~\eqref{eq:hamiltonian_toriccode}, one can see that any state featuring a closed string is an eigenvector with eigenvalue +1 of the star operators $V_+$ in Eq.~\eqref{eq:hamiltonian_toriccode}. The action of the plaquette operators $P_\square$ on a closed-string configuration generates  another closed-string configuration. As Eq.~\eqref{eq:groundstate00_toriccode} features all the closed-string configurations with equal weights, it is thus an eigenvector of $P_\square$ with eigenvalue +1, and, as a consequence, the ground state of Eq.~\eqref{eq:hamiltonian_toriccode}.


\subsection{Ground state degeneracy and string operators} \label{sec:GSdegeneracytoric}

The four-fold degeneracy of the toric code is of topological nature~\cite{Kitaev2003,Simon_TopologicalQuantum}, as it depends on the genus of the manifold on which the Hamiltonian~\eqref{eq:hamiltonian_toriccode} is defined~\cite{wen2004quantum} -- in this case the torus. The nature of this topological degeneracy becomes apparent when we construct the remaining three ground states: $\ket{01}_{\rm tc},\ket{10}_{\rm tc}$, and $\ket{11}_{\rm tc}$. 

\begin{figure}[t]
    \centering
    \hspace{-8.7cm}
    \includegraphics[width=\linewidth]{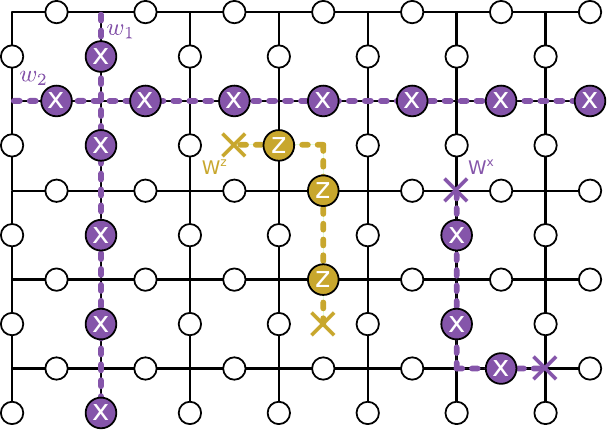}
    \caption{String operators $W^x$ (purple) and $W^{z}$ (yellow) applied to the toric code, according to the definitions~\eqref{eq:xstring_def} and~\eqref{eq:zstring_def}. We also illustrate the special case of non-contractible $x$-string operators $w_{1,2}$, winding around the loops of the torus.}
    \label{fig:toriccode_chainoperators}
\end{figure}

To do this, we consider the string operators depicted in Fig.~\ref{fig:toriccode_chainoperators}~\cite{Hamma, Wen_QuantumInfo}. 
The $x$-type string (or simply $x$-string) operators $W^{x}$ are tensor products of Pauli $\sigma_{x}$ matrices 
\begin{equation}
    \label{eq:xstring_def}
    W^{x} = \prod_{i\in \mathcal C_{x}} \sigma_{x}^{i}\,,
\end{equation}
acting on $x$-paths  $\mathcal C_x$ defined by connecting lattice vertices. Remarkably, when the path $\mathcal C_x$ is closed and winds around one of the two non-contractible cycles of the torus, see Fig.~\ref{fig:toriccode_chainoperators}, the associated $x$-string operators, that we label $w_{1,2}$, commute with the Hamiltonian~\eqref{eq:hamiltonian_toriccode}. The four degenerate ground states of the toric code can be constructed by either applying or not each $w_{i}$ on the ground state~\eqref{eq:groundstate00_toriccode}~\cite{Hamma}: 
\begin{align}\label{eq:groundstates_basis_toriccode}
    \ket{ij}_{\rm tc} &=  w_{1}^{i}w_{2}^{j}\ket{00}_{\rm tc}\,, & i,j &\in \{0,1\}\,.
\end{align}
Note that $i$ and $j$ in the right-hand side denote exponents, not super-indices.
This construction showcases  a key topological property of the toric code: the degenerate ground states are exclusively connected by a global perturbation winding around the entire torus.

As discussed in Section~\ref{sec:spurious}, general superpositions of the four ground states
\begin{equation}
    \label{eq:groundspace_toriccode}
    \ket{\psi} = \sum_{i,j=0,1} c_{ij}\ket{ij}_{\rm tc}\,, 
\end{equation}
where $c_{ij} \in \mathbb{C}$ and $\sum_{ij} |c_{ij}|^{2} = 1$, may cause spurious contributions to the entanglement entropy~\cite{Ashvin_MES}. In Section~\ref{sec:results_xmas_toriccode}, we will show that the ZX representation of the contour diagram and of the Pauli tree discussed in the main text do not depend  on the superposition~\eqref{eq:groundspace_toriccode}.

For completeness, we mention that, in analogy to the $x$-string operators~\eqref{eq:xstring_def}, we can define $z$-string operators 
\begin{equation}
    \label{eq:zstring_def}
    W^{z} = \prod_{i\in \mathcal C_{z}} \sigma_{z}^{i}\,, 
\end{equation}
where the $z$-paths $\mathcal C_z$ connect plaquette centers, see Fig.~\ref{fig:toriccode_chainoperators}. Also in this case, the two $z$-string operators $w^z_{1,2}$ winding around the torus commute with the Hamiltonian~\eqref{eq:hamiltonian_toriccode}. However, the states $\ket{ij}_{\rm tc}$ are eigenstates of $w^z_{1,2}$, with eigenvalues $-1$ or $+1$. They thus detect the parity of the number of non-trivial $x$-strings $w_{1,2}$  applied to the $\ket{00}_{\rm tc}$ state Eq.~\eqref{eq:groundstate00_toriccode}.


\subsection{Topological Entanglement Entropy for the Toric Code ground states  \label{sec:gamma_toriccode}}

In this Section, we  discuss the spurious contributions to the topological entanglement entropy $\gamma$ for the toric code caused by different bipartitions. Our discussion largely follows Ref.~\cite{Hamma}, which develops a theory to compute the entanglement entropy $S$ for spin systems from group theory calculations.
We  use these results to benchmark our computations with ZX calculus in Section~\ref{sec:results_gamma_toriccode} and in Section~\ref{sec:results_xmas_toriccode}, where we establish the connection between ZX contour diagrams, Pauli trees and $\gamma_{\rm top}$.

We start by defining $G$ as the group generated by the plaquette operators.
The elements of $G$ are products of $x$-string operators, see Eq.~\eqref{eq:xstring_def}, following contractible, closed paths.
For plaquette operators,  the property~\eqref{eq:constraints_operators_toriccode_P}, $\prod_{\square} P_{\square} = I$, implies that one plaquette operator depends on all the others.
Thus, the order of $G$ is $|G|$ = $2^{n_{p} - 1}$, where $n_{p}$ is the number of plaquette operators.

Given a lattice bipartition $(A,B)$, we define two subgroups of $G$.
$G_{A}$ and $G_{B}$ are the groups formed by elements of $G$ purely acting on qubits of $A$ and $B$, respectively.
We denote the order of $G_{A}$ and $G_{B}$ as $d_{A}$ and $d_{B}$. They   indicate the number of closed $x$-string operators acting purely on $A$ and $B$.
If we denote $\Sigma_{A}$ and $\Sigma_{B}$ the number of independent elements of $G_{A}$ and $G_{B}$, then $d_{A} = 2^{\Sigma_{A}}$ and $d_{B} = 2^{\Sigma_{B}}$.
Finally, $\Sigma_{AB}$ denotes the number of independent elements in $G$ that act both on $A$ and $B$.
We have the following relation~\cite{Hamma}
\begin{equation}\label{eq:sigmas_relation}
    \Sigma_{A} + \Sigma_{B} + \Sigma_{AB} = n_{p}\,.
\end{equation}
Note that the independent elements of $G_{A}$ and $G_{B}$ are not always simple plaquettes.
For example, imagine two horizontally adjacent plaquettes whose common qubit belongs to $A$, and all their other qubits are in $B$.
Then, neither of the plaquette operators belongs to $G_{B}$, but their product, which is a rectangular operator, does.

\begin{figure}[t]
    \centering
    \hspace{-8.7cm}
    \includegraphics[width=\columnwidth]{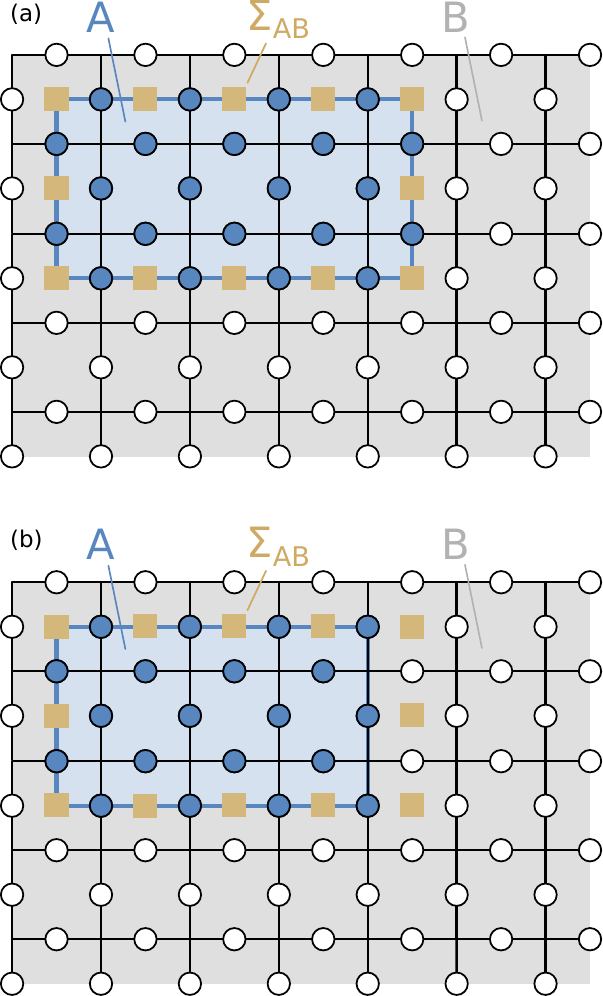}\\
    \caption{
    (a) Example of a rectangular bipartition $(A,B)$ for the calculation of the entanglement entropy and $\gamma$. The cut is performed along a $z$-path.
    The blue region with blue qubits designates $A$, while the grey region with white qubits belongs to subsystem $B$.
    The orange squares designate the $\Sigma_{AB}$ independent plaquette operators acting both on qubits  of the regions $A$ and $B$, see also discussion before Eq.~\eqref{eq:sigmas_relation}. According to Eq.~\eqref{eq:entropy_hamma}, the number of plaquettes $\Sigma_{AB}$ determines the entanglement entropy between regions $A$ and $B$.     (b) An alternative bipartition mixing $z$- and $x$-paths. It reduces the number of boundary qubits by one, but  keeps $\Sigma_{AB}$ unchanged. 
    }
    \label{fig:toriccode_gammatheory_rectangle}
\end{figure}

For our purposes, the relevant results in Ref.~\cite{Hamma} relate to the partial density matrices $\rho_{A}^{n}$ of the toric-code  ground states 
and  associated entanglement entropies. 
For a given lattice bipartition $(A,B)$ along a contractible cut:
\begin{itemize}
\item all powers of the reduced density matrix $\rho_A$ are proportional to each other
\begin{equation}\label{eq:rhoApower_toric}
        \rho_{A}^{n} = \xi^{n-1}\rho_{A}\,,
        \end{equation}
        where $\xi$ is a constant and $n>0$;
\item the entanglement entropy is the same for any superposition of states~\eqref{eq:groundspace_toriccode};

\item both the von Neumann~\eqref{eq:entanglement_entropy_app} and the Renyi~\eqref{eq:renyi_entropy_app} entanglement entropies, $S_{\rm vN}$ and $S_n$ respectively, are equal to (for all $n$)
        \begin{equation}\label{eq:entropy_hamma}
        \begin{split}
            S_{\mathrm{vN}}&=S_n= \log_{2} \left(\frac{|G|}{d_{A}d_{B}}\right) =\\
            & = n_{p} - 1 - \Sigma_{A} - \Sigma_{B} = \Sigma_{AB} - 1\,. 
        \end{split}
        \end{equation}
\end{itemize}        
Comparing the area law Eq.~\eqref{eq:topologicalarealaw} with Eq.~\eqref{eq:entropy_hamma}, the sub-additive constant $-1$ and $\Sigma_{AB}$ play the role of $\gamma$ and $L$, respectively, where we recall that $L$ is the length of the boundary between $A$ and $B$. 
Note that $\gamma$ is related to the constraint on the number of independent plaquette operators in Eq.~\eqref{eq:constraints_operators_toriccode_P}. 
We will rely on the extension of Eq.~\eqref{eq:entropy_hamma} to deduce $\gamma$ for the hexagonal toric and color codes, discussed in more detail in Sections~\ref{sec:hextoriccode} and~\ref{sec:colorcode}.

Moreover, all partial density matrices are equivalent except for a proportionality constant.
This will be particularly important in  Section~\ref{sec:rhoAb} , where we discuss how to extract non-local spiders and  Pauli trees in ZX.

Finally, Eq.~\eqref{eq:entropy_hamma} states that  $S_{\mathrm{vN}} $ and all the Renyi entropies $S_n$ coincide, simplifying  calculations as argued in Section~\ref{sec:TEE}.
Thus, calculating $\gamma$ and $S$ amounts to counting independent plaquettes (or combinations of them) that completely act on $A$ ($\Sigma_{A}$), on $B$ ($\Sigma_{B}$) or on both ($\Sigma_{AB}$).
These results apply to all the ground states of the basis.

Let us illustrate these results with the example in Fig.~\ref{fig:toriccode_gammatheory_rectangle}a.
There, the edges of $A$ follow a $z$-path, defined before Eq.~\eqref{eq:zstring_def}.
However, the calculations would be analogous for a rectangle following an $x$-path, defined before Eq.~\eqref{eq:xstring_def}, or any shape which is topologically equivalent.
For such bipartitions, we can count the number of shared plaquettes $\Sigma_{AB}=12$, indicated as orange squares in Fig.~\ref{fig:toriccode_gammatheory_rectangle}.
Thus, according to Eq.~\eqref{eq:entropy_hamma}, the entropy equals $S = \Sigma_{AB} - 1=11$, and, therefore, $\gamma = 1$. 

In the particular case of the rectangular cut discussed above, the number $\Sigma_{AB}$ of  plaquettes shared between $A$ and $B$ equals the number of qubits $L$ delimiting region $A$. To illustrate the ambiguities impacting on the area law~\eqref{eq:topologicalarealaw}, we consider in Fig.~\ref{fig:toriccode_gammatheory_rectangle}b  a cut  mixing $z$- and $x$-paths. This path removes two qubits from  the $A$ region and assigns them to $B$, changes the number of qubits at the boundary of $A$ from $L$ to $L'=L-1$, but preserves the number of plaquettes  acting on both $A$ and $B$, $\Sigma_{AB}'=\Sigma_{AB}$. According to Eq.~\eqref{eq:entropy_hamma}, the entanglement entropy does not change,  even though the number of qubits $L'=L-1$ at the border of $A$ was changed with respect to the previous bipartition. As a consequence, attempting to estimate $\gamma$ based on counting the number $L$ of qubits at the boundary, as required by the area law~\eqref{eq:topologicalarealaw}, would lead to a misleading result, $\gamma = 0$. However, Eq.~\eqref{eq:entropy_hamma}, which counts   the number $\Sigma_{AB}$ of shared plaquettes, still delivers the correct result  $\gamma=\gamma_{\rm top}= 1$.


\begin{figure*}[t]
    \centering
    \hspace{-18cm}
    \includegraphics[width=\linewidth]{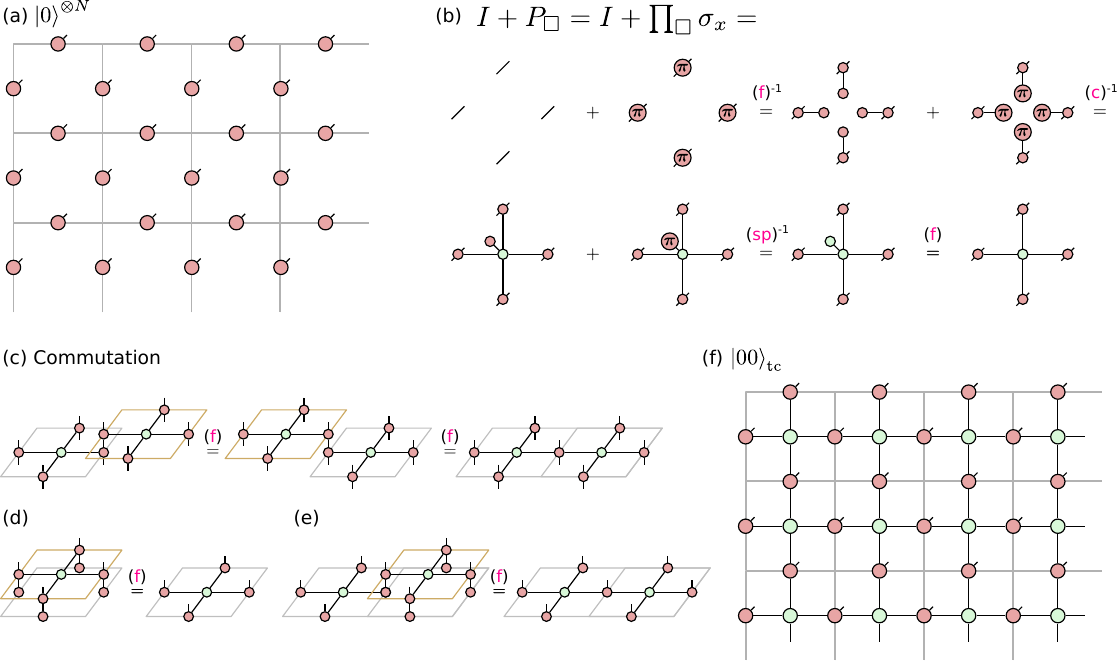}
    \caption{
Construction of the ZX diagram of the $\ket{00}_{\rm tc}$ toric-code ground state, Eq.~\eqref{eq:groundstate00_toriccode},  on  a $4 \times 3$ lattice, with periodic boundary conditions.     (a) ZX-representation of the product state $\ket{0}^{\otimes N}$. Each qubit in the $\ket{0} $  state is represented  by the  
\begin{tikzpicture}
	\begin{pgfonlayer}{nodelayer}
		\protect\node [style=X dot] (0) at (0, 0) {};
		\protect \node [style=none] (1) at (1, 0) {};
	\end{pgfonlayer}
	\begin{pgfonlayer}{edgelayer}
		\protect\draw (0) to (1.center);
	\end{pgfonlayer}
\end{tikzpicture} diagram with an output line.   The gray lines correspond to the original toric code square lattice. 
    (b) The ZX diagram for $I + P_{\square}$ is given by  the sum of the identity operator $I = $ \begin{tikzpicture}
	\begin{pgfonlayer}{nodelayer}
		\protect\node [style=none] (1) at (.5, 0) {};
		\protect\node [style=none] (3) at (-.5, 0) {};
	\end{pgfonlayer}
	\begin{pgfonlayer}{edgelayer}
		\protect\draw (3.center) to (1.center);
	\end{pgfonlayer}
\end{tikzpicture} with a product of four Pauli operators $\sigma_{x} = $ \begin{tikzpicture}
	\begin{pgfonlayer}{nodelayer}
		\protect\node [style=none] (1) at (1, 0) {};
		\protect\node [style=X phase dot] (2) at (0, 0) {$\pi$};
		\protect\node [style=none] (3) at (-1, 0) {};
	\end{pgfonlayer}
	\begin{pgfonlayer}{edgelayer}
		\protect\draw (2) to (1.center);
		\protect\draw (2) to (3.center);
	\end{pgfonlayer}
\end{tikzpicture}, see Eq.~\eqref{eq:sigmax_zx}.
    Applying the  \FusionRule and \CopyRule rules in Fig.~\ref{fig:zx-rules} and \SupRule in Eq.~\eqref{eq:ZXsuperposition} leads to a more compact representation of the operator.
    (c) In ZX, the commutation of the $I + P_{\square}$ operators directly follows from the fusion rule \FusionRule of ZX calculus.
    (d) ZX diagram issued from the application of $I + P_{\square}$ on a single plaquette, where the qubits are in the $\ket{0}$ state. By applying the \FusionRule rule, the diagrams merge and the projector looses the output legs.
    (e) Extension of panel (d) to  a neighboring plaquette. 
    (f) The final ZX diagram corresponding to $\ket{00}_{\rm tc}$  is obtained by extension of the procedure in (e) to the entire lattice. 
    }\label{fig:toriccode_protocol}
\end{figure*}

\section{ZX representation of the Toric Code Ground States} \label{sec:ToricCode_ZX}

In this section, we describe how to construct the ZX diagram for the $\ket{00}_{\rm tc}$ ground state of the toric code, Eq.~\eqref{eq:groundstate00_toriccode}~\cite{kissinger2022phasefreezxdiagramscss,khesin2024equivalenceclassesquantumerrorcorrecting}.
We then generalize the construction to 
the other three ground states $\ket{01}_{\rm tc}$, $\ket{10}_{\rm tc}$ and $\ket{11}_{\rm tc}$ in Eq.~\eqref{eq:groundstates_basis_toriccode}, 
as well as for the arbitrary superposition~\eqref{eq:groundspace_toriccode}.

\subsection{ZX-calculus representation of $\ket{00}_{\rm tc}$ \label{sec:00groundstateZX}}

The ground state $\ket{00}_{\rm tc}$ of the toric code, Eq.~\eqref{eq:groundstate00_toriccode}, results from the action of the product of projectors $\prod_\square (I + P_{\square})/2$, which act upon all the square plaquettes $\square$ of an ensemble of $N$ qubits initialized in the  $\ket{0}$ state.
Figure~\ref{fig:toriccode_protocol}a shows the ZX diagram for the initialization state $\ket{0}^{\otimes N}$. It is directly obtained by using the equivalence $\ket0=$
\begin{tikzpicture}
    \begin{pgfonlayer}{nodelayer}
        \node [style=X phase dot] (0) at (-0.5, 0) {};
        \node [style=none] (1) at (0.5, 0) {};
    \end{pgfonlayer}
    \begin{pgfonlayer}{edgelayer}
        \draw (0) to (1.center);
    \end{pgfonlayer}
\end{tikzpicture}, Eq.~\eqref{eq:0_zx},  for each qubit on the lattice.
Recall from Section~\ref{sec:intro_ZX}, that this is not a strict equality, but rather a relation of proportionality, because in Eq.~\eqref{eq:0_zx} we have also a scalar factor $\sqrt{2}$.
In what follows, we will not keep track of overall prefactors (and they will all be non-zero), as we can normalize the state at the end.
Thus, unless otherwise stated, the equalities should be considered as proportionality relations.

Figure~\ref{fig:toriccode_protocol}b illustrates the procedure to create the ZX-representation of an operator  $I+P_\square$. Also in this case, in the ZX language, there is no need to keep track of the factor $1/2$, normalizing this operator to be a projector.
As explained in the main text, from a product of four identities and four $\sigma_{x}$ Pauli operators, applying the \FusionRule and \CopyRule rules in Fig.~\ref{fig:zx-rules} (some applied from right to left) and the \SupRule rule in Eq.~\eqref{eq:ZXsuperposition} leads to the compact representation of $I + P_{\square}$ as a ZX diagram.

According to Eq.~\eqref{eq:groundstate00_toriccode}, the state $\ket{00}_{\rm tc}$ requires applying the operators $I + P_{\square}$ onto each plaquette.
These operators commute with each other and Fig.~\ref{fig:toriccode_protocol}c shows the commutation relation using ZX rules.
Thus, the $I + P_{\square}$ operators can be applied in any order.

Figure~\ref{fig:toriccode_protocol}d shows the ZX diagram resulting from applying $I+ P_{\square}$ to a plaquette where all the qubits are in the $\ket{0}$ state.
The resulting diagram is similar to the operator  in Fig.~\ref{fig:toriccode_protocol}b, but the red spiders have each one output, rather than two, because the diagram represents a state, rather than an operator.
Figure~\ref{fig:toriccode_protocol}e shows the result of applying $I+ P_{\square}$ to a plaquette when one of the qubits is already entangled with the qubits of an other plaquette, due to the previous application of the same operator to that plaquette.

By repeating the steps in Fig.~\ref{fig:toriccode_protocol}e for all the lattice plaquettes,  we obtain the ZX diagram corresponding to  $\ket{00}_{\rm tc}$  in Fig.~\ref{fig:toriccode_protocol}f. The diagram clearly shows that the qubits initially in the $\ket0=$
\begin{tikzpicture}
    \begin{pgfonlayer}{nodelayer}
        \node [style=X phase dot] (0) at (-0.5, 0) {};
        \node [style=none] (1) at (0.5, 0) {};
    \end{pgfonlayer}
    \begin{pgfonlayer}{edgelayer}
        \draw (0) to (1.center);
    \end{pgfonlayer}
\end{tikzpicture} state are now connected by green Z-spiders following the dual lattice of the toric code, which connects the centers of the plaquettes~\cite{kissinger2022phasefreezxdiagramscss}.

\subsection{ZX-calculus representation of arbitrary ground state superpositions }\label{sec:genericgroundstateZX}

The three remaining ground states~\eqref{eq:groundstates_basis_toriccode} are obtained by applying the non-contractible string operators $w_{1,2}$ onto the $\ket{00}_{\rm tc}$ state.

Recalling from Section~\ref{sec:intro_ZX} that, in ZX calculus,  $\sigma_{x} = $ 
\begin{tikzpicture}
	\begin{pgfonlayer}{nodelayer}
		\node [style=X phase dot] (0) at (0, 0) {$\pi$};
		\node [style=none] (1) at (-1.25, 0) {};
		\node [style=none] (2) at (1.25, 0) {};
	\end{pgfonlayer}
	\begin{pgfonlayer}{edgelayer}
		\draw (1.center) to (0);
		\draw (0) to (2.center);
	\end{pgfonlayer}
\end{tikzpicture}, see Eq.~\eqref{eq:sigmax_zx}, 
by virtue of Eq.~\eqref{eq:groundstates_basis_toriccode}, the $\ket{01}_{\rm tc}$  state in ZX is derived by applying  \begin{tikzpicture}
	\begin{pgfonlayer}{nodelayer}
		\node [style=X phase dot] (0) at (0, 0) {$\pi$};
		\node [style=none] (1) at (-1.25, 0) {};
		\node [style=none] (2) at (1.25, 0) {};
	\end{pgfonlayer}
	\begin{pgfonlayer}{edgelayer}
		\draw (1.center) to (0);
		\draw (0) to (2.center);
	\end{pgfonlayer}
\end{tikzpicture} along a horizontal loop of the torus
\begin{equation}
    \label{eq:toriccode_groundstate00ZX}
    \includegraphics[width=0.5\linewidth]{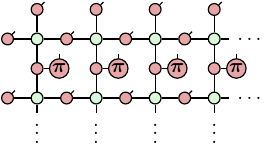}
\end{equation}
Analogously, by applying a non-contractible string operator $w$ along the vertical direction, we generate $\ket{10}_{\rm tc}$, or $\ket{11}_{\rm tc}$ by applying non-contractible string operators  both in the vertical and horizontal direction.

\begin{figure*}[t]
    \centering
    \hspace{-15.5cm}
    \includegraphics[width=0.9\linewidth]{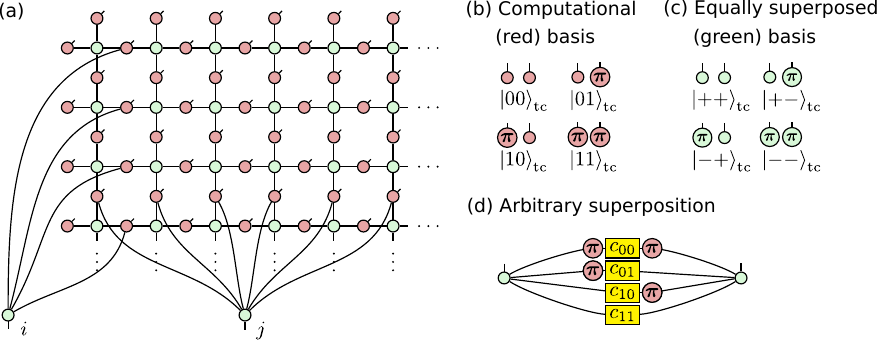}
    \caption{(a) ZX diagram for a generic ground state of the toric code.
Two non-local input green spiders act on the ZX diagram of $\ket{00}_{\rm tc}$ shown in Fig.~\ref{fig:toriccode_protocol}f along the  two non trivial loops of the torus. They are labeled $i,j$ as in the ground-state notation $\ket{ij}_{\rm tc}$. 
To obtain the diagram of a particular ground state, we input the corresponding spiders as shown in the panels (b)-(d).
(b) Spiders to select the basis $\ket{ij}_{\mathrm{tc}},$ with $i,j \in \{0,1\}$.
(c) Spiders to select the basis $\ket{ij}_{\mathrm{tc}}$, with $i,j \in \{+,-\}$.
(d) Spider to generate the arbitrary superposition $\ket\psi=\sum_{ij \in \{0,1\}}c_{ij}\ket{ij}_{\mathrm{tc}}$.
    }
    \label{fig:toriccode_groundstate_complete}
\end{figure*}

An equivalent, and more compact, representation of  the diagram~\eqref{eq:toriccode_groundstate00ZX} is obtained by applying backwards the copy rule \CopyRule in Fig.~\ref{fig:zx-rules} (excluding scalars)
\begin{equation}\label{eq:toriccode_groundstate10_transf}
    \includegraphics[width=\linewidth]{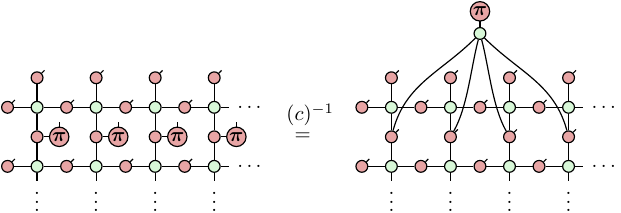}
\end{equation}
Remarkably, if in Eq.~\eqref{eq:toriccode_groundstate10_transf} we substitute 
\begin{tikzpicture}
	\begin{pgfonlayer}{nodelayer}
		\node [style=X phase dot] (0) at (0, 0.25) {$\pi$};
		\node [style=none] (1) at (0, -0.25) {};
	\end{pgfonlayer}
	\begin{pgfonlayer}{edgelayer}
		\draw (0) to (1.center);
	\end{pgfonlayer}
\end{tikzpicture}
by 
\begin{tikzpicture}
	\begin{pgfonlayer}{nodelayer}
		\node [style=X phase dot] (0) at (0, 0.25) {};
		\node [style=none] (1) at (0, -0.25) {};
	\end{pgfonlayer}
	\begin{pgfonlayer}{edgelayer}
		\draw (0) to (1.center);
	\end{pgfonlayer}
\end{tikzpicture}, 
we immediately recover the $\ket{00}_{\rm tc}$ ground state:
\begin{equation}\label{eq:toriccode_groundstate10_transf00}
    \includegraphics[width=\linewidth]{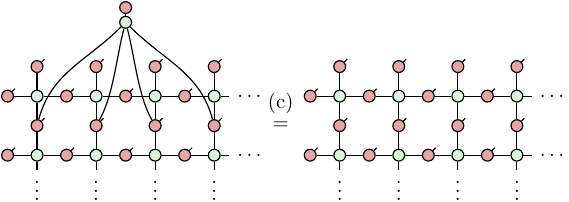}
\end{equation}
As a consequence, the input of 
\begin{tikzpicture}
	\begin{pgfonlayer}{nodelayer}
		\node [style=X phase dot] (0) at (0, 0.25) {$~$};
		\node [style=none] (1) at (0, -0.25) {};
	\end{pgfonlayer}
	\begin{pgfonlayer}{edgelayer}
		\draw (0) to (1.center);
	\end{pgfonlayer}
\end{tikzpicture}
or
\begin{tikzpicture}
	\begin{pgfonlayer}{nodelayer}
		\node [style=X phase dot] (0) at (0, 0.25) {$\pi$};
		\node [style=none] (1) at (0, -0.25) {};
	\end{pgfonlayer}
	\begin{pgfonlayer}{edgelayer}
		\draw (0) to (1.center);
	\end{pgfonlayer}
\end{tikzpicture}
onto a single non-local input green spider controls whether or not we apply the non-contractible string operators. Thus, these inputs  switch between the states $\ket{00}_{\rm tc}$ and $\ket{01}_{\rm tc}$ and, similarly, $\ket{10}_{\rm tc}$ and $\ket{11}_{\rm tc}$, see Fig.~\ref{fig:toriccode_groundstate_complete}a and b.

This construction readily extends to the construction of the rotated ground states $\ket{ij}_{\rm tc}$, with  $i,j \in \{+,-\}$ instead.
If we input 
\begin{tikzpicture}
	\begin{pgfonlayer}{nodelayer}
		\node [style=Z dot] (0) at (0, 0.25) {};
		\node [style=none] (1) at (0, -0.25) {};
	\end{pgfonlayer}
	\begin{pgfonlayer}{edgelayer}
		\draw (0) to (1.center);
	\end{pgfonlayer}
\end{tikzpicture}
to the non-local input spider, by using the \SupRule rule, Eq.~\eqref{eq:ZXsuperposition}, we can write
\begin{equation}\label{eq:toriccode_groundstate_+}
    \includegraphics[width=\linewidth]{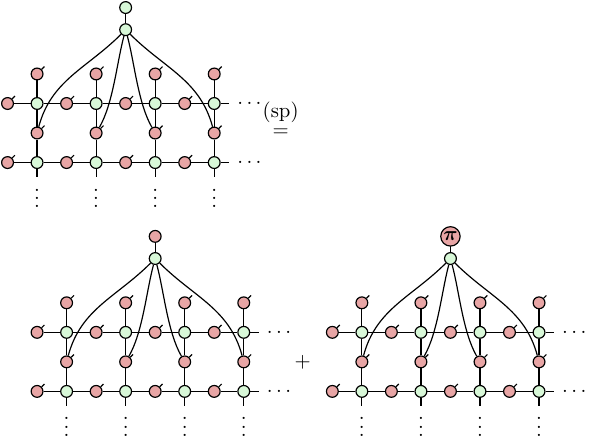}
\end{equation}
which corresponds to the state $\ket{0+}_{\rm tc}$. 
Similarly, applying
\begin{tikzpicture}
	\begin{pgfonlayer}{nodelayer}
		\node [style=Z phase dot] (0) at (0, 0.25) {$\pi$};
		\node [style=none] (1) at (0, -0.25) {};
	\end{pgfonlayer}
	\begin{pgfonlayer}{edgelayer}
		\draw (0) to (1.center);
	\end{pgfonlayer}
\end{tikzpicture} 
leads to the $\ket{0-}_{\rm tc}$ state.
See Fig.~\ref{fig:toriccode_groundstate_complete}c for the spiders to generate the ``$+,-$'' basis.

Introducing the non-local green spiders allows us to represent compactly also  all the different ground state superpositions of the toric code of Eq.~\eqref{eq:groundspace_toriccode}. To do so, we rely upon  the H-box defined in Eq.~\eqref{eq:Hbox_def} and the generic input diagram shown in  Fig.~\ref{fig:toriccode_groundstate_complete}d.

To show the validity of this ZX representation,  first recall that the matrix elements of an H-box are all one except for the last diagonal element, corresponding to $c$:
\begin{equation}    \label{eq:Hbox_GS}
    \begin{tikzpicture}
	\begin{pgfonlayer}{nodelayer}
		\node [style=H] (0) at (0, 0) {$c$};
		\node [style=X phase dot] (1) at (1, 0) {$~q\pi~$};
		\node [style=X phase dot] (2) at (-1, 0) {$~p\pi~$};
		\node [style=none] (3) at (2, 0) {=};
		\node [style=none] (4) at (4, .2) {$c^{\delta_{p,1}\delta_{q,1}}$};
	\end{pgfonlayer}
	\begin{pgfonlayer}{edgelayer}
		\draw (0) to (2);
		\draw (0) to (1);
	\end{pgfonlayer}
\end{tikzpicture}\,,
\end{equation} 
where $p,q =0,1$.  We then apply  the inverse of the fusion rule \FusionRule in Fig.~\ref{fig:zx-rules} to duplicate the number of green spiders on both sides of the diagram in Fig.~\ref{fig:toriccode_groundstate_complete}d, by substituting 
\begin{tikzpicture}
	\begin{pgfonlayer}{nodelayer}
		\node [style=Z dot] (0) at (0, 0) {};
		\node [style=none] (1) at (0, 0.5) {};
		\node [style=none] (2) at (0.5, -0.375) {};
		\node [style=none] (3) at (0.5, -0.125) {};
		\node [style=none] (4) at (0.5, 0.125) {};
		\node [style=none] (5) at (0.5, 0.375) {};
	\end{pgfonlayer}
	\begin{pgfonlayer}{edgelayer}
		\draw (1.center) to (0);
		\draw [bend left=15] (0) to (5.center);
		\draw (0) to (4.center);
		\draw [in=165, out=-15] (0) to (3.center);
		\draw [bend right=15] (0) to (2.center);
	\end{pgfonlayer}
\end{tikzpicture}=
\begin{tikzpicture}
	\begin{pgfonlayer}{nodelayer}
		\node [style=Z dot] (0) at (0, 0) {};
		\node [style=none] (1) at (0, 0.5) {};
		\node [style=none] (2) at (0.5, -0.375) {};
		\node [style=none] (3) at (0.5, -0.125) {};
		\node [style=none] (4) at (0.5, 0.125) {};
		\node [style=none] (5) at (0.5, 0.375) {};
	\end{pgfonlayer}
	\begin{pgfonlayer}{edgelayer}
		\draw (1.center) to (0);
		\draw [bend left=15] (0) to (5.center);
		\draw (0) to (4.center);
		\draw [in=165, out=-15] (0) to (3.center);
		\draw [bend right=15] (0) to (2.center);
	\end{pgfonlayer}
    	\begin{pgfonlayer}{nodelayer}
		\node [style=Z dot] (11) at (-.7, 0) {};
		\node [style=none] (12) at (-.7, 0.5) {};
	\end{pgfonlayer}
	\begin{pgfonlayer}{edgelayer}
		\draw (11.center) to (0);
	\end{pgfonlayer}
\end{tikzpicture}
, 
\begin{tikzpicture}
	\begin{pgfonlayer}{nodelayer}
		\node [style=Z dot] (0) at (0, 0) {};
		\node [style=none] (1) at (0, 0.5) {};
		\node [style=none] (2) at (-0.5, -0.375) {};
		\node [style=none] (3) at (-0.5, -0.125) {};
		\node [style=none] (4) at (-0.5, 0.125) {};
		\node [style=none] (5) at (-0.5, 0.375) {};
	\end{pgfonlayer}
	\begin{pgfonlayer}{edgelayer}
		\draw (1.center) to (0);
		\draw [bend left=15] (0) to (2.center);
		\draw (0) to (3.center);
		\draw (0) to (4.center);
		\draw [bend left=15] (5.center) to (0);
	\end{pgfonlayer}
\end{tikzpicture}=
\begin{tikzpicture}
	\begin{pgfonlayer}{nodelayer}
		\node [style=Z dot] (0) at (0, 0) {};
		\node [style=none] (1) at (0, 0.5) {};
		\node [style=none] (2) at (-0.5, -0.375) {};
		\node [style=none] (3) at (-0.5, -0.125) {};
		\node [style=none] (4) at (-0.5, 0.125) {};
		\node [style=none] (5) at (-0.5, 0.375) {};
	\end{pgfonlayer}
	\begin{pgfonlayer}{edgelayer}
		\draw (1.center) to (0);
		\draw [bend left=15] (0) to (2.center);
		\draw (0) to (3.center);
		\draw (0) to (4.center);
		\draw [bend left=15] (5.center) to (0);
	\end{pgfonlayer}
    \begin{pgfonlayer}{nodelayer}
		\node [style=Z dot] (11) at (.7, 0) {};
		\node [style=none] (12) at (.7, 0.5) {};
	\end{pgfonlayer}
	\begin{pgfonlayer}{edgelayer}
		\draw (11.center) to (0);
	\end{pgfonlayer}
\end{tikzpicture}.
In analogy to Eq.~\eqref{eq:toriccode_groundstate_+}, we apply to both of these diagrams  the \SupRule rule, Eq.~\eqref{eq:ZXsuperposition}. We   obtain an equal superposition of four diagrams like the  one in Fig.~\ref{fig:toriccode_groundstate_complete}d, but where \begin{tikzpicture}
	\begin{pgfonlayer}{nodelayer}
		\node [style=Z dot] (0) at (0, 0) {};
		\node [style=none] (1) at (0, 0.5) {};
		\node [style=none] (2) at (0.5, -0.375) {};
		\node [style=none] (3) at (0.5, -0.125) {};
		\node [style=none] (4) at (0.5, 0.125) {};
		\node [style=none] (5) at (0.5, 0.375) {};
	\end{pgfonlayer}
	\begin{pgfonlayer}{edgelayer}
		\draw (1.center) to (0);
		\draw [bend left=15] (0) to (5.center);
		\draw (0) to (4.center);
		\draw [in=165, out=-15] (0) to (3.center);
		\draw [bend right=15] (0) to (2.center);
	\end{pgfonlayer}
\end{tikzpicture} is replaced by 
\begin{tikzpicture}
	\begin{pgfonlayer}{nodelayer}
		\node [style=Z dot] (0) at (0, 0) {};
		\node [style=none] (1) at (0, 0.5) {};
		\node [style=none] (2) at (0.5, -0.375) {};
		\node [style=none] (3) at (0.5, -0.125) {};
		\node [style=none] (4) at (0.5, 0.125) {};
		\node [style=none] (5) at (0.5, 0.375) {};
	\end{pgfonlayer}
	\begin{pgfonlayer}{edgelayer}
		\draw (1.center) to (0);
		\draw [bend left=15] (0) to (5.center);
		\draw (0) to (4.center);
		\draw [in=165, out=-15] (0) to (3.center);
		\draw [bend right=15] (0) to (2.center);
	\end{pgfonlayer}
    	\begin{pgfonlayer}{nodelayer}
		\node [style=X dot] (11) at (-.7, 0) {};
		\node [style=none] (12) at (-.7, 0.5) {};
	\end{pgfonlayer}
	\begin{pgfonlayer}{edgelayer}
		\draw (11.center) to (0);
	\end{pgfonlayer}
\end{tikzpicture} or \begin{tikzpicture}
	\begin{pgfonlayer}{nodelayer}
		\node [style=Z dot] (0) at (0, 0) {};
		\node [style=none] (1) at (0, 0.5) {};
		\node [style=none] (2) at (0.5, -0.375) {};
		\node [style=none] (3) at (0.5, -0.125) {};
		\node [style=none] (4) at (0.5, 0.125) {};
		\node [style=none] (5) at (0.5, 0.375) {};
	\end{pgfonlayer}
	\begin{pgfonlayer}{edgelayer}
		\draw (1.center) to (0);
		\draw [bend left=15] (0) to (5.center);
		\draw (0) to (4.center);
		\draw [in=165, out=-15] (0) to (3.center);
		\draw [bend right=15] (0) to (2.center);
	\end{pgfonlayer}
    	\begin{pgfonlayer}{nodelayer}
		\node [style=X phase dot] (11) at (-.7, 0) {$\pi$};
		\node [style=none] (12) at (-.7, 0.5) {};
	\end{pgfonlayer}
	\begin{pgfonlayer}{edgelayer}
		\draw (11.center) to (0);
	\end{pgfonlayer}
\end{tikzpicture}, and \begin{tikzpicture}
	\begin{pgfonlayer}{nodelayer}
		\node [style=Z dot] (0) at (1, 0) {};
		\node [style=none] (1) at (1, 0.5) {};
		\node [style=none] (2) at (0.5, -0.375) {};
		\node [style=none] (3) at (0.5, -0.125) {};
		\node [style=none] (4) at (0.5, 0.125) {};
		\node [style=none] (5) at (0.5, 0.375) {};
	\end{pgfonlayer}
	\begin{pgfonlayer}{edgelayer}
		\draw (1.center) to (0);
		\draw [bend right=15] (0) to (5.center);
		\draw (0) to (4.center);
		\draw [in=15, out=-165] (0) to (3.center);
		\draw [bend left=15] (0) to (2.center);
	\end{pgfonlayer}
\end{tikzpicture}  by 
\begin{tikzpicture}
	\begin{pgfonlayer}{nodelayer}
		\node [style=Z dot] (0) at (1, 0) {};
		\node [style=none] (1) at (1, 0.5) {};
		\node [style=none] (2) at (0.5, -0.375) {};
		\node [style=none] (3) at (0.5, -0.125) {};
		\node [style=none] (4) at (0.5, 0.125) {};
		\node [style=none] (5) at (0.5, 0.375) {};
		\node [style=X dot] (6) at (1.7, 0) {};
		\node [style=none] (7) at (1.7, 0.5) {};
	\end{pgfonlayer}
	\begin{pgfonlayer}{edgelayer}
		\draw (1.center) to (0);
		\draw [bend right=15] (0) to (5.center);
		\draw (0) to (4.center);
		\draw [in=15, out=-165] (0) to (3.center);
		\draw [bend left=15] (0) to (2.center);
		\draw (6) to (0);
	\end{pgfonlayer}
\end{tikzpicture} or \begin{tikzpicture}
	\begin{pgfonlayer}{nodelayer}
		\node [style=Z dot] (0) at (1, 0) {};
		\node [style=none] (1) at (1, 0.5) {};
		\node [style=none] (2) at (0.5, -0.375) {};
		\node [style=none] (3) at (0.5, -0.125) {};
		\node [style=none] (4) at (0.5, 0.125) {};
		\node [style=none] (5) at (0.5, 0.375) {};
		\node [style=X phase dot] (6) at (1.7, 0) {$\pi$};
		\node [style=none] (7) at (1.7, 0.5) {};
	\end{pgfonlayer}
	\begin{pgfonlayer}{edgelayer}
		\draw (1.center) to (0);
		\draw [bend right=15] (0) to (5.center);
		\draw (0) to (4.center);
		\draw [in=15, out=-165] (0) to (3.center);
		\draw [bend left=15] (0) to (2.center);
		\draw (6) to (0);
	\end{pgfonlayer}
\end{tikzpicture}. 
Each of these diagrams selects one of the  $c_{ij}$ amplitudes. Consider for instance
\begin{equation}\label{eq:toriccode_proof_input_superposition}
    \includegraphics[width=\linewidth]{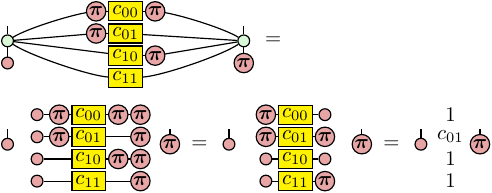}
\end{equation}
Note that the rightmost diagram in Eq.~\eqref{eq:toriccode_proof_input_superposition} features the same red spiders used as an input on the leftmost diagram.
According to the previous discussion, these spiders generate the ground state $\ket{01}_{\rm tc}$, see also Fig.~\ref{fig:toriccode_groundstate_complete}b.
However, we have kept track of the scalars resulting from Eq.~\eqref{eq:Hbox_GS}.
The scalars are arranged vertically for clarity, but, according to the ZX-calculus rules, they are multiplied together. 
We see that all the scalars are one except for $c_{01}$, which is the desired amplitude of $\ket{01}_{\rm tc}$ in Eq.~\eqref{eq:groundspace_toriccode}. Similar considerations hold for the other three diagrams, leading to the correct ZX representation of the arbitrary superposition of ground states, Eq.~\eqref{eq:groundspace_toriccode}.


\begin{figure*}[t]
    \centering
    \hspace{-16.2cm}
    \includegraphics[width=0.9\linewidth]{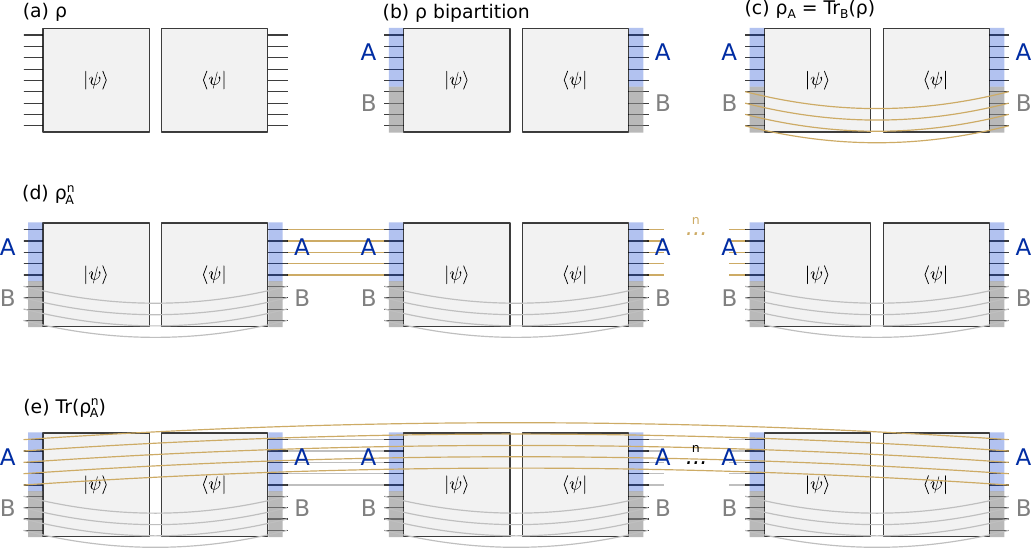}
    \caption{ 
    ZX calculation of  $\rho_{A}$, $\rho_{A}^{n}$ and related Renyi entanglement entropies $S_{n}$. (a) The density matrix $\rho=\ket \psi\bra\psi$  is represented as a ZX diagram with the convention of inputs to the left for kets and outputs to the right for bras. (b) After assigning inputs and outputs to the regions $A$ and $B$, (c) the reduced density matrix $\rho_A=\mbox{Tr}_B(\rho)$ is derived by connecting the  input and output lines corresponding to region $B$ (we take the convention to highlight all new connection in orange). (d) $\rho_A^n$ is  obtained by contracting the output lines of a ZX representation of $\rho_A$ with the input lines of a following one. (e) The final trace $\mbox{Tr}(\rho_A^n)$ is obtained by connecting the remaining inputs of the leftmost diagram to the outputs of the rightmost one. 
    }
    \label{fig:steps_calculations_rhoA_rhoAn_S_rhoAb}
\end{figure*}

\section{Reduced density matrices, entanglement entropies, contour diagrams and Pauli trees in ZX} \label{sec:results}

\subsection{Topological Entanglement Entropy Benchmarks} \label{sec:results_gamma_toriccode}

In this Section, we calculate entanglement entropies of toric-code ground states using ZX diagrams. These calculations benchmark our  ZX derivations   and provide a first step towards the derivation of the contour diagrams and Pauli trees discussed in the main text.

As discussed in Section~\ref{sec:TEE}, for long-range topologically entangled states, Renyi entanglement entropies scale as $S_n = \alpha_n L -\gamma$, Eq.~\eqref{eq:topologicalarealaw}, where $L$ is the number of qubits at the boundary, $\alpha_{n}$ is a coefficient and $\gamma$ is the topological entanglement entropy. %
We start by choosing a particular bipartition  $(A,B)$, where $A$ is a rectangle  as in Fig.~\ref{fig:toriccode_gammatheory_rectangle}a. We recall that, in this case, the number of shared plaquettes $\Sigma_{AB}$ between the regions $A$ and $B$ coincides with the number of qubits $L$ delimiting  region $A$, see Section~\ref{sec:gamma_toriccode}. Thus, for this specific choice, the topological entanglement $\gamma$ does not suffer from spurious contributions ($\gamma= \gamma_{\rm topo}$), see also Section~\ref{sec:spurious}. 

In Fig.~\ref{fig:steps_calculations_rhoA_rhoAn_S_rhoAb}, we specify the steps to calculate the Renyi entropies $S_{n} = -\frac{1}{n-1}\log(\Tr(\rho_{A}^{n}))$, see Eq.~\eqref{eq:renyi_entropy_app}, based on the ZX-calculus rules introduced in Section~\ref{sec:intro_ZX}. 
For practical purposes, we focus on the second Renyi entropy $S_2$. This choice is convenient for two reasons. First, compared to the von Neumann entanglement entropy Eq.~\eqref{eq:entanglement_entropy_app}, the Renyi entanglement entropies $S_{n}$ avoid calculating the logarithm of a diagram or a matrix.
Second, compared to other Renyi entranglement entropies, $S_{2}$ depends only on the square of $\rho_{A}$, reducing operations and, consequently, computational complexity.

\begin{figure}[t]
    \centering
    \hspace{-8.7cm}
    \includegraphics[width=0.97\linewidth]{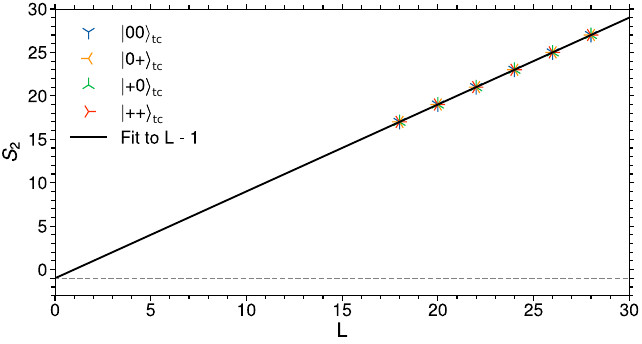}
    \caption{
    Second Renyi entropy $S_{2}$,  Eq.~\eqref{eq:renyi_entropy_app}, in units of $\log_{2}$, as a function of the bipartition boundary length $L$. Here $L$ is the number of boundary qubits, which for these cuts coincides with $\Sigma_{AB}$, see text. The points correspond  to calculations in ZX for the ground states $\ket{00}_{\rm tc}$ (blue), $\ket{0+}_{\rm tc}$ (yellow), $\ket{+0}_{\rm tc}$ (green) and $\ket{++}_{\rm tc}$ (red). According to Eq.~\eqref{eq:entropy_hamma}, all the points fall on top of $S=L-\gamma$, with $\gamma=1$. The dashed horizontal line  intersects the vertical axis at $-1$. 
    }
    \label{fig:gamma_plots}
\end{figure}

To construct and efficiently simplify ZX diagrams, we use the python package \textsc{PyZX}~\cite{kissinger2019Pyzx}.
\textsc{PyZX} allows to define diagrams specifying the location of the nodes (spiders), their phases and the connections between them (wires).
It provides routines for the ZX simplifying rules that can be applied to  a given diagram.

To calculate the Renyi entropies $S_n$, we strongly rely on the \textsc{full\_reduce()} routine. It takes diagrams as inputs and it outputs the most simplified version of the diagram, with the least possible number of nodes.
Notice that, within ZX calculus, one may find more than one fully simplified diagram and that full reduction may turn into an NP hard problem, as it is equivalent to efficiently contract a tensor network~\cite{de2022circuit}. 
However, for the states considered in this work, \textsc{full\_reduce()} seems  efficient for the calculation of $S$, and we will provide an explicit procedure to derive $\rho_A$ and $S_2$ for the toric code in Section~\ref{sec:results_rhoA_toriccode}.

Figure~\ref{fig:gamma_plots} shows the entropy $S_{2}$ calculated with \textsc{PyZX} as a function of the boundary length $L$ and for four different ground states, $\ket{00}_{\rm tc}$, $\ket{+0}_{\rm tc}$, $\ket{0+}_{\rm tc}$ and $\ket{++}_{\rm tc}$, whose expression as ZX diagrams was given in Fig.~\ref{fig:toriccode_groundstate_complete}.
From Fig.~\ref{fig:gamma_plots}, we verify that $\gamma$ does not depend on the ground state for the rectangular bipartition.
In particular, we observe that $\gamma = 1$ for all the ground states, in agreement with the results exposed in Section~\ref{sec:gamma_toriccode}.
As a final comment, we have performed analogous calculations for other kinds of bipartitions, including bipartitions where we expect $\gamma$ to depend on the ground state. 
We have verified that our ZX calculations led to the expected $\gamma$ obtained from the theory of Ref.~\cite{Hamma} explained in Section~\ref{sec:gamma_toriccode}.
%


\subsection{Reduced density matrices as ZX diagrams}\label{sec:results_rhoA_toriccode}

In this Section, we provide an explicit derivation of the reduced density matrix $\rho_{A}$ of the toric code as a ZX diagram. This calculation breaks down the steps  in Fig.~\ref{fig:steps_calculations_rhoA_rhoAn_S_rhoAb} and it illustrates the ability of ZX calculus to identify visually the Pauli trees contained in $\rho_A$.
Later in Section \ref{sec:rhoAb} we will use steps of this warm-up calculation to derive $\mathcal{D}_{\partial A}$, the robust contour diagram we discuss in the main text.
We initially focus on the rectangular bipartitions of Fig.~\ref{fig:toriccode_gammatheory_rectangle}. We  discuss irregular and non-contractible bipartitions in Sec.~\ref{sec:results_xmas_toriccode}. The derivations have been performed with the python package \textsc{ZXLive}~\cite{Zxlive}, which is a  graphical tool for ZX calculus. Here, we will just outline the most important steps of the derivation of $\rho_A$. The entire proof is included in a supplemental \textsc{ZXLive} file \textit{toric\_code/rhoA.zxp}, in Ref.~\cite{Mas_mendoza_zenodo}.

\subsubsection{The ZX diagram for $\rho_A$}

The starting object is the ZX diagram representing 
$\rho = \ket{00}_{\rm tc}\bra{00}_{\rm tc}$, where we assign qubits to each partition $A$ (in blue) and $B$ (in gray)
        \begin{equation}\label{eq:proof_rhoA_0}
            \centering
            \includegraphics[width=\linewidth]{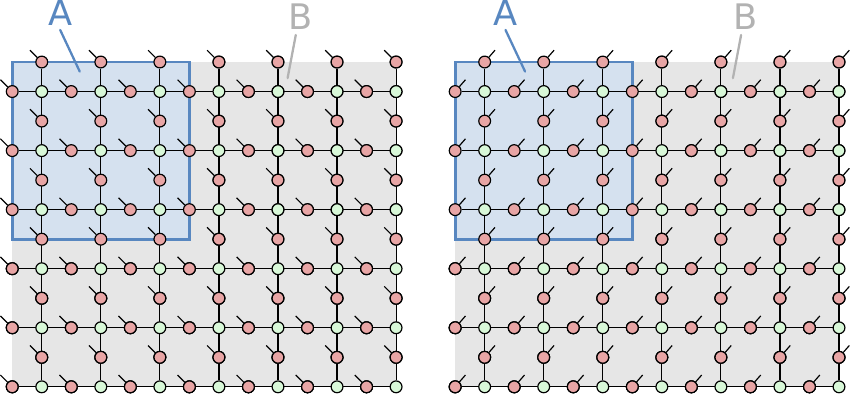}
        \end{equation}
Recall that $\bra{00}_{\rm tc}$ is obtained by switching wires from outputs to inputs in the ZX diagram representing $\ket{00}_{\rm tc}$ in Fig.~\ref{fig:toriccode_protocol}f, see also Section~\ref{sec:intro_ZX}.  Since all phases in the diagram are zero, they do not change after conjugation.

To trace  the qubits in subsystem $B$, we connect the inputs and outputs of this region as specified in Section~\ref{sec:ZXtranspositionconjugationtracing}
\begin{equation}\label{eq:proof_rhoA_1}
    \centering
    \includegraphics[width=\linewidth]{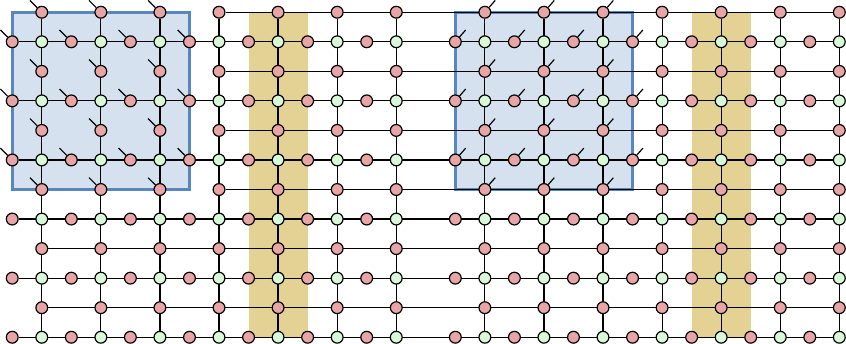}
\end{equation}
This diagram provides a formal expression of $\rho_A=\mbox{Tr}_B(\rho)$. However, it can be further simplified using ZX calculus as follows. Consider the ensemble of red X spiders highlighted in the orange region  in Eq.~\eqref{eq:proof_rhoA_1}. According to the $\FusionRule$ rule, they can be fused leading to 
\begin{equation}\label{eq:proof_rhoA_2}
    \centering
    \includegraphics[width=\linewidth]{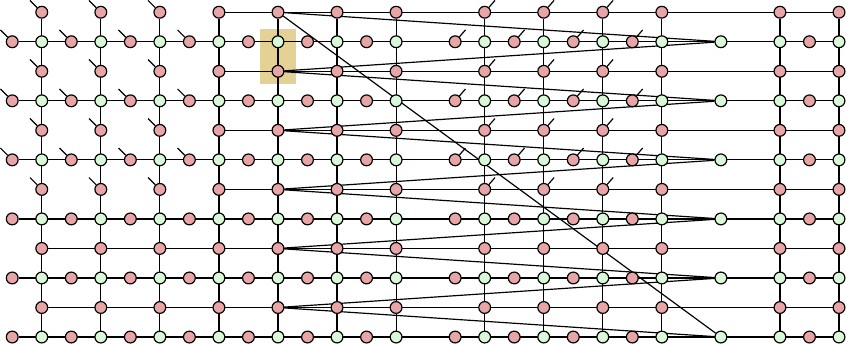}
\end{equation}
Next, we use the bialgebra rule $\BialgRule$ on the pair of red X and green Z spiders highlighted in orange in Eq.~\eqref{eq:proof_rhoA_2}.     The resulting diagram reads
        \begin{equation}\label{eq:proof_rhoA_3-1}
            \centering
            \includegraphics[width=\linewidth]{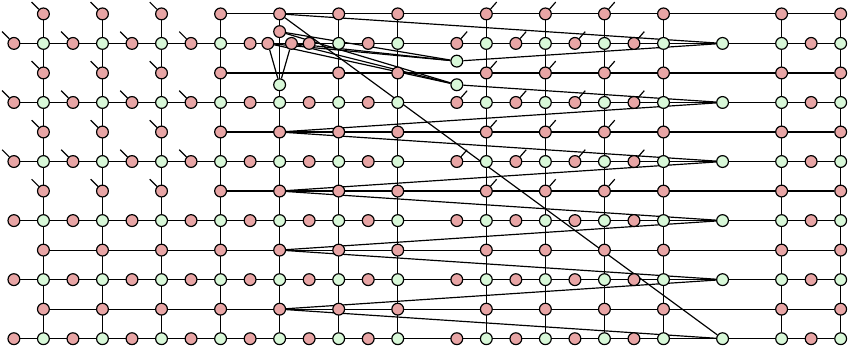}
        \end{equation}
where  additional red X and green Z spiders were generated. They can be also fused via the $\FusionRule$ rule. We obtain 
\begin{equation}\label{eq:proof_rhoA_3-2}
    \centering
    \includegraphics[width=\linewidth]{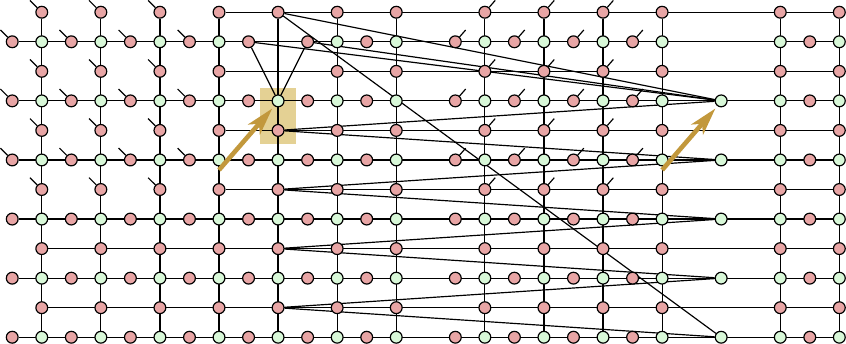}
\end{equation}
These manipulations have reduced the number of spiders in the diagram. Remarkably, the connections of the simplified green Z spiders  are now inherited by the green Z spiders below the simplified pair, which are indicated by arrows. 
This operation can be repeated for the pair below, which is again highlighted in orange in Eq.~\eqref{eq:proof_rhoA_3-2}, leading to 
        \begin{equation}\label{eq:proof_rhoA_4}
            \centering
            \includegraphics[width=\linewidth]{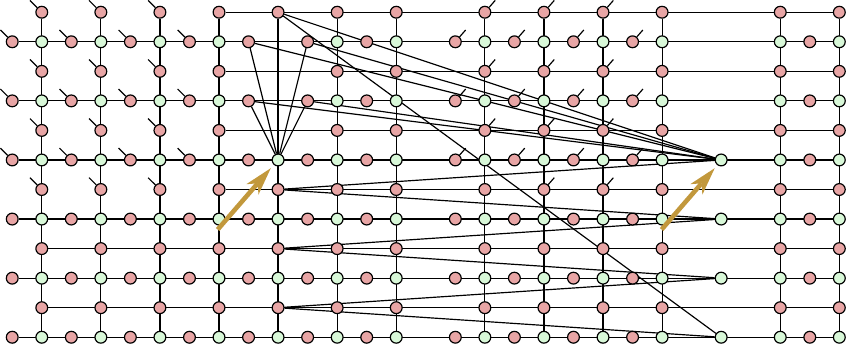}
        \end{equation}
Also in this case, the green Z spiders  pointed by arrows, have  inherited all the connections of the two simplified green spiders. This operation can be thus repeated all along the pairs of red X and green Z spiders of the column:
\begin{equation}\label{eq:proof_rhoA_5}
    \centering
    \includegraphics[width=\linewidth]{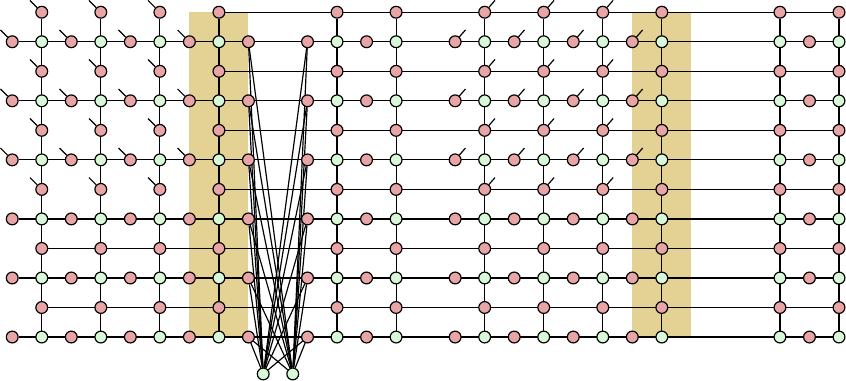}
\end{equation}
This diagram already showcases non-trivial topological properties of the state $\ket{00}_{\rm tc}$. By partially tracing  region $B$ along a non-trivial loop of the torus, we observe the emergence of two non-local green Z spiders, which are connected to all the red X spiders neighboring the traced region. They are a preliminary signature -- even though not conclusive, as we are going to discuss --  of the non-trivial topology diagnosed by Pauli trees, see also main text. 

We repeat the previous steps for the neighboring column highlighted in orange in Eq.~\eqref{eq:proof_rhoA_5}
        \begin{equation}\label{eq:proof_rhoA_6}
            \centering
            \includegraphics[width=\linewidth]{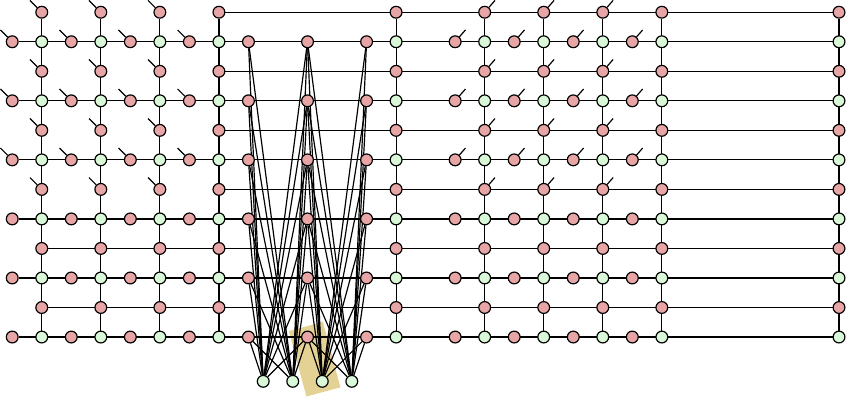}
        \end{equation}
Now the number of non-local green Z spiders has doubled. However, by performing a bialgebra $\BialgRule$ between the red X and green Z spider pair highlighted in orange at the bottom of Eq.~\eqref{eq:proof_rhoA_6},  and then fusing $\FusionRule$ all the remaining spiders, we obtain a simplified version of the diagram, where the number of non-local green Z spiders comes back to two 
        \begin{equation}\label{eq:proof_rhoA_7}
            \centering
            \includegraphics[width=\linewidth]{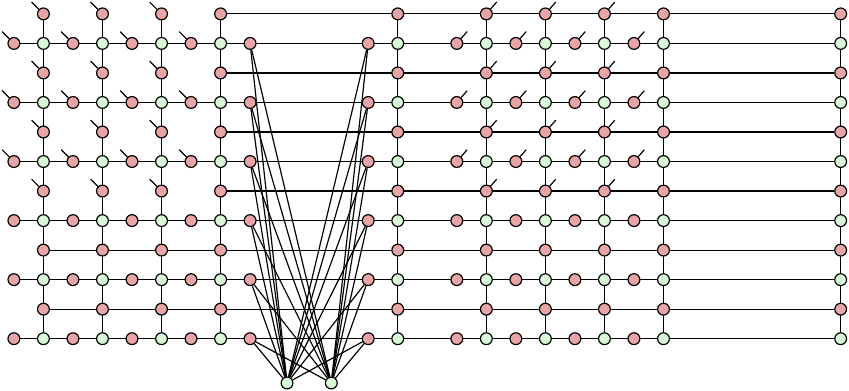}
        \end{equation}
Notice that performing the first bialgebra on any other pair of green Z  and red X spider leads to the same simplification. Remarkably, the diagram~\eqref{eq:proof_rhoA_7} has the same structure of Eq.~\eqref{eq:proof_rhoA_5}, while featuring fewer red spiders.

This simplification can be thus extended  for every column which is entirely in $B$, leading to 
    \begin{equation}\label{eq:proof_rhoA_8}
        \centering
        \includegraphics[width=\linewidth]{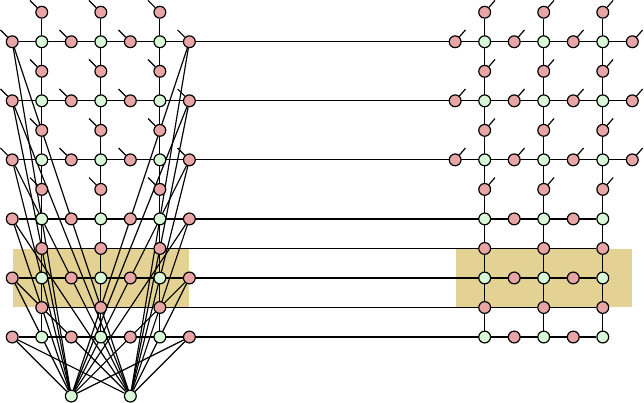}
    \end{equation}

Now that we have  simplified the columns, we perform the same procedure along the rows of $B$, starting from the  column highlighted in orange in~\eqref{eq:proof_rhoA_8}. We obtain:
        \begin{equation}\label{eq:proof_rhoA_9}
            \centering
            \includegraphics[width=\linewidth]{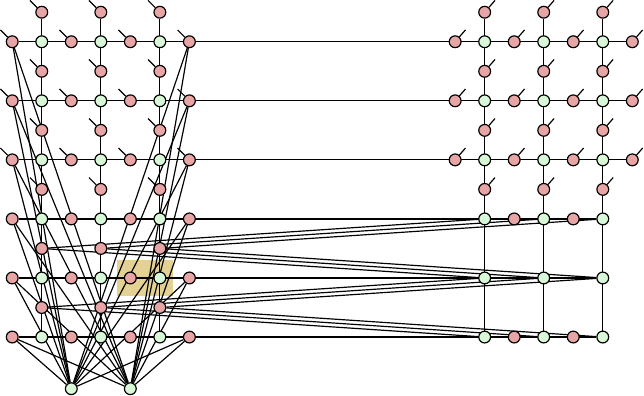}
        \end{equation}
Notice that the  conjugated green Z spiders in the simplified row are now connected to the same four red X spiders of the plaquettes. In analogy to the simplification applied to Eq.~\eqref{eq:proof_rhoA_2} that results in Eqs.~\eqref{eq:proof_rhoA_3-1} and ~\eqref{eq:proof_rhoA_3-2}, we simplify the highlighted pair of green Z and red X spiders in Eq.~\eqref{eq:proof_rhoA_9} using the bialgebra rule $\BialgRule$, and then fusing $\FusionRule$ the spiders. We obtain:
        \begin{equation}\label{eq:proof_rhoA_10}
            \centering
            \includegraphics[width=\linewidth]{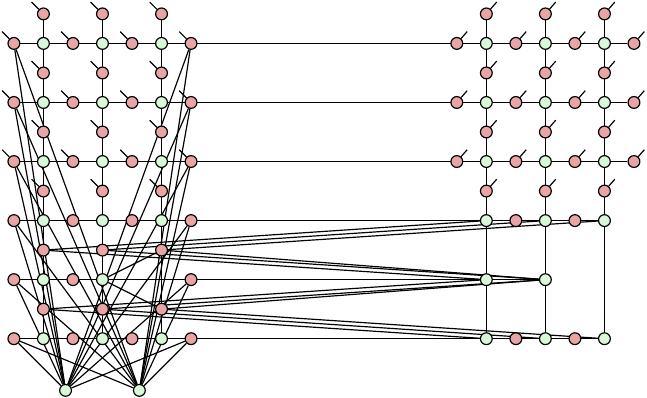}
        \end{equation}
and repeat the same operation for all pairs of green Z and red X spiders in the row, leading to 
        \begin{equation}\label{eq:proof_rhoA_11}
            \centering
            \includegraphics[width=\linewidth]{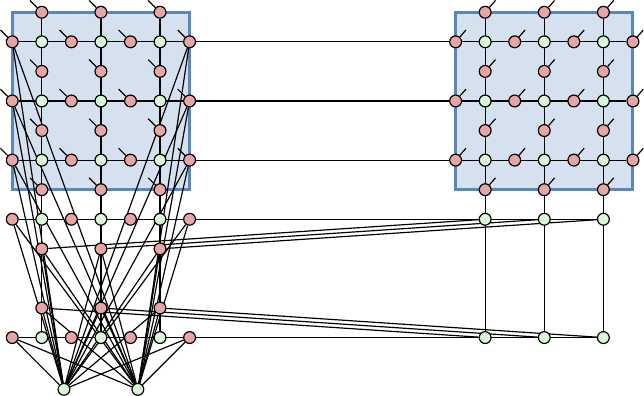}
        \end{equation}
where we have retraced region $A$ for the reader's convenience. 
By repeating analogous simplifications for all the remaining rows entirely in $B$, we obtain a first version of the reduced density matrix  $\rho_A$, where all the red spiders associated to qubits in region $B$ have been traced out
        \begin{equation}\label{eq:proof_rhoA_12}
            \centering
            \includegraphics[width=\linewidth]{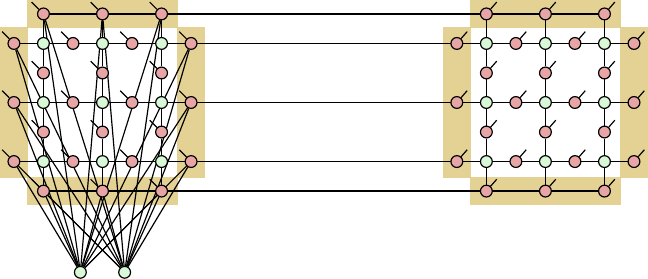}
        \end{equation}    
The fact that $\rho_A$ is a mixed state is signaled by the fact that the left and right part of the diagram cannot be disconnected. This is not the case for the diagram~\eqref{eq:proof_rhoA_0}, describing the pure state $\rho= \ket{00}_{\rm tc}\bra{00}_{\rm tc}$.  It is now  a matter of further simplifications to highlight the hidden presence of the non-local Pauli trees in $\rho_A$.


\subsubsection{Appearance of generalized  Pauli projectors in $\rho_A$}
The diagram~\eqref{eq:proof_rhoA_12} can be further simplified by fusing $\FusionRule$ the red X spiders highlighted in orange in Eq.~\eqref{eq:proof_rhoA_12}. They are located at the boundary of $A$ and, to facilitate our discussion, we place them symmetrically in the middle, together with the two non-local green Z spiders 
 \begin{equation}\label{eq:proof_rhoA_14}
            \centering
            \includegraphics[width=\linewidth]{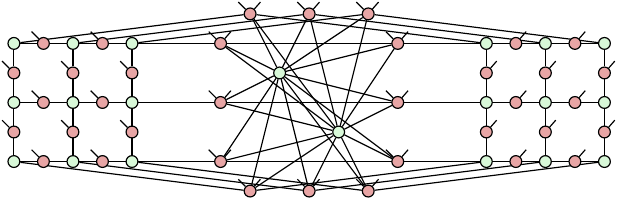}
        \end{equation}
This form clearly shows that the interior of the region $A$ preserves its original form as in Eq.~\eqref{eq:proof_rhoA_0}, both in its bra and ket components. The tracing of region $B$ has led to  two non-local green Z spiders, which connect exclusively to the qubits at the boundary of $A$.

However, the number of non-local green spiders can be further reduced by applying the fusion \FusionRule, bi-algebra $\BialgRule$ and copy $\CopyRule$ rules as follows:
\begin{equation}\label{eq:laststep_proof_rhoA}
            \centering
            \includegraphics[width=\linewidth]{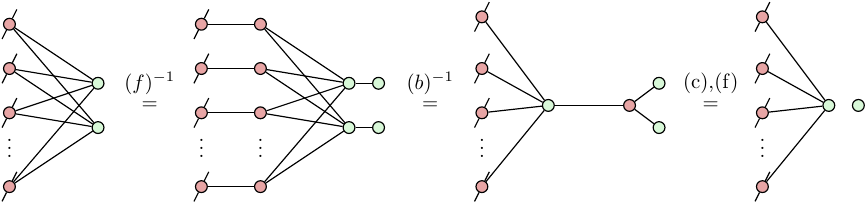}
        \end{equation}
leading to \footnote{We thank Norbert Schuch for suggesting to us the possibility of this simplification. }
\begin{equation}\label{eq:rhoA_in_its_full_beauty}
            \centering
            \includegraphics[width=\linewidth]{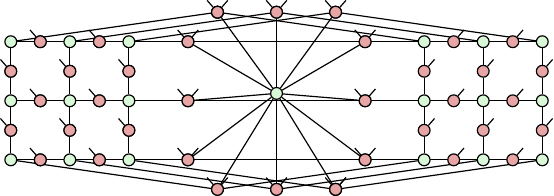}
        \end{equation}
We see that the operator on the right hand side of Eq.~\eqref{eq:laststep_proof_rhoA} acts on the qubits at the boundary of region $A$. This operator is a generalization to $L$ qubits of the operator in Fig.~\ref{fig:toriccode_protocol}b, which appears in the toric code ground state $\ket{00}_{\rm tc}$, Eq.~\eqref{eq:groundstate00_toriccode}. As discussed in the main text, when suitably normalized, such operators are a generalization of Pauli projectors~\cite{KissingerWetering2024Book} and will lead to the  {\it Pauli trees} discussed in Section~\ref{sec:results_xmas_toriccode}. In the case of the diagram~\eqref{eq:rhoA_in_its_full_beauty}, their tree structure  features exactly $L$ red X spiders and a single, non-local Z green spider.   This structure is interesting as it seems to reflect the expression~\eqref{eq:entropy_hamma} for the entaglement entropy. Recall that it equals $S=L-1=\Sigma_{AB}-1$  for regular cuts along $z$- or $x$-paths, see Section~\ref{sec:gamma_toriccode}, and it counts exactly the difference between the number of red X (`$L$') and green Z (`$1$') spiders composing the ZX representation of the  Pauli projector.

\begin{figure}[t]
    \centering
    \hspace{-9cm}
    \includegraphics[width=\linewidth]{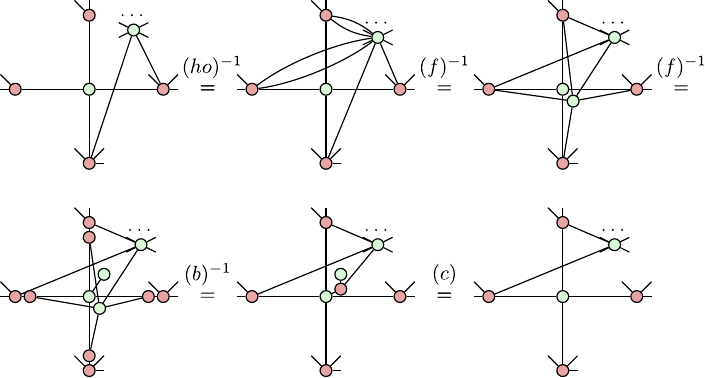}
    \caption{Diagrammatic equalities showing how  to displace the connections of the non-local green Z spider between two different couples of red X spiders of the same plaquette. As a consequence, one can displace the non-local green spider in Eq.~\eqref{eq:rhoA_in_its_full_beauty} to the bulk in Eq.~\eqref{eq:green_displaced}.}
    \label{fig:rule_no_green}
\end{figure}

However, relying on the diagrammatic equalities  in Fig.~\ref{fig:rule_no_green},  the non-local green Z spider in the diagram~\eqref{eq:rhoA_in_its_full_beauty} can be displaced inside the bulk of $A$ and  connected to the four red spider of a single plaquette as follows
\begin{equation}\label{eq:green_displaced}
\centering
\includegraphics[width=\linewidth]{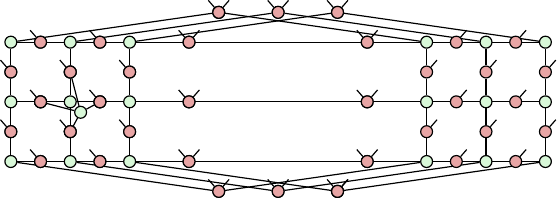}
\end{equation}
In this diagram, the operator  $I + P_{\square}=$
    \begin{tikzpicture}
	\begin{pgfonlayer}{nodelayer}
		\node [style=Z dot] (0) at (0, 0) {};
		\node [style=X dot] (1) at (0.75, 0) {};
		\node [style=X dot] (2) at (0, 0.75) {};
		\node [style=X dot] (3) at (-0.75, 0) {};
		\node [style=X dot] (4) at (0, -0.75) {};
		\node [style=none] (5) at (1, 0.25) {};
		\node [style=none] (6) at (0.5, -0.25) {};
		\node [style=none] (7) at (0.25, 1) {};
		\node [style=none] (8) at (-0.25, 0.5) {};
		\node [style=none] (9) at (-0.5, 0.25) {};
		\node [style=none] (10) at (-1, -0.25) {};
		\node [style=none] (11) at (0.25, -0.5) {};
		\node [style=none] (12) at (-0.25, -1) {};
	\end{pgfonlayer}
	\begin{pgfonlayer}{edgelayer}
		\draw (0) to (4);
		\draw (0) to (1);
		\draw (0) to (2);
		\draw (0) to (3);
		\draw (3) to (9.center);
		\draw (3) to (10.center);
		\draw (4) to (12.center);
		\draw (4) to (11.center);
		\draw (1) to (6.center);
		\draw (1) to (5.center);
		\draw (2) to (7.center);
		\draw (2) to (8.center);
	\end{pgfonlayer}
\end{tikzpicture} is applied twice to the same plaquette. However, $(I+P_\square)^2=2(I+P_\square)$, which can be also shown in ZX by applying  the equalities in  Eq.~\eqref{eq:laststep_proof_rhoA} (ignoring the factor 2).
Thus, the non-local green Z spider in Eq.~\eqref{eq:rhoA_in_its_full_beauty}, can be eliminated leading to the diagram 
\begin{equation}\label{eq:rhoA_with_less_beauty}
    \centering
    \includegraphics[width=\linewidth]{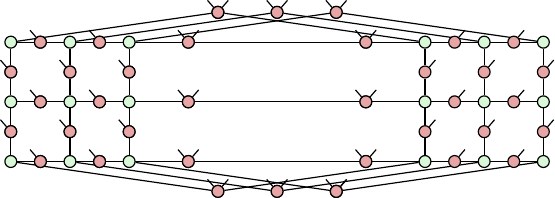}
\end{equation}
It is important to note that this simplification implies  a scalar factor that should be accounted for when calculating the entanglement entropy.

These considerations  extend to all superposition of ground-state diagrams collected in Fig.~\ref{fig:toriccode_groundstate_complete}a. We also append in Ref.~\cite{Mas_mendoza_zenodo} the \textsc{ZXLive} file \textit{toric\_code/rhoA\_superposition.zxp} of the proof of how to get to the diagram Eq.~\eqref{eq:proof_rhoA_14} for a general ground state choice, Eq.~\eqref{eq:groundspace_toriccode}. The related derivation of Eq.~\eqref{eq:rhoA_with_less_beauty} for a general ground-state superposition follows the same logic as outlined above.

As a final comment, recall from Eq.~\eqref{eq:rhoApower_toric} that all the $\rho_{A}^{n}$ are proportional to each other for the toric-code ground state.
We can represent $\rho_{A}^{n}$ as a ZX diagram following the procedure  described in Fig.~\ref{fig:steps_calculations_rhoA_rhoAn_S_rhoAb}. By performing the contractions, one can show that all the density matrices are described by the same ZX diagram and differ only by a multiplicative constant, as expected.

\subsection{Contour diagram $\mathcal D_{\partial A}$ and Pauli tree } \label{sec:rhoAb}

As we have seen in the previous Section, $\rho_{A}$ may be put in a form featuring a non-local spider connected to its boundary. This non-local spider has interesting properties, which directly relate to the long-range topological entanglement properties of the system.
However, the fact that this non-local spider can be simplified in the bulk of region $A$ prevents the possibility to isolate such object and  test its robust connection to entanglement. 

In this Section, we provide a recipe to unambiguously eliminate the bulk of region $A$ in such a way to derive  a {\it countour diagram} $\mathcal D_{\partial A}$. This diagram isolates the non-local spider, can be simplified to a Pauli-tree form  and correctly diagnoses long-range topological order, by isolating the topological entanglement entropy $\gamma_\mathrm{top}$, without suffering from spurious contributions.

\subsubsection{Definition and motivation for $\mathcal D_{\partial A}$} \label{sec:def_rhoAb}

To start, we notice that eliminating the non-local spider in Eq.~\eqref{eq:rhoA_in_its_full_beauty} --  leading to Eq.~\eqref{eq:rhoA_with_less_beauty} -- is not possible if region $A$ is chosen topologically equivalent to an annulus, namely by tracing part of its bulk. 

To unambiguously isolate this non-local spider, irrespective to the bulk and contour properties of  region  $A$, we get inspiration from Eq.~\eqref{eq:entropy_hamma} of the entanglement entropy $S$  derived in Ref.~\cite{Hamma}, and that we reproduced numerically with ZX in Section~\ref{sec:results_gamma_toriccode}, see Fig.~\ref{fig:gamma_plots}. Recall that  the von Neumann and the Renyi entropies equal $S=\Sigma_{AB}-\gamma_\mathrm{top}$, where the topological contribution is $\gamma_\mathrm{top}=1$ and $\Sigma_{AB}$ is the number of plaquette operators acting simultaneously on the partitions $A$ and $B$.

This motivates the strategy we follow in the main text, which is to pin \emph{only} the boundary degrees of freedom associated to $\Sigma_{AB}$ and read off the entanglement between regions $A$ and $B$. 
We remind the reader that pinning refers to the act of doubling the spiders associated to qubits connected by ZX edges which link the qubits of region $A$ to qubits in region $B$. This is practically achieved by relying on the inverse of the ZX \FusionRule rule 
\begin{tikzpicture}
	\begin{pgfonlayer}{nodelayer}
		\node [style=X dot] (0) at (1, 0) {};
		\node [style=none] (1) at (0.5, 0.5) {};
		\node [style=none] (2) at (0.5, -0.5) {};
		\node [style=none] (3) at (0.5, 0.04) {\rotatebox[origin=c]{90}{\tiny \dots}};
	\end{pgfonlayer}
	\begin{pgfonlayer}{edgelayer}
		\draw (1.center) to (0);
		\draw (0) to (2.center);
	\end{pgfonlayer}
\end{tikzpicture} = 
\begin{tikzpicture}
	\begin{pgfonlayer}{nodelayer}
		\node [style=X dot] (0) at (1, 0) {};
		\node [style=X dot] (1) at (2, 0) {};
		\node [style=none] (2) at (0.5, 0.5) {};
		\node [style=none] (3) at (0.5, -0.5) {};
		\node [style=none] (4) at (0.5, 0.04) {\rotatebox[origin=c]{90}{\tiny \dots}};
	\end{pgfonlayer}
	\begin{pgfonlayer}{edgelayer}
		\draw (2.center) to (0);
		\draw (0) to (3.center);
		\draw (0) to (1);
	\end{pgfonlayer}
\end{tikzpicture}
We refer to this operation as \textit{pinning the spider}, as it indicates which doubled qubits to leave unsimplified.
Pinning is merely a computational tool allowing us to keep track of the entanglement across the boundary while using the existing python packages \textsc{PyZX}~\cite{kissinger2019Pyzx} and \textsc{ZX-Live}~\cite{Zxlive} for diagrammatic simplification.

\begin{figure}[t]
    \centering
    \hspace{-9cm}
    \includegraphics[width=\linewidth]{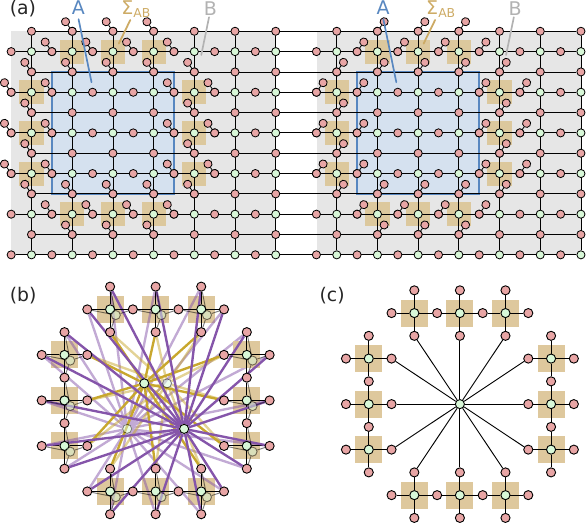}
    \caption{
    Derivation of the contour diagram $\mathcal D_{\partial A}$ of the toric code.
    (a) Representation in ZX of $\Tr(\rho)$.
    We choose a bipartition $(A,B)$ ($A$ in blue, $B$ in grey) and highlight in orange the $\Sigma_{AB}$  green spiders corresponding to the plaquettes acting on both $A$ and $B$.
    We then double (pin) the red X spiders of each of those plaquettes using the inverse of the fusion \FusionRule rule.
    (b) Simplified version of the ZX diagram in (a). As a visual aid, we thicken the connections of the outer and inner boundary non-local spiders in purple and yellow, respectively. We have also reduced the opacity of one of the conjugated copies of all the green spiders (either local or non-local) since we know that, in virtue of Eq.~\eqref{eq:laststep_proof_rhoA}, they can be simplified away, see also step~\eqref{eq:step1}.
    (c) Further simplification of the diagram in (b), following the steps described in the text.
    }
    \label{fig:rhoAb}
\end{figure}

We  explain in this and the next section two complementary calculations that define two representations of the contour diagram $\mathcal D_{\partial A}$, which are summarized in Figs.~\ref{fig:rhoAb} and \ref{fig:xmastree}. 
First, in Fig.~\ref{fig:rhoAb}a we show the unsimplified ZX diagram of $\mbox{Tr}(\rho)=1$ where we highlight with orange squares the plaquettes that belong to $\Sigma_{AB}$. 
Here, we have pinned the qubits that belong to $\Sigma_{AB}$ plaquettes by using the inverse fusion \FusionRule rule to double (pin) the red X spiders with connections to $A$ and $B$. Note that, in contrast to the previous section, we now trace \textit{all} the degrees of freedom, while simultaneously avoiding to simplify the pinned spiders.

We simplify the diagram in  Fig.~\ref{fig:rhoAb}a following a procedure which is entirely analogous to the one leading to the reduced density matrix $\rho_A$,  Eq.~\eqref{eq:proof_rhoA_14}.
The only difference with respect to the calculation presented in Section~\ref{sec:results_rhoA_toriccode} is that we now trace also qubits in the interior of $A$.
The \textsc{ZXLive} file of this calculation, \textit{toric\_code/contour\_diagram.zxp}, is appended in Ref.~\cite{Mas_mendoza_zenodo}.

We thus obtain the diagram in Fig.~\ref{fig:rhoAb}b.
It features exactly $\Sigma_{AB}$ plaquettes (highlighted by orange squares), each with two green Z spiders, one from each conjugated copy.
In addition, the diagram features four non-local green Z spiders, positioned at the center. 
Two of them, with connections highlighted in purple, are connected to the red X spiders of the external boundary and coincide with those derived in Eq.~\eqref{eq:proof_rhoA_14}, resulting from the tracing of region $B$, external to $A$. 
The two additional ones, with connections highlighted in yellow, connect instead to the interior of the boundary and result from tracing the bulk region inside $A$.

Exactly as for the simplification leading from Eq.~\eqref{eq:proof_rhoA_14} to Eq.~\eqref{eq:rhoA_in_its_full_beauty}, applying Eq.~\eqref{eq:laststep_proof_rhoA} halves the number of non-local green spiders, leading to 
\begin{equation}\label{eq:step1}
    \centering
    \includegraphics[width=0.5\linewidth]{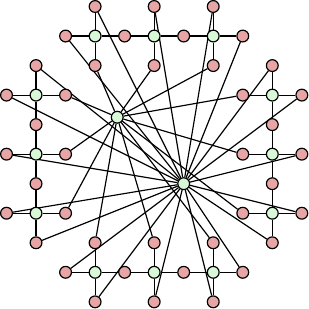}
\end{equation}
We further simplify the diagram by  applying the equalities in Fig.~\ref{fig:rule_no_green}, moving the connections  of the non-local green Z spider from  the outer boundary to the inner one \footnote{we could equivalently move the inner spider to the outer boundary}:
\begin{equation}\label{eq:step2}
    \centering
    \includegraphics[width=0.5\linewidth]{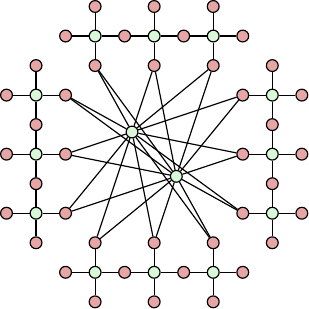}
\end{equation}
We observe that now the two non-local spiders connect to the same red spiders as in Fig.~\ref{fig:rhoAb}b.
Applying Eq.~\eqref{eq:laststep_proof_rhoA}  simplifies one of the non-local spiders, leading to the final result, the diagram in Fig.~\ref{fig:rhoAb}c, reported in the main text.

Comparing this derivation to the procedure converting the diagram in~\eqref{eq:rhoA_in_its_full_beauty}, with one non-local green Z spider, into the diagram in~\eqref{eq:green_displaced}, without non-local green spiders, clearly shows that the non-local green spider cannot be simplified if region $A$ is ring-shaped.

The number of non-local green spiders diagnoses the topological nature of the toric-code ground state with $\gamma=\gamma_{\rm topo} = 1$, as discussed in the main text. 

\subsubsection{Pauli trees  and connection to topological entanglement entropy in the toric code} \label{sec:results_xmas_toriccode}

\begin{figure}[t]
    \centering
    \hspace{-8.7cm}
    \includegraphics[width=\linewidth]{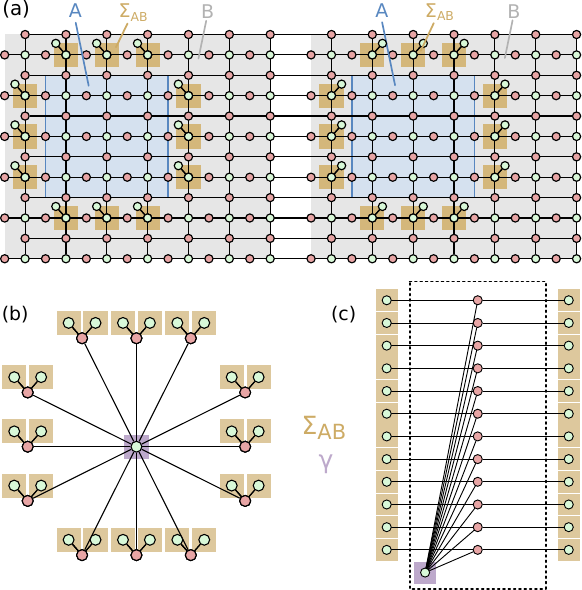}
    \caption{ 
    Pauli tree representation of the contour diagram $\mathcal{D}_{\partial A}$
    (a) Representation in ZX of $\mbox{Tr}(\rho)$=1 for the toric-code ground state. All the green spiders in the $\Sigma_{AB}$ orange boxes act on qubits (red spiders) of both the $A$ and $B$ regions. They have been doubled to show the emergence of Pauli trees, see text.
    (b) Simplified version of the ZX diagram in (a), where an additional green spider, highlighted in purple, non-locally connects to $\Sigma_{AB}$ red X spiders. 
    (c) Equivalent rearrangement of the spiders in the ZX diagram (b). The resulting diagram has a Pauli tree form, see main text, with $\Sigma_{AB}$ red X spiders and $\gamma=\gamma_{\rm topo}=1$ non-local green Z spiders.
    The dashed rectangle encloses a Pauli projector, as discussed in the main text.
    }
    \label{fig:xmastree}
\end{figure}

We now show that, in the particular case of the toric code, it is possible to further simplify the diagram in Fig.~\ref{fig:rhoAb}c to a {\it Pauli tree}, introduced in the main text. 
The Pauli tree is a representation of the contour diagram $\mathcal D_{\partial A}$ that is useful to directly compare the results of the toric code in the square lattice with those for the hexagonal toric and  color codes discussed below.
Combining these examples, we will see that the Pauli tree remains the same except for the number of non-local spiders that track the class of topological order.

To obtain the Pauli tree, we follow a procedure which is entirely analogous to the one sketched in Fig.~\ref{fig:rhoAb}. However, this time we pin the green spiders corresponding to the $\Sigma_{AB}$ shared plaquettes, as we show in Fig.~\ref{fig:xmastree}a, instead of the corresponding red spiders~\footnote{It is of course possible to simplify directly the diagram in Fig.~\ref{fig:rhoAb}c, obtaining the same result, but it turns out to be more involved.}.
The \textsc{ZXLive} file of this calculation, \textit{toric\_code/pauli\_tree.zxp}, is appended in Ref.~\cite{Mas_mendoza_zenodo} 

Proceeding this way, we obtain the diagram in Fig.~\ref{fig:xmastree}b, that we  rearrange in the equivalent form in Fig.~\ref{fig:xmastree}c.
It features exactly $\Sigma_{AB}$ red X spiders, acting on the pinning green Z spiders introduced in panel (a).
Additionally, a single green Z spider is connected to each of the $\Sigma_{AB}$ red spiders, signaling the topological nature of the toric code ground state with $\gamma=\gamma_{\rm topo} = 1$, see main text. 
This diagram is similar to the one in Fig.~\ref{fig:rhoAb}c, but the boundary plaquettes are simplified to a collection of green Z spiders connected to a single red X spider. 

Lacking outputs, the diagram in Fig.~\ref{fig:xmastree}c represents a number, more precisely $2^{\Sigma_{AB}/2 + 1}$.
This number differs from the expected result $\mbox{Tr}(\rho) = 1$, or equivalently for the toric-code ground state, $\mbox{Tr}(\rho^n)=\xi^{n-1}$, where $\xi=2^{1-\Sigma_{AB}}$, leading to $S_n=\Sigma_{AB}-1$.
However, as mentioned above, this apparent ``discrepancy'' is expected, as we have deliberately neglected all the multiplicative constants generated by the ZX simplifications.
The bookkeeping of these constants leads to the correct result, as we have shown in Section~\ref{sec:results_gamma_toriccode}.
However, as stressed in the main text,  we aim to show that topological long-range order can be diagnosed directly from a  ZX simplification of the density matrix $\rho$, rather than from complex functionals of $\rho_A$, such as entanglement entropies, mutual informations or similar quantities.

As discussed in the main text, the diagram in Fig.~\ref{fig:xmastree}c,  
acting on the green Z spiders pinning the $\Sigma_{AB}$ plaquettes shared between the regions $A$ and $B$, is the ZX representation of what we call a {\it Pauli tree}. Its central part, highlighted within a dashed rectangle, features a Pauli projector~\cite{KissingerWetering2024Book}.

We show  now the robustness of the non-local green spider in the ZX representation of the Pauli tree to generic superposition of toric code ground states~\eqref{eq:groundspace_toriccode} and arbitrary contour deformations, including non-contractible ones winding around the torus. 

{\it Arbitrary superpositions --}   The contour diagram $\mathcal D_{\partial A}$ and the associated Pauli tree do not depend on the superposition between toric-code ground states~\eqref{eq:groundspace_toriccode}. 

This is the case for all the  ground states  $\ket{01}_{\mathrm{tc}}$, $\ket{10}_{\mathrm{tc}}$ and $\ket{11}_{\mathrm{tc}}$ individually. They differ from $\ket{00}_{\mathrm{tc}}$ by the application of the string operator $w_{i}$ looping around the torus.
Since the derivation of  $\mathcal D_{\partial A}$ requires to trace all the qubits, and $w_{i} = \prod \sigma_{x}$, we have $w_{i}\bar{w}_{i} = I$ and the string operator  annihilates.
Thus, the contour diagram $\mathcal D_{\partial A}$ coincides for all the ground states. 

When considering a general superposition $\ket\psi$ of the form~\eqref{eq:groundspace_toriccode}, since all the states are  are orthogonal, $\Tr(\ket\psi\bra\psi) = \sum_{i,j} |c_{ij}|^2\Tr(\ket{ij}_{\mathrm{tc\,tc}}\hspace{-1mm}\bra{ij}))$. As each term in the sum leads to the same contour diagram $\mathcal D_{\partial A}$ of the form in Fig.~\ref{fig:rhoAb}c, it will also be the case for $\ket\psi$.

The same conclusion can be derived within the ZX formalism 
by applying exactly the same procedure discussed above. 
Specifically, we build Fig.~\ref{fig:xmastree}a this time using the generic diagram shown in Fig.~\ref{fig:toriccode_groundstate_complete}a. 
Using the inverse \FusionRule rule we double (pin) the green Z spiders and simplify, and obtain two disconnected diagrams.
The first diagram features the Pauli tree as in Figs.~\ref{fig:xmastree}b and c.
The second diagram is a result of simplifying the non-local input spiders in Fig.~\ref{fig:toriccode_groundstate_complete} and leads to multiplicative scalars. 
These scalars can be ignored, as they do not contribute to the structure of the diagram, and we obtain the same Pauli tree as in Figs.~\ref{fig:xmastree}b and c. See \textsc{ZXlive} file \textit{toric\_code/pauli\_tree\_superposition.zxp} in Ref.~\cite{Mas_mendoza_zenodo} for the proof.

\begin{figure}[t]
    \centering
    \hspace{-8.5cm}
    \includegraphics[width=\linewidth]{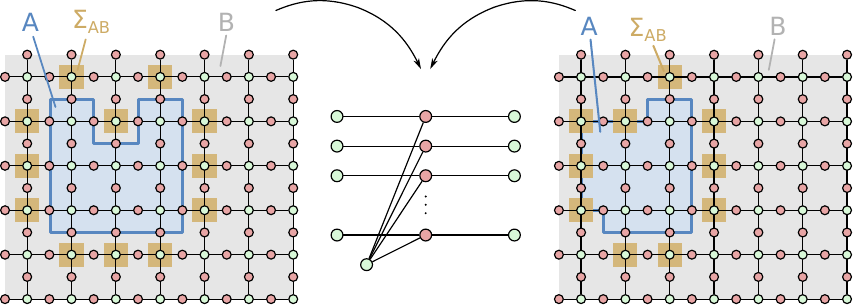}
    \caption{Robustness of the Pauli tree to boundary contour deformations. In both the left and right example, we consider  bipartitions between the $A$ and $B$ regions which deform the rectangular one chosen in Fig.~\ref{fig:xmastree}. However, an analogous simplification of the diagrams leads to a Pauli tree featuring a non-local green spider connected to $\Sigma_{AB}$ red spiders.	}
    \label{fig:robustness_xmas}
\end{figure}

{\it Arbitrary contours --}  In Fig.~\ref{fig:robustness_xmas}, we  provide two additional examples, where the boundary between the $A$ and $B$ regions has been deformed, also mixing $z$- and $x$-paths.
In both cases, the same steps leading to the Pauli tree in Fig.~\ref{fig:xmastree}c always lead to an equivalent  Pauli tree,
see \textsc{ZXLive} files \textit{toric\_code/pauli\_tree\_boundary1.zxp} and \textit{toric\_code/pauli\_tree\_boundary2.zxp} in Ref.~\cite{Mas_mendoza_zenodo} for a proof.

The same conclusion applies to the form of $\mathcal D_{\partial A}$ shown in Fig.~\ref{fig:rhoAb}c.
Regardless of the shape of the boundary, we always recover the diagram where the boundary plaquettes are attached to a single non-local spider, Fig.~\ref{fig:rhoAb}c.

We also stress that the form of the contour diagram $\mathcal D_{\partial A}$ and of the associated Pauli tree remain unchanged also in the case where region $A$ is non-contractible, namely when it winds around the torus. We remind the reader that also such bipartitioning gives rise to spurious contributions, see Section~\ref{sec:spurious}, including to Eq.~\eqref{eq:entropy_hamma} for the topological entanglement entropy~\cite{Hamma}.

These examples provide an additional confirmation of the robustness of the non-local spider in the contour diagram $\mathcal D_{\partial A}$ and related Pauli tree, as opposed to the entanglement-based diagnostics~\cite{Cano2015,Bravyi,Zou2016,Fliss2017,Santos2018,Williamson2019,Kato2020,Stephen2019}, see also discussion in Section~\ref{sec:TEE}.


\section{General recipe to derive the contour diagram $\mathcal D_{\partial A}$} \label{sec:general_framework}

The derivation of the Pauli trees from simplifying the contour diagram $\mathcal D_{\partial A}$ relies on the  knowledge of the fact that the topological entanglement entropy~\eqref{eq:entropy_hamma} is (\textit i)  directly related to the shared plaquette operators between regions $A$ and $B$ and (\textit{ii}) that,  in the ZX representation of the toric-code ground state, such plaquette operators are directly related to the green spiders in their middle. 

However, it is important to provide a diagnostic which does not rely on any preliminary knowledge about the state.
The contour diagram $\mathcal D_{\partial A}$ offers a more general, robust and potentially agnostic tool to diagnose long-range topological order, or its absence. The  advantage of $\mathcal D_{\partial A}$ is that it is conceived  to directly track entanglement, carried in ZX by connections among qubits. Such connections are kept by pinning the spiders corresponding to those qubits in $A$ and those in $B$ that share ZX connections.

In this Section, we give  the general recipe to derive the contour diagram $\mathcal D_{\partial A}$ and, based on its form, conjecture how to diagnose topological order. Given the ZX representation of a state $\ket\psi$:
\begin{enumerate}
    \item Define a bipartition $(A,B)$ of qubits and assign to each bipartition the corresponding  spiders in the ZX representation of $\ket\psi$;
    \item Construct the ZX representation of $\mbox{Tr}(\rho)$ and, before starting the simplification, pin the spiders corresponding to qubits which display ZX connections (which may include spiders  without output  wires) between regions $A$ and $B$;
    \item Simplify the diagram without fusing the pinned spiders;
    \item Interrogate the simplified diagram for non-local spiders connected to the pinned spiders.
\end{enumerate}

We conjecture that the non-local spiders connected to the pinned spiders found in this way on the simplified diagram diagnose the presence of long-range topological order.
For the examples we consider in this paper, the number of non-local spiders  coincides with the topological entanglement entropy $\gamma_\mathrm{top}$.  
As we will show below, the contour diagram so obtained correctly diagnoses long-range topological order without suffering from spurious contributions, also when we do not  input information about the stabilizers, see Section~\ref{sec:universality_gamma}.

\section{Extension to the Hexagonal Toric Code and the Color Code}\label{sec:hextoriccolorcode}

In this Section, we briefly introduce two models defined on hexagonal lattices, where we can define an extension of the toric code with the same topological entanglement entropy $\gamma_{\rm top}=1$, but also the color code~\cite{Bombin_colorcode}, which has a different topological entanglement entropy $\gamma_{\rm top}=2$~\cite{kargarian_tee_colorcode}. 
In Section \ref{sec:hexTC_CC_ZX}, we represent the ground states of these models using ZX diagrams and derive the corresponding Pauli trees.


%
\subsection{Hexagonal Toric Code}\label{sec:hextoriccode}

\begin{figure}
    \centering
    \hspace{-8.7cm}
    \includegraphics[width=\linewidth]{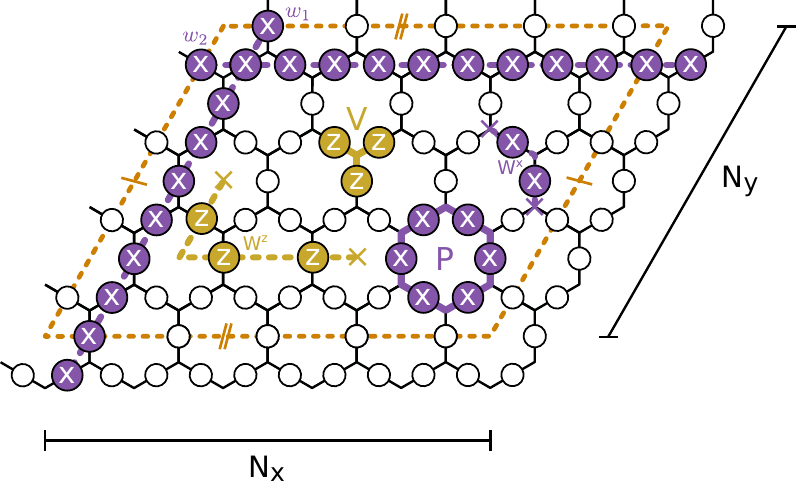}\quad
    \caption{
    Toric code on the hexagonal lattice. In full analogy to the toric code on the square lattice in Figs.~\ref{fig:toriccode_def} and~\ref{fig:toriccode_chainoperators}, we modify plaquette, vertex and non-contractible operators to obtain the same topological phase. 
    }
    \label{fig:hextoriccode_def}
\end{figure}

The hexagonal toric code is defined on a $N_{x} \times N_{y}$ hexagonal lattice, where $N_{x,y}$ label the number of hexagonal plaquettes in the $x$ and $y$ directions of the lattice,  see Fig.~\ref{fig:hextoriccode_def}.
As for the toric code on the square lattice, discussed in Section~\ref{sec:toriccode}, the qubits sit at each edge of a plaquette and the Hamiltonian has the  form~\eqref{eq:hamiltonian_toriccode},  namely
\begin{equation} \label{eq:hamiltonian_hextoriccode}
  H_{\mathrm{hTC}} = -\sum_{\Yup} V_{\Yup} -\sum_{\hexagon}P_{\hexagon}\,.
\end{equation}
The main difference consists in the form of the vertex and   plaquette operators, $V_{\Yup}$ and  $P_{\hexagon}$~\cite{Simon_TopologicalQuantum}: 
\begin{align} \label{eq:vertexop_def_hextoriccode}
    V_{\Yup} &= \prod_{i \in \Yup} \sigma_{z}^{(i)}\,, & P_{\hexagon} &= \prod_{i \in \hexagon} \sigma_{x}^{(i)}\,,
\end{align}
where $\Yup$ denotes any  (downward- or upward-pointing) vertex  and $\hexagon$ the hexagonal plaquettes. These  operators are schematically shown in Fig.~\ref{fig:hextoriccode_def}. 

As for the square toric code in Section~\ref{sec:toriccode}, the vertex and plaquette operators all commute with each other, they square to the identity and thus have $\pm 1$ eigenvalues. When assuming periodic boundary conditions, an additional constraint, equivalent to Eq.~\eqref{eq:constraints_operators_toriccode_P} in the square lattice, is enforced to the vertex and plaquette operators
\begin{align}\label{eq:restrictions_hextoriccode}
    \prod_{\Yup}V_{\Yup} &= I\,,&
    \prod_{\hexagon}P_{\hexagon} &= I\,.
\end{align}
As in the square toric code, the number of plaquette and vertex  operators commuting with the Hamiltonian equals the number of qubits in the systems. As a consequence, the constraint~\eqref{eq:restrictions_hextoriccode} leaves two  qubits unconstrained, which result in a $2^{2}=4$-fold  degeneracy of all the eigenstates.

The simplest ground state of $N=3N_xN_y$ qubits  has exactly the form of Eq.~\eqref{eq:groundstate00_toriccode} for the square lattice
\begin{equation}\label{eq:groundstate00_hextoriccode}
    \ket{00}_{\rm htc} = \frac{1}{2^{(n_p+1)/2}}\prod_{\hexagon} (I + P_{\hexagon})\ket{0}^{\otimes N}\,,
\end{equation}
with the difference that the plaquette operators $P_{\hexagon}$ are now defined on hexagonal plaquettes and   $n_p=N/3$ indicates the total number of plaquettes. The remaining three ground states are also generated by applying non-contractible string operator of the form~\eqref{eq:xstring_def}, looping around the whole system as shown in Fig.~\ref{fig:hextoriccode_def}.
Recall from Section~\ref{sec:gamma_toriccode} that the constrain $\prod_{\square}P_{\square} = I$ relates to $\gamma$.
Since in the case of the hexagonal toric code we  also have the condition $\prod_{\hexagon}P_{\hexagon} = I$, we  have $\gamma =\gamma_{\rm top}= 1$.
%


\subsection{Color Code}\label{sec:colorcode}

\begin{figure}
    \centering
    \hspace{-8.7cm}
    \includegraphics[width=\linewidth]{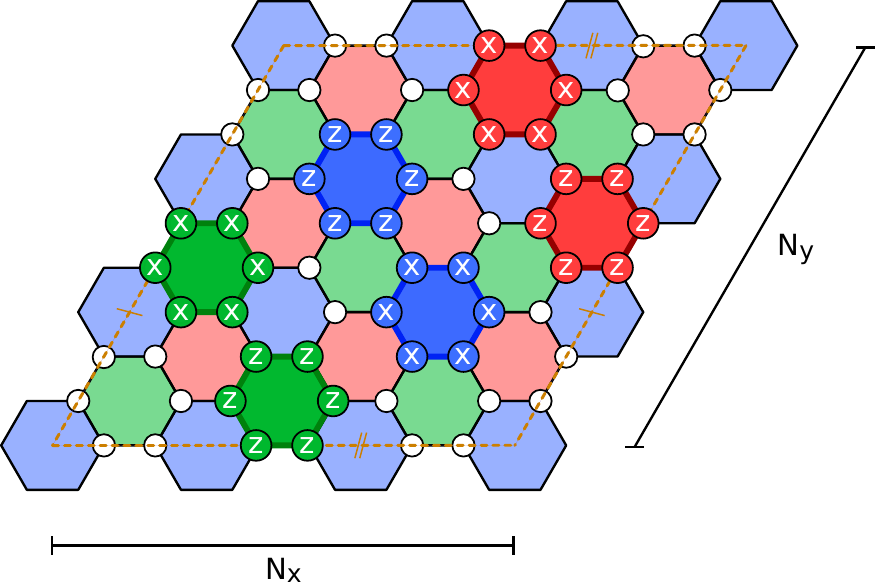}
    \caption{
    Schematic representation of the color code.
    The code is defined on an hexagonal lattice, where the qubits sit on the vertices (white circles). A  red, green and blue color is assigned to each hexagonal plaquette.
    The figure also depicts the six types of plaquette operators, defined in Eq.~\eqref{eq:operators_colorcode}, which are also colored according to the plaquette they act upon.}
    \label{fig:colorcode_def}
\end{figure}

The color code is another example of an exactly solvable model with topological order~\cite{Bombin_colorcode}, but with higher total
quantum dimension. 
It can be defined on any tri-colorable lattice and, without loss of generality, we focus here on the hexagonal one. This choice also allows us to directly  compare with the toric code on the same lattice.

The color code features three types of hexagonal plaquettes colored in red (R), green (G), and blue (B), see Fig.~\ref{fig:colorcode_def}.
We define the lattice to have $N_{x}\times N_{y}$ unit cells, with three plaquettes (one per colour) for each unit cell.
Differently from the toric code, the qubits are now placed on the  vertices of the lattice, and the code Hamiltonian
\begin{equation} \label{eq:hamiltonian_colorcode}
  H_{\mathrm{cc}} = -\sum_{\hexagon}\Big[B_{\hexagon}^{x}+B_{\hexagon}^{z}\Big],
\end{equation}
features exclusively  $x$- and $z$-plaquette operators
\begin{align}\label{eq:operators_colorcode}
        B_{\hexagon}^{x} &= \prod_{i \in \hexagon} \sigma_{x}^{(i)}\,,&
        B_{\hexagon}^{z} &= \prod_{i \in \hexagon} \sigma_{z}^{(i)}\,,
\end{align}
where $\hexagon$ denotes the set of qubits on the vertices of a specific plaquette.
The $x$- and $z$-plaquette operators are sketched in  Fig.~\ref{fig:colorcode_def}, and we assign to them  the color of the plaquette they act upon.

As for the toric code, the plaquette operators~\eqref{eq:operators_colorcode} square to the identity,  commute with each other and with the Hamiltonian~\eqref{eq:hamiltonian_colorcode}. They  are as many as the number of qubits in the system and, in the case of periodic boundary conditions, are subjected to four constraints~\cite{Bombin_colorcode}
\begin{subequations}\label{eq:restrictions_colorcode}
\begin{equation}\label{eq:restrctions_colorcode_x}
    \prod_{{\hexagon_R}}B_{{\hexagon_R}}^{x} = \prod_{{\hexagon_G}}B_{{\hexagon_G}}^{x} = \prod_{{\hexagon_B}}B_{{\hexagon_B}}^{x}\,,
\end{equation}
\begin{equation}\label{eq:restrctions_colorcode_z}
    \prod_{{\hexagon_R}}B_{{\hexagon_R}}^{z} = \prod_{{\hexagon_G}}B_{{\hexagon_G}}^{z} = \prod_{{\hexagon_B}}B_{{\hexagon_B}}^{z}\,,
\end{equation}
\end{subequations}
where the products run over the red, green and blue plaquettes, respectively. 
These four constraints leave as many qubits unconstrained, which results in a $2^{4}=16$-fold degeneracy of the states. 
The simplest ground state of the color code Hamiltonian~\eqref{eq:hamiltonian_colorcode} of $N = 6N_{x}N_{y}$ qubits has the same form as the ground state $\ket{00}_{\rm tc}$, Eq.~\eqref{eq:groundstate00_toriccode} of the toric code, namely
\begin{equation}\label{eq:groundstate0000_colorcode}
    \ket{0000}_{\rm cc} =\frac{1}{2^{(n_p+2)/2}} \prod_{\hexagon} \left(I + B_{\hexagon}^{x}\right)\ket{0}^{\otimes N} \,,
\end{equation}
with $n_p=N/2$  the total number of $x$ plaquette operators.

\begin{figure}
    \centering
    \hspace{-8.7cm}
    \includegraphics[width=\linewidth]{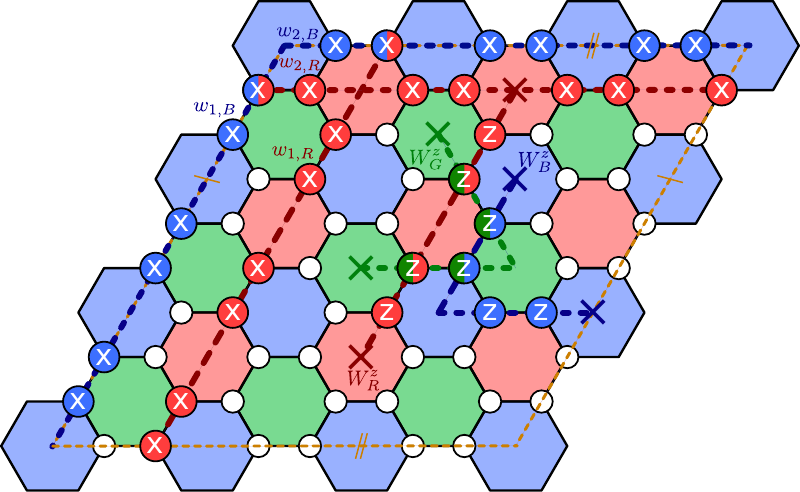}
    \caption{Schematic representation of the string operators $W_s^{x,z}$ of the color code, defined in Eq.~\eqref{eq:stringoperators_colorcode}. When these operators loop around the torus, we refer to them as $w_{p,s}$ operators, where $p\in\{1,2\}$  denotes the direction of the winding  ($p=1$ for $y$ and $p=2$ for $x$). The figure also illustrates an example showing how any string operator of a specific color (here green) can be constructed from string operators of the other two colors (here red and blue). 
 }
    \label{fig:colorcode_stringoperators}
\end{figure}

The procedure to construct the remaining 15 ground states is entirely similar to the one exposed in Section~\ref{sec:GSdegeneracytoric} for the toric code. The main difference resides in the colored nature of the code, which requires to distinguish among red, green and blue paths. As illustrated in Fig.~\ref{fig:colorcode_stringoperators}, these paths are defined over edges that connect centers of plaquettes of the same  color. Thus, we label the paths by the color of the plaquettes they connect: red $\mathcal C_{R}$, green $\mathcal C_{G}$ and $\mathcal C_{B}$ paths.

The $x$- and $z$-type string  operators $W^{x,z}_s$ are a product of Pauli $\sigma_{x,z}$ operators acting on any of the three colored paths $\mathcal C_{s}$, with $s \in \{R,G,B\}$,
\begin{equation}\label{eq:stringoperators_colorcode}
    W^{x,z}_s = \prod_{i\in \mathcal C_{s}} \sigma_{x,z}^{(i)}\,.
\end{equation}

However, we can always construct a string operator of one color as a product of the other two.
In Fig.~\ref{fig:colorcode_stringoperators}, we show an example by constructing a green string operator as a product of  red and  blue strings. 
Thus, in practice, there are two independent colors for each type of string operator.
Without loss of generality, we choose them to be red and blue,  leaving four types of independent string operators $W^{x}_B,\,W^{x}_R,\,W^{z}_B,\,W^{z}_R$.

As for the toric code in Section~\ref{sec:GSdegeneracytoric}, the other ground states are generated by applying these string operators along non-contractible paths wrapping around the torus.
We denote such  non-contractible string  operators $w_{p,s}$, where the  label $p\in\{1,2\}$  denotes the direction of the winding  ($p=1$ for $y$ and $p=2$ for $x$) and the label  $s$  the color, see Fig.~\ref{fig:colorcode_stringoperators}.
The sixteen possible ground states are thus generated by applying any of the $w_{p,s}$ to \eqref{eq:groundstate0000_colorcode}
\begin{equation}\label{eq:groundstates_basis_colorcode}
    \ket{i_{ 1}i_{2}j_{1}j_{2}}_{\rm cc} 
    = w_{1B}^{i_{1}}w_{1R}^{i_{2}}w_{2B}^{j_{1}}w_{2R}^{j_{2}}\ket{0000}_{\rm cc}\,,
\end{equation}
with $i_{1},i_{2},j_{1},j_{2} \in \{0,1\}$.

Notice that we can define other string operators, which are  products of $\sigma_{z}$.
When they act along non-contractible loops, they change the phase of the ground state adding a $\pm1$ depending on whether a non-contractible string operator $w_{p,s}$ has been applied to the state $\ket{0000}_{\rm cc}$.

We conclude our introduction to the color code mentioning that the expression~\eqref{eq:entropy_hamma} for the entanglement entropy $S=\Sigma_{AB}-\gamma$ extends to the color code, where $\Sigma_{AB}$ is the number of $x$-plaquette operators  acting on both bipartitions $A$ and $B$. The important difference with the toric code is that the color code has a higher total quantum dimension, corresponding to a topological entanglement entropy of $\gamma_\mathrm{top} = 2$ \cite{Bombin_colorcode, kargarian_tee_colorcode}. 
Recall from Sections~\ref{sec:gamma_toriccode} and~\ref{sec:hextoriccode} that the constraint $\prod_{\square}P_{\square} = I$ relates to $\gamma_\mathrm{top}=1$ for the toric code.
Since in the case of the color code we have two constraints~\eqref{eq:restrctions_colorcode_x}, $\gamma_{\rm top} = 2$.
%

\begin{figure*}
    \centering
    \hspace{-16.2cm}
    \includegraphics[width=0.9\linewidth]{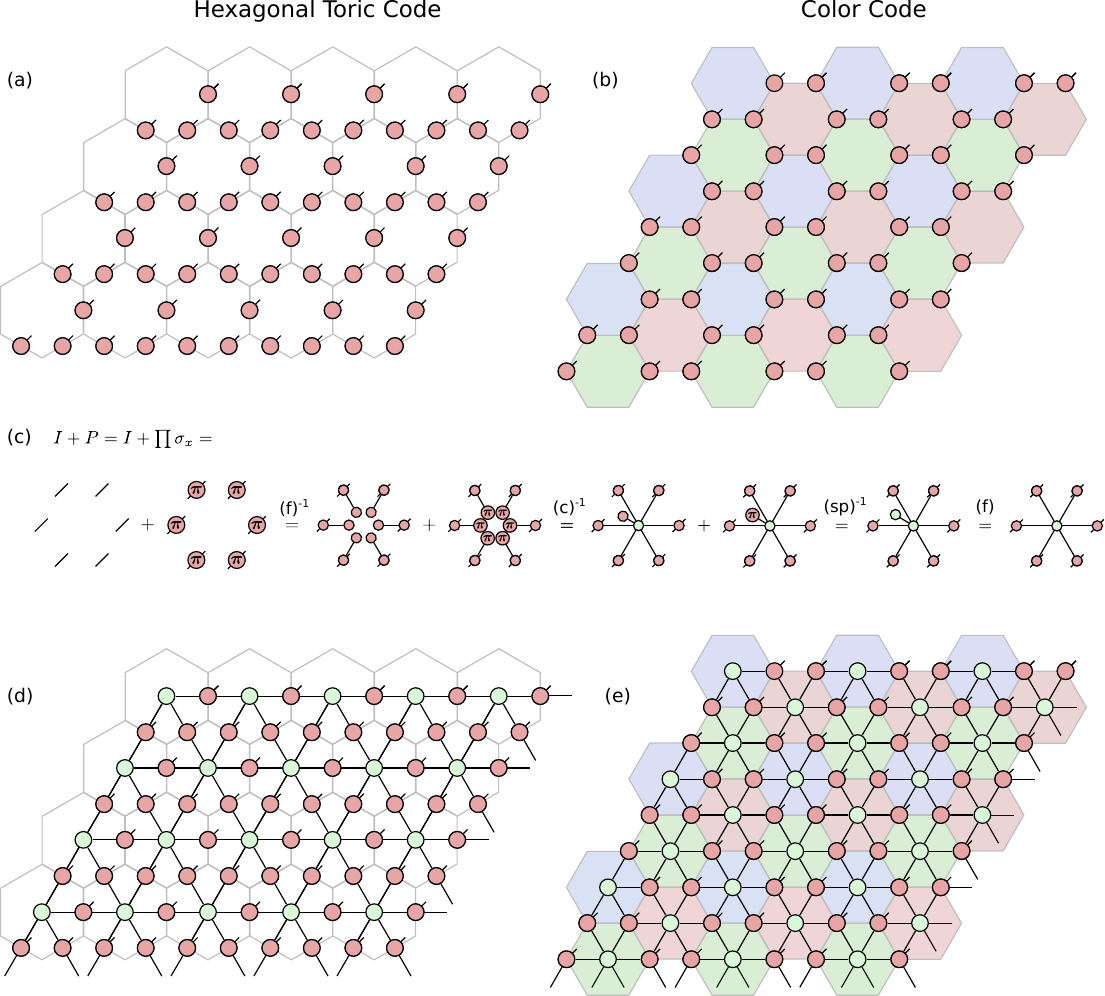}
    \caption{{ ZX diagrams for the hexagonal toric and color code ground states.} The procedure is entirely analogous to the one presented in Fig.~\ref{fig:toriccode_protocol}.  
    (a-b) Initialization state $\ket{0}^{\otimes N}$ for the hexagonal toric and color codes. We recall that $\ket 0$=  \begin{tikzpicture}
    \begin{pgfonlayer}{nodelayer}
        \protect\node [style=X dot] (0) at (-0.5, 0) {};
        \protect\node [style=none] (1) at (0.5, 0) {};
    \end{pgfonlayer}
    \begin{pgfonlayer}{edgelayer}
        \protect\draw (0) to (1.center);
    \end{pgfonlayer}
    \end{tikzpicture}, see  Eq.~\eqref{eq:0_zx}.
    The underlying hexagonal lattice is indicated in pale grey. 
    (c) ZX diagram for the Pauli projectors $I + P_{\hexagon}$ and $I + B_{\hexagon}^{x}$ acting on the hexagonal plaquettes of the toric and color code.  Their representation as ZX diagrams coincides, but they act on qubits either on the edges or vertices of the hexagonal lattice, respectively.
    (d-e) ZX diagrams for the $\ket{00}_{\mathrm{htc}}$ and $\ket{0000}_{\mathrm{cc}}$ ground states in Eqs.~\eqref{eq:groundstate00_hextoriccode} and ~\eqref{eq:groundstate0000_colorcode}, respectively.
    They result from applying the projector diagram in (c) to each  plaquette.
}
    \label{fig:hexTC_CC_protocol}
\end{figure*}

\section{Pauli trees in the hexagonal toric and color codes}\label{sec:hexTC_CC_ZX}

In this Section, we extend  the results of  Sections~\ref{sec:ToricCode_ZX} and \ref{sec:results} to the cases of the hexagonal toric and color codes. 
Our goal is to show that a generalized form of the Pauli tree, discussed in the main text and in Fig.~\ref{fig:xmastree}, distinguishes robustly among different topological orders. More precisely, the number of non-local green spiders connected to the boundary equals the topological entanglement entropy $\gamma_{\rm top}$, regardless of microscopic details. We show that  the  Pauli trees in the hexagonal toric code feature  one  non-local green spider, reflecting the topological entanglement entropy $\gamma_{\rm top}=1$. For the color code, we find instead a generalized form of the Pauli tree with $two$ non-local green spiders, corresponding to the value of its topological entanglement entropy $\gamma_{\rm top}=2$.
We will restrict our discussion to the Pauli tree representation of the contour diagrams here, as they carry equivalent information compared to more general contour diagrams, see the case of the square toric code discussed in Section \ref{sec:results}.

\subsection{ZX diagrams for the ground states}\label{sec:zxhexcc}

The ground state $\ket{00}_{\rm tc}$ of the toric code on the square lattice, Eq.~\eqref{eq:groundstate00_toriccode}, has a very similar form as  the ground states of the hexagonal toric and color codes -- Eqs.~\eqref{eq:groundstate00_hextoriccode} and~\eqref{eq:groundstate0000_colorcode} respectively. The only difference resides in substituting the square plaquettes operators $P_\square$, see Eq.~\eqref{eq:plaquetteop_def_toriccode}, with their hexagonal counterparts $P_{\hexagon}$ and $B^x_{\hexagon}$ defined in Eqs.~\eqref{eq:vertexop_def_hextoriccode} and~\eqref{eq:operators_colorcode}. Thus, the ZX diagram   which represents the $\ket{00}_{\rm tc}$ state on the square lattice in Fig.~\ref{fig:toriccode_protocol}  can be readily modified to obtain  the ZX diagrams for the ground states $\ket{00}_{\rm htc}$ and $\ket{0000}_{\rm cc}$ of the hexagonal toric and color codes~\cite{kissinger2022phasefreezxdiagramscss,colorcode_zx_arxiv}, see Fig.~\ref{fig:hexTC_CC_protocol}.

The ZX diagrams representing $\ket{00}_{\rm htc}$ and $\ket{0000}_{\rm cc}$ visually show  the structural differences between these states. The spider connections in  $\ket{00}_{\rm htc}$ follow a triangular lattice, which is dual to the hexagonal lattice. For  $\ket{0000}_{\rm cc}$, they reproduce instead a dice lattice. This structural difference corresponds to a different configuration of entanglement, which is diagnosed by the emergence of different Pauli trees, as we discuss shortly. 

\begin{figure}
    \centering
    \hspace{-8.5cm}
    \includegraphics[width=\linewidth]{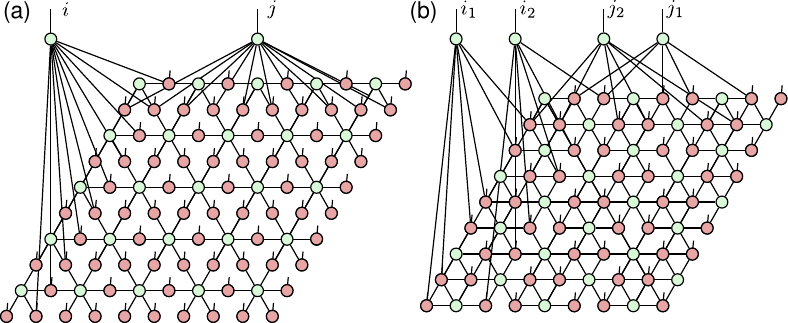}
    \caption{{ ZX diagram for generic ground states of the hexagonal toric and color code}. 
    This figure shows the generalization of Fig.~\ref{fig:toriccode_groundstate_complete}a to the hexagonal toric code  (a) and to the color code (b).
    We omit the connections implementing periodic boundary conditions for clarity.
    As for the toric code, we add 2 and 4 non-local input spiders to the $\ket{00}_{\mathrm{htc}}$ and $\ket{0000}_{\mathrm{cc}}$ diagrams in Fig.~\ref{fig:hexTC_CC_protocol}d and e, respectively.
    These non-local input spiders connect to the position of the non-contractible string operators in Figs.~\ref{fig:hextoriccode_def} and~\ref{fig:colorcode_stringoperators}.
    To these non-local input spiders, we input specific combinations of red and green spiders, or generic diagrams, as in Fig.~\ref{fig:toriccode_groundstate_complete}b-d, to select a specific ground state, or a linear combination.
    For a linear combination we would need 4 H-boxes in the hexagonal toric code and 16 for the color code.
    We omit this part of the diagram.
    }
    \label{fig:hextoriccode_generalgroundstates_zx}
\end{figure}

For completeness, Fig.~\ref{fig:hextoriccode_generalgroundstates_zx}  shows the procedure to construct in ZX the additional 3 and 15 ground states of the hexagonal toric and of the color code, and of  generic superpositions of them. 
As is Fig.~\ref{fig:toriccode_groundstate_complete}a, we need to add non-local input spiders to the ground state diagrams representing  $\ket{00}_{\mathrm{htc}}$ and $\ket{0000}_{\mathrm{cc}}$.
To build  specific states, or some superposition of them, one has to input the corresponding  combinations of red and green spiders shown in Fig.~\ref{fig:toriccode_groundstate_complete}b-d.

\subsection{{Numerical benchmarks}}

To  benchmark the validity  of our diagrammatic  calculations in ZX, we have  performed the same numerical analysis  of Section~\ref{sec:results_gamma_toriccode} for the hexagonal toric and color codes.  We have considered the bipartitions shown in Figs.~\ref{fig:htc_results}a and c. 
To benchmark them, we numerically verified that $S=\Sigma_{AB}-\gamma$, just as for the toric code, see Eq.~ \eqref{eq:entropy_hamma}.
Here, $\Sigma_{AB}$ counts the number of hexagonal and $x$ hexagonal plaquette operators shared between the bipartitions $A$ and $B$ for the hexagonal toric code and the color codes, respectively. 
We obtain $\gamma=\gamma_\mathrm{top}=1$  for the hexagonal toric code and $\gamma=\gamma_\mathrm{top}=2$ for the color code, which distinguishes the topological class of both models.


\subsection{The Pauli tree distinguishes the topological order of the toric code from the color code}

\begin{figure*}[t]
    \centering
    \hspace{-18cm}
    \includegraphics[width=\linewidth]{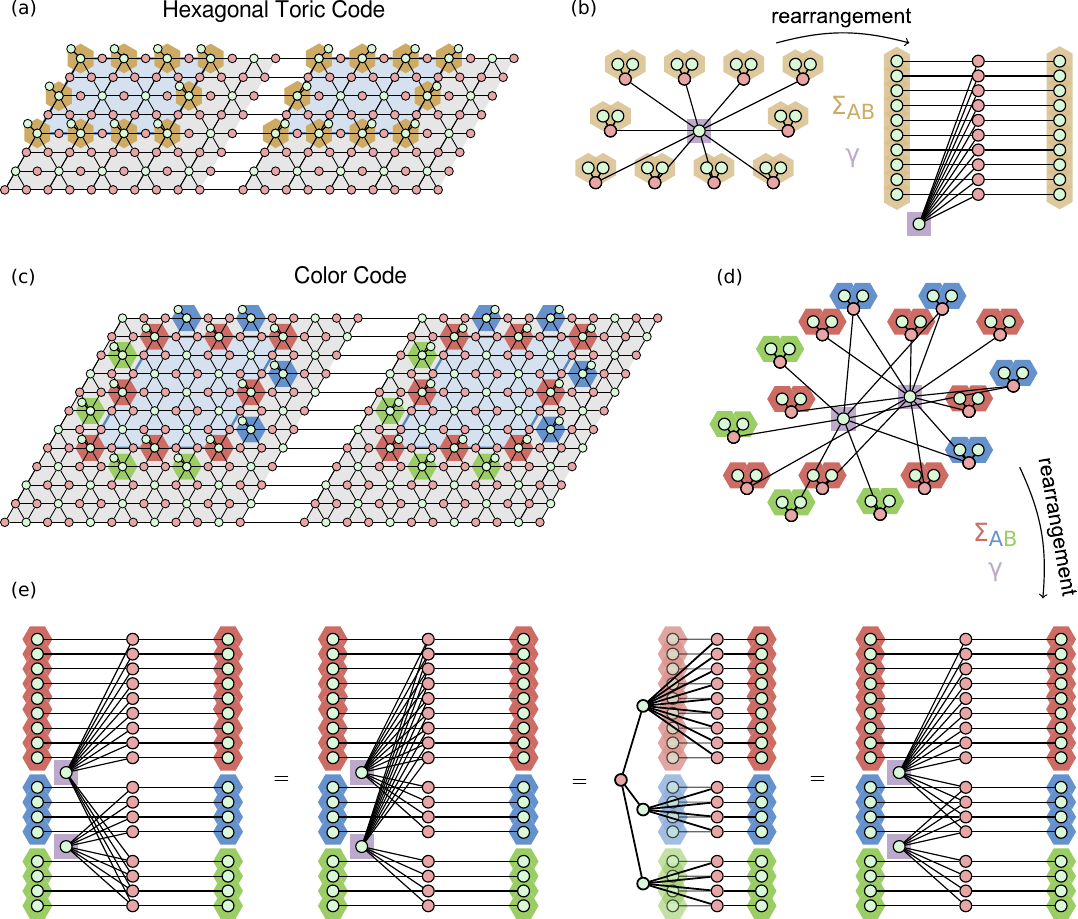}
    \caption{
    Pauli trees in the hexagonal toric (a-b) and color codes (c-e). 
    (a)  ZX representation of $\mbox{Tr}(\rho)$ with pinned plaquettes as in Fig.~\ref{fig:xmastree}a for the $\ket{00}_{\rm htc}$ ground state of the hexagonal toric code,  Eq.~\eqref{eq:groundstate00_hextoriccode}.
    We doubled (pinned) all green Z spiders  at the center of the $\Sigma_{AB}$ boundary plaquettes using the inverse \FusionRule rule. 
    (b) ZX-simplified version of the diagram in (a), leading to the same Pauli tree as in Fig.~\ref{fig:xmastree}.
    The doubled green Z spiders pinning the green spiders in panel (a) are connected as inputs and outputs of the Pauli projector. 
    (c) Same as (a) but for  the $\ket{0000}_{\rm cc}$ ground state of the color code, Eq.~\eqref{eq:groundstate0000_colorcode}.
    (d) ZX-simplified diagram in panel (c), leading to a generalized Pauli tree.
    This diagram features {\it two} non-local green spiders which cross couple to the spiders associated to the plaquette of a specific color.
    (e) The rightmost diagram is a rearrangement of panel (d).
    This diagram can be symmetrized by introducing  one red X spider connected to three green Z spiders, one for each color, leading to the second diagram from the right.
    From this symmetrized form, we can obtain any other permuted diagram (the two diagrams from the left) by applying the bialgebra \BialgRule rule.
   }
    \label{fig:htc_results}
\end{figure*}

Figure~\ref{fig:htc_results} shows the generalization of Fig.~\ref{fig:xmastree} from the toric code on the square lattice to the  toric and to the color codes on the hexagonal lattice.  The details of the derivation of these diagram can be found in the  \textsc{ZXLive} files \textit{hexagonal\_toric\_color\_codes/pauli\_tree\_htc.zxp} and \textit{hexagonal\_toric\_color\_codes/pauli\_tree\_cc.zxp} in Ref.~\cite{Mas_mendoza_zenodo}.

In the case of the hexagonal toric code, see Fig.~\ref{fig:htc_results}a-b, we isolate exactly the same  Pauli  tree of the square toric code in  Fig.~\ref{fig:xmastree}. It features {\it one} single green non-local spider coupled to the red spiders related to  the boundary  separating the bipartition between $A$ and $B$. This allows us to conclude that the Pauli tree is robust to microscopic (lattice) deformations of the model and that the number of non-local green spiders is associated to the long-range topological order.

This conjecture is further substantiated by looking at the results of the same procedure when applied to the color code, which is shown in Fig.~\ref{fig:htc_results}c-e. In this case, we obtain a different diagram, where {\it two} non-local green spiders  connect  to the red spiders  associated to the plaquettes of a specific color (say blue in Fig.~\ref{fig:htc_results}d), and  each one  individually to the plaquettes of the two remaining colors (say green and red in Fig.~\ref{fig:htc_results}d). The symmetry among colors and connections can be restored by applying the inverted $\FusionRule$ and $\BialgRule$ rules, leading to a less compact diagram with additional green and red spiders that we show in Fig.~\ref{fig:htc_results}e.
This diagram mediates the permutation of the connections among colors.

We stress again that we could have equally focused on the representation of the contour diagram $\mathcal D_{\partial A}$ that chooses to pin {spiders associated to }the qubits instead of the plaquettes. As in the case of the toric code on the square lattice discussed in Section~\ref{sec:rhoAb}, it contains equivalent information about the long-range topological order of the state.


\section{Absence of spurious contributions to the contour diagram of trivial states}  \label{sec:universality_gamma}

\begin{figure*}[t]
    \centering
    \hspace{-18cm}
    \includegraphics[width=\linewidth]{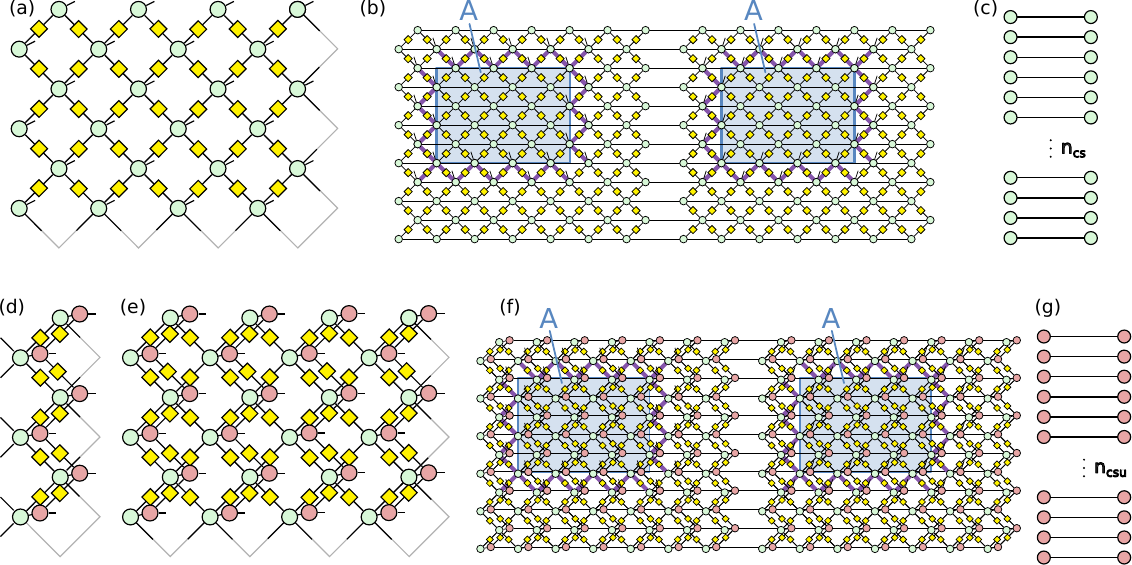}
    \caption{
    Cluster states and related contour diagrams $\mathcal D_{\partial A}$ in ZX.
    (a) ZX diagram of the cluster state introduced in Section~\ref{sec:clusterstate}.
    We initialize the qubits in the $\ket{+} =$
\begin{tikzpicture}
	\begin{pgfonlayer}{nodelayer}
		\protect\node [style=Z dot] (0) at (0, 0) {};
		\protect\node [style=none] (1) at (0, 0.5) {};
	\end{pgfonlayer}
	\begin{pgfonlayer}{edgelayer}
		\protect\draw (1.center) to (0);
	\end{pgfonlayer}
\end{tikzpicture} 
state and apply a  CZ =
\begin{tikzpicture}
	\begin{pgfonlayer}{nodelayer}
		\protect\node [style=Z dot] (0) at (0.5, 0) {};
		\protect\node [style=none] (1) at (0.5, 0.5) {};
		\protect\node [style=Z dot] (2) at (2, 0) {};
		\protect\node [style=none] (3) at (2, 0.5) {};
		\protect\node [style=none] (4) at (2, -0.5) {};
		\protect\node [style=none] (5) at (0.5, -0.5) {};
		\protect\node [style=hadamard] (6) at (1.25, 0) {};
		\protect\node [style=hadamard] (7) at (1.25, 0) {};
	\end{pgfonlayer}
	\begin{pgfonlayer}{edgelayer}
		\protect\draw (1.center) to (0);
		\protect\draw (0) to (5.center);
		\protect\draw (3.center) to (2);
		\protect\draw (2) to (4.center);
		\protect\draw [style=none] (2) to (7);
		\protect\draw [style=none] (7) to (0);
	\end{pgfonlayer}
\end{tikzpicture} gate to each pair of neighboring qubits.
    (b) Contour diagram $\mathcal D_{\partial A}$  for the cluster state in (a), before simplification. We highlight the segments connecting $A$ and $B$ as thick purple lines. These segments define which $n_{\mathrm{cs}}$ qubits we pin. To keep the diagram readable, we adopt the convention
    of marking the pinned boundary qubits by input wires, instead of unfusing the green spiders.
    These inputs should be thought of as having an additional green Z spider at their end in (b) and red X spider in (f).
    We highlight $A$ in blue while keeping $B$ in white. 
    (c) Diagram obtained after simplifying the diagram in (b) leading to the identity $I$  operator with $n_{\mathrm{cs}} = 28$ lines between boundary spiders.
    (d) Vertical line of unitaries $(H \otimes H)CZ(H \otimes H)$ as in Eq.~\eqref{eq:HHCZHH} applied to the diagram of the cluster state in~\ref{fig:clusterstates}a.
    (e) Application of the vertical line of unitaries in (d) to  every vertical line of the cluster state.
    (f) Contour diagram $\mathcal D_{\partial A}$ for the cluster state in panel (e), before simplification.
    (g) Diagram obtained after simplifying the diagram in (f) leading to the same trivial diagram as in (c) with $n_\mathrm{csu}$ lines.
    }
    \label{fig:clusterstates}
\end{figure*}

As discussed in Section~\ref{sec:spurious}, a major issue concerning the diagnosis of topological order via entanglement measures is the presence of spurious contributions to the subleading correction to the area law. Because of these spurious contributions $\gamma=\gamma_\mathrm{top}+\gamma_{\rm spur}$, and thus $\gamma$ can be finite even in the absence of topological order. 

A notable example is the cluster state discussed in Ref.~\cite{Williamson2019}, which is  topologically trivial, but yet generates $\gamma_{\mathrm{spur}} > 0$. We are going to show that the ZX contour diagram $\mathcal D_{\partial A}$, derived for the cluster state according to the prescriptions given in Section~\ref{sec:general_framework},   does not feature non-local spiders, supporting that  $\mathcal D_{\partial A}$ correctly diagnoses  the presence or absence of long-range entanglement. 

\subsection{Cluster state} \label{sec:clusterstate}
The cluster state is defined as follows.
The qubits sit on the vertices of a square lattice in a diamond fashion as in Fig.~\ref{fig:clusterstates}a.
We initialize each of the qubits in the $\ket +=(\ket0+\ket1)/\sqrt2$ state.
Recall from Eq.~\eqref{eq:+_zx} in Section~\ref{sec:intro_ZX} that in ZX calculus $\ket{+} =$
\begin{tikzpicture}
	\begin{pgfonlayer}{nodelayer}
		\node [style=Z dot] (0) at (0, 0) {};
		\node [style=none] (1) at (0, 0.5) {};
	\end{pgfonlayer}
	\begin{pgfonlayer}{edgelayer}
		\draw (1.center) to (0);
	\end{pgfonlayer}
\end{tikzpicture}.
Then, we apply a controlled Z (CZ) gate to each pair of neighbouring qubits.
Recall from Eq.~\eqref{eq:CZZX} that CZ =
\begin{tikzpicture}
	\begin{pgfonlayer}{nodelayer}
		\node [style=Z dot] (0) at (0.5, 0) {};
		\node [style=none] (1) at (0.5, 0.5) {};
		\node [style=Z dot] (2) at (2, 0) {};
		\node [style=none] (3) at (2, 0.5) {};
		\node [style=none] (4) at (2, -0.5) {};
		\node [style=none] (5) at (0.5, -0.5) {};
		\node [style=hadamard] (6) at (1.25, 0) {};
		\node [style=hadamard] (7) at (1.25, 0) {};
	\end{pgfonlayer}
	\begin{pgfonlayer}{edgelayer}
		\draw (1.center) to (0);
		\draw (0) to (5.center);
		\draw (3.center) to (2);
		\draw (2) to (4.center);
		\draw [style=none] (2) to (7);
		\draw [style=none] (7) to (0);
	\end{pgfonlayer}
\end{tikzpicture}.
Thus, when applied to a pair of qubits of a square plaquette, CZ$(\ket{+}\otimes\ket{+}) =$
\begin{equation} \label{eq:step_cluster}
    \centering
    \includegraphics[width=0.5\linewidth]{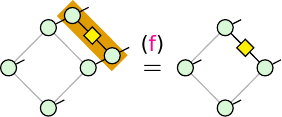}\,.
\end{equation}
By  repeating Eq.~\eqref{eq:step_cluster} for every edge, we obtain the diagram of the cluster state, shown in Fig.~\ref{fig:clusterstates}a.

Applying a tensor product of CZ gates to a product state is a constant depth unitary circuit~\cite{Williamson2019}. 
Thus, the state remains topologically trivial ($\gamma_\mathrm{top} = 0$).
However, different direct calculations of the topological entanglement entropy $\gamma$ leads to $\gamma = \gamma_\text{spur}\neq 0$, in direct contradiction with the triviality of the state.
For example, using finite-size scaling on an infinite cylinder (non-contractible) bipartition leads to $\gamma = \gamma_\text{spur}= 1$~\cite{Williamson2019}.
We have confirmed that this is indeed the case by calculating the entropy using ZX calculus directly.
Additionally, Ref.~\cite{Williamson2019} also shows that using the Kitaev-Preskill prescription~\cite{KitaevPreskill2006} also leads to spurious results.
These results show that some entanglement cuts of such trivial system lead to $\gamma=\gamma_{\mathrm{spur}} > 0$ instead of $\gamma=0$.

As discussed in the main text, we show now that the contour diagram $\mathcal D_{\partial A}$, introduced in Section~\ref{sec:rhoAb}, does not suffer from such spurious contributions.
Specifically, $\mathcal D_{\partial A}$ does not feature non-local spiders, independently of the chosen bipartition, consistent with the absence of long-range topological order.
We present the case of the contractible bipartition depicted in Fig.~\ref{fig:clusterstates}b, but we have checked that our results also apply to the non-contractible cylinder bipartition. This latter result can be also viewed as a consequence of the case discussed in Section \ref{sec:bravyi}.

Recall that once we define the bipartition $(A,B)$, we focus on connections between the two partitions $A$ and $B$.
Differently from the cases of the square and hexagonal topological codes, where the connections between the two regions were ``+''- and  ``\raisebox{-1ex}{\large *}''-shaped,
in the case of the cluster state they are simple lines, indicated in  Fig.~\ref{fig:clusterstates}b as thick purple lines.
We proceed by doubling (pinning) the two green spiders at their ends using the inverse \FusionRule rule.
To keep the diagram readable, especially considering the following example, we do not show the doubled green Z spiders but rather indicate them as inputs. 
These inputs should be thought of as having an additional green Z spider at their end in Fig.~\ref{fig:clusterstates}b.
We denote the number of unfused spiders in one of the conjugated copies of the state as $n_{cs}$. 
Hence we have $2n_{cs}$ in total.
In the particular case of Fig.~\ref{fig:clusterstates}b, $n_{cs} = 28$. 

The resulting contour diagram $\mathcal D_{\partial A}$ obtained from simplifying Fig.~\ref{fig:clusterstates}b is shown in Fig.~\ref{fig:clusterstates}c. We observe $n_{\mathrm{cs}}$ pairs of green Z spiders, each one corresponding to the unfused green Z spiders in~\ref{fig:clusterstates}b marked by an input wire. These are connected together via a trivial operator proportional to the identity, which as a ZX diagram is given by a collection of $n_\mathrm{sc}$ disconnected lines without non-local spiders. See the  \textsc{ZXLive} file \textit{cluster\_states/cluster\_state.zxp}  in Ref.~\cite{Mas_mendoza_zenodo} for the explicit proof. We stress that we obtain the same result also for the case when $A$ is non-contractible and winds around an infinite cylinder, see Section~\ref{sec:bravyi}.
We interpret this lack of non-local connections as a direct diagnose of the absence of long range-topological order.


\subsection{Robustness of the the contour diagram $\mathcal D_{\partial A}$ against local unitaries} \label{sec:clusterstate_unitaries}

As a further demonstration of the robustness of the ZX representation of the contour diagram $\mathcal D_{\partial A}$ to diagnose presence or absence of topological order, we repeat the previous calculation but after applying a layer of local unitary gates to the system. 
This example is interesting because such operations modify the value of $\gamma$ obtained through entanglement entropy calculations~\cite{Williamson2019}.
However, we are going to show that the contour diagram $\mathcal D_{\partial A}$ does not change when we apply such local unitaries.  

Following Ref.~\cite{Williamson2019}, we apply unitary gates of the form $(H \otimes H)CZ(H \otimes H)$, to pairs of qubits that are nearest neighbors along vertical lines
through the lattice~\cite{Williamson2019}, see Fig.~\ref{fig:clusterstates}d.
As ZX diagrams, $(H \otimes H)CZ(H \otimes H) =$
\begin{equation}
\label{eq:HHCZHH}
\begin{tikzpicture}
	\begin{pgfonlayer}{nodelayer}
		\node [style=Z dot] (0) at (0.5, 0) {};
		\node [style=Z dot] (1) at (2, 0) {};
		\node [style=hadamard] (2) at (2, 0.5) {};
		\node [style=hadamard] (3) at (2, -0.5) {};
		\node [style=hadamard] (4) at (0.5, -0.5) {};
		\node [style=hadamard] (5) at (0.5, 0.5) {};
		\node [style=none] (6) at (2, 1) {};
		\node [style=none] (7) at (0.5, 1) {};
		\node [style=none] (8) at (0.5, -1) {};
		\node [style=none] (9) at (2, -1) {};
		\node [style=hadamard] (10) at (1.25, 0) {};
		\node [style=none] (11) at (5.5, 1) {};
		\node [style=none] (12) at (4, 1) {};
		\node [style=none] (13) at (4, -1) {};
		\node [style=none] (14) at (5.5, -1) {};
		\node [style=X dot] (15) at (4, 0) {};
		\node [style=X dot] (16) at (5.5, 0) {};
		\node [style=hadamard] (17) at (4.75, 0) {};
		\node [style=none] (18) at (3, 0) {=};
		\node [style=none] (19) at (3, 0.5) {$\hcolorchange$};
	\end{pgfonlayer}
	\begin{pgfonlayer}{edgelayer}
		\draw [style=none] (6.center) to (2);
		\draw [style=none] (2) to (1);
		\draw [style=none] (1) to (3);
		\draw [style=none] (3) to (9.center);
		\draw [style=none] (4) to (8.center);
		\draw [style=none] (0) to (4);
		\draw [style=none] (5) to (0);
		\draw [style=none] (7.center) to (5);
		\draw [style=none] (0) to (10);
		\draw [style=none] (1) to (10);
		\draw [style=none] (12.center) to (15);
		\draw [style=none] (15) to (17);
		\draw [style=none] (17) to (16);
		\draw [style=none] (16) to (11.center);
		\draw [style=none] (16) to (14.center);
		\draw [style=none] (15) to (13.center);
	\end{pgfonlayer}
\end{tikzpicture}. 
\end{equation}

While these local unitaries change $\gamma$~\cite{Williamson2019}, as we have confirmed with a direct entropy calculation using the ZX calculus, when we calculate the contour diagram $\mathcal D_{\partial A}$, we find again a trivial $\mathcal D_{\partial A}$, see Figs.~\ref{fig:clusterstates}f-g. See also the simplification in the \textsc{ZXLive} file \textit{cluster\_states/cluster\_state\_unitaries.zxp} in Ref.~\cite{Mas_mendoza_zenodo}.
The contour diagram still lacks the non-local spiders that we obtained for the toric and color codes, indicating that the cluster state is a trivial state, despite the fact that entanglement entropy diagnostics suffer from $\gamma_\mathrm{spur} \neq 0$.
Hence our diagrammatic diagnostic of topological order appears immune to spurious effects when computing $\gamma$.

\subsection{Bravyi's state} \label{sec:bravyi}

\begin{figure*}[t]
    \centering
    \hspace{-18cm}
    \includegraphics[width=\linewidth]{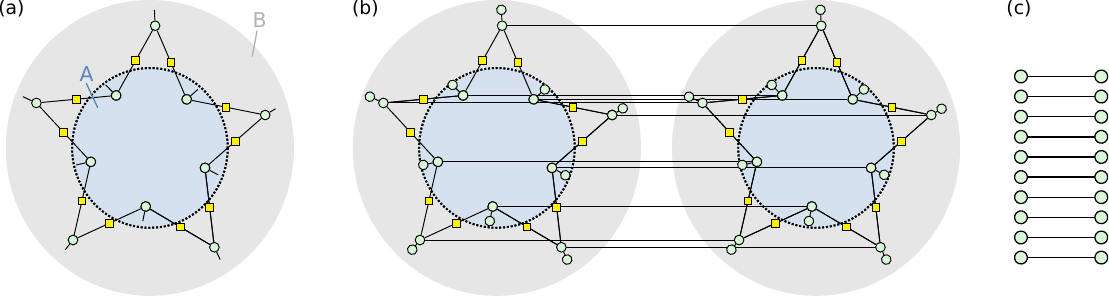}
    \caption{ 
    ZX diagram and contour diagram $\mathcal D_{\partial A}$ representing Bravyi's state.Bravyi
    (a) ZX diagram of the Bravyi state. 
    To generate it, we initialize the qubits in a chain to $\ket{+} =$
\begin{tikzpicture}
	\begin{pgfonlayer}{nodelayer}
		\protect\node [style=Z dot] (0) at (0, 0) {};
		\protect\node [style=none] (1) at (0, 0.5) {};
	\end{pgfonlayer}
	\begin{pgfonlayer}{edgelayer}
		\protect\draw (1.center) to (0);
	\end{pgfonlayer}
\end{tikzpicture} and apply a  CZ =
\begin{tikzpicture}
	\begin{pgfonlayer}{nodelayer}
		\protect\node [style=Z dot] (0) at (0.5, 0) {};
		\protect\node [style=none] (1) at (0.5, 0.5) {};
		\protect\node [style=Z dot] (2) at (2, 0) {};
		\protect\node [style=none] (3) at (2, 0.5) {};
		\protect\node [style=none] (4) at (2, -0.5) {};
		\protect\node [style=none] (5) at (0.5, -0.5) {};
		\protect\node [style=hadamard] (6) at (1.25, 0) {};
		\protect\node [style=hadamard] (7) at (1.25, 0) {};
	\end{pgfonlayer}
	\begin{pgfonlayer}{edgelayer}
		\protect\draw (1.center) to (0);
		\protect\draw (0) to (5.center);
		\protect\draw (3.center) to (2);
		\protect\draw (2) to (4.center);
		\protect\draw [style=none] (2) to (7);
		\protect\draw [style=none] (7) to (0);
	\end{pgfonlayer}
\end{tikzpicture} gate  to each pair of neighboring qubits.
    (b) Contour diagram $\mathcal D_{\partial A}$ for the cluster state in (a).
    We highlight region $A$ in blue, and region $B$ in gray.
    (c) The simplified diagram of (b) results in a trivial contour diagram $\mathcal D_{\partial A}$.}
    \label{fig:bravyi}
\end{figure*}

In Sections~\ref{sec:clusterstate} and~\ref{sec:clusterstate_unitaries}, we discussed examples of trivial states that should be topologically trivial ($\gamma = 0$) but actually give $\gamma > 0$.
In this Section, we  discuss a last example of a pathological cluster state without long-range topological order, but leading to a spurious contribution such that $\gamma=\gamma_{\rm spur} = 1$. This state was the first one found to display spurious contributions. It was introduced by Bravyi, remaining unpublished until it was discussed in Ref.~\cite{Zou2016}.

As sketched in Fig.~\ref{fig:bravyi}a with a ZX diagram, the Bravy state is a one dimensional cluster state embedded in an otherwise trivial two-dimensional system.
Hence, as a two-dimensional state, it is not translational invariant. 
Choosing a bipartition ($A,B$) cutting along this cluster state, see Fig.~\ref{fig:bravyi}a, leads to spurious contributions, even though $\gamma_{\mathrm{top}} = 0$~\cite{Bravyi,Zou2016}.
Specifically, the bipartition in Fig.~\ref{fig:bravyi}a leads to $\gamma = 1$~\cite{Zou2016}.
Alternatively, we could think of the boundary of this bipartition as a zig-zag vertical line looping around the torus of the cluster state in Fig.~\ref{fig:clusterstates}a when considering a cylinder bipartition.

As shown in Fig.~\ref{fig:bravyi}b-c, the contour diagram $\mathcal D_{\partial A}$ associated to this bipartition is once again a trivial ZX diagram, without non-local spiders. Thus, also in this case, the contour diagram correctly diagnoses the absence of long-range topological order. The proof can be found in the \textsc{ZXLive} file \textit{cluster\_states/bravyi.zxp} in Ref.~\cite{Mas_mendoza_zenodo}.


\end{document}